\documentclass[aps,onecolumn,showpacs]{revtex4}
\usepackage{dcolumn}
\usepackage{graphicx}
\usepackage{amsmath}
\usepackage{amsfonts}
\usepackage{amssymb}
\usepackage{psfrag}
\usepackage{wrapfig}
\usepackage{subfigure}
\usepackage{makeidx}
\usepackage{bm}
\usepackage{epsf}

\newtheorem{lem}{Lemma}
\newtheorem{prop}{Proposition}

\begin{document}
\title{Multi-soliton, multi-breather and higher order rogue wave solutions to
the complex short pulse equation}
\author{Liming Ling$^{1}$}
\author{Bao-Feng Feng$^{2}$}
\email{baofeng.feng@utrgv.edu, linglm@scut.edu.cn, znzhu@sjtu.edu.cn}
\author{Zuonong Zhu$^{3}$}
\affiliation{$^1$Department of Mathematics, South China University of
Technology,Guangzhou 510640, China}
\affiliation{$^2$ School of Mathematical and Statistical Sciences, The University of
Texas Rio Grande Valley, Edinburg Texas, 78541, USA}
\affiliation{$^3$Department of Mathematics, Shanghai Jiaotong University, Shanghai, China}

\begin{abstract}
In the present paper, we are concerned with the general localized solutions for the complex short pulse equation
including soliton, breather and rogue wave solutions. With the aid of a generalized Darboux transformation, we construct the $N$-bright soliton solution in a compact determinant form,
%, which is equivalent to the one constructed by one of the authors in ({\bf Physica D 297 (2015) 62-75 }) in the form of paffians.
then the $N$-breather solution including the Akhmediev breather and a general higher order rogue wave solution. The first- and second-order rogue wave solutions are given explicitly and illustrated by graphs.
The asymptotic analysis is performed rigourously for both the $N$-soliton and the $N$-breather solutions.
All three forms of the localized solutions admit either smoothed-, cusped- or looped-type ones for the CSP equation
depending on the parameters. It is noted that, due to the reciprocal (hodograph) transformation, the rogue wave solution to the CSP equation is different from the one to the nonlinear Schr\"odinger (NLS) equation, which could be a cusponed- or a looped one.  \newline
\end{abstract}

\maketitle

%%%%%%%%%%%%%%%%%%%%%%%%%%%%%%%%%%%%%%%%%%%%%%%%%

%\date{Sept 22, 2015}
\section{Introduction}
The nonlinear Schr\"odinger (NLS) equation, as one of the universal models
that describe the evolution of slowly varying packets of quasi-monochromatic
waves in weakly nonlinear dispersive media, plays an key role in nonlinear
optics \cite{Hasegawa,Agrawal}. Recently, there are several experiments
reported related to the modulational instability (MI) and the breather solution \cite{MI,Zakh}
of the NLS equation in nonlinear optics. The Akhmediev breather (periodic in
space but localized in time) \cite{AB}, the Peregrine soliton or rogue wave
(RW) solution (time and space homoclinic) \cite{Pregr} and the
Kuznetsov-Ma soliton (periodic in time but localized in space) \cite{K-M}
have recently been experimentally observed in optical fibers \cite%
{Dudley,Kibler,Kibler1} in succession. Beside the experimental observation in optical fibers,
the RWs have also been observed in water-wave tanks \cite{Chabchoub} and plasmas
\cite{Bailung}.

However, in the regime of ultra-short pulses where the width of optical
pulse is in the order of femtosecond ($10^{-15}$ s), the quasi-monochromatic assumption
to derive the NLS equation is not valid anymore \cite{Roth}. Description of ultra-short processes requires a
modification of standard slow varying envelope models based on the NLS
equation. There are usually two ways to satisfy this requirement in the
literature. The first one is to add several higher-order dispersive terms to
yield higher-order NLS equation \cite{Agrawal}. The second one is to
construct a suitable fit to the frequency-dependent dielectric constant $%
\varepsilon(\omega)$ in the desired spectral range. Several models have been
proposed by the latter approach such as the short-pulse (SP) equation \cite%
{Sch,Sko,Kim,Amir} and the complex short pulse (CSP) equation \cite{Feng2}.

Recently, Sch\"{a}fer and Wayne derived a short pulse (SP) equation \cite{Sch}
\begin{equation}  \label{SP}
u_{xt}=u+\frac{1}{6}(u^3)_{xx}
\end{equation}
to describe the propagation of ultra-short optical pulses in nonlinear
media. Here, $u=u(x,t)$ is a real-valued function, representing the
magnitude of the electric field. The SP equation \eqref{SP} has been shown
to be completely integrable \cite{Robelo,Beals,Sako,Brun,Brun1}. The
periodic and soliton solutions of the SP equation \eqref{SP} were found in
\cite{Sako1,Kuet,Parkes}. The connection between the SP equation \eqref{SP}
and the sine-Gordon equation through the reciprocal transformation was
clarified, and then the $N$-soliton solutions including multi-loop and
multi-breather ones were given in \cite{Matsuno,Matsuno1} by using the
Hirota's bilinear method \cite{Hirota}. The integrable discretization and the
geometric interpretation of the SP equation were given in \cite{Feng,Feng1}.

Most recently, one of the authors  proposed a complex short
pulse (CSP) equation \cite{Feng2}
\begin{equation}  \label{CSP}
q_{xt}+q+\frac{1}{2}(|q|^2q_x)_x=0
\end{equation}
that governs the propagation of ultra short pulse packet along optical
fibers.
%It is known that the complex-valued function is useful in the
%description of wave phenomenon, especially the optical waves .
There are several advantages in using complex representation description
of wave phenomenon, especially of the optical waves \cite{Yariv}. Firstly,
amplitude and phase are two fundamental characteristics for a wave packet,
the information of these two factors are nicely combined into a single
complex-valued function. Secondly, the use of complex representation can
make a lot of manipulations including soliton interactions much easier. Such
advantages can be observed in many analytical results related to the NLS
equation, the complex short pulse equation and their coupled models.
As is shown in \cite{Feng2,shen},
%the CSP equation \eqref{CSP} has explicit expressions for one- and multi-soliton
%solutions with physical interpretation.
in contrast with the fact that one-soliton solution to the SP equation is always a loop soliton without physical
meaning \eqref{SP}, the one-soliton solution to the CSP equation \eqref{CSP} is an envelope soliton
with a few optical cycles.

Compared to the SP equation, few results are known to the CSP equation \eqref{CSP}.
It is necessary to study the CSP equation mathematically, as well as its applications in
nonlinear optics. Therefore, it is the aim of the present paper to investigate all kinds of solutions of
the CSP equation by Darboux transformation.

Based on the previous study \cite{Feng2,shen}, it is known that the CSP equation \eqref{CSP} is
linked to a complex coupled dispersionless (CCD) equation \cite%
{cd}
\begin{equation}
\begin{split}
& q_{ys}=\rho q, \\
& \rho _{s}+\frac{1}{2}(|q|^{2})_{y}=0,
\end{split}
\label{cd}
\end{equation}%
through the following reciprocal (hodograph) transformation
\begin{equation}
\mathrm{d}x=\rho \mathrm{d}y-\frac{1}{2}|q|^{2}\mathrm{d}s,\,\,\mathrm{d}t=-%
\mathrm{d}s,  \label{reciproc1}
\end{equation}%
 The CCD equation \eqref{cd} is the first
negative flow of the Landau-Lifshitz hierarchy，while the SP and the CSP equations being the first negative flow of    
Wadati-Konno-Ichikawa (WKI) hierarchy \cite{WKI,Qiao,Zimerman}.
%So we could study the CCD equation \eqref{cd} through the theory of Landau-Lifshitz hierarchy. But
%this is not our required.
By constructing a generalized Darboux transformation to the CCD equation and integrating the integrals exactly involved in the reciprocal (hodograph) transformation, we are able to construct the general analytical solutions to the CSP equation including the $N$-bright soliton, $N$-breather solution and higher order rogue wave solutions.

%To obtain the CSP solutions, we need to use the
%reciprocal transformation. Nevertheless, the reciprocal transformation
%involves the integration. This is a easy work to obtain the explicit
%expression for these integrals. To solve this problem, we reobtain some
%useful formulas from the CCD equation \eqref{cd}. With these formulas, the
%integration can be solved automatically. Then the general solitonic formulas
%for CSP equation \eqref{CSP} are obtained. By using these solitonic
%formulas,  can be constructed.
It should be pointed out that the compact formulas for these solutions are more convenient for us to perform the
asymptotic analysis.
%With the aid of asymptotic analysis method and complex
%operation of linear algebra, we establish the asymptotic analysis for the $N$%
%-bright soliton, $N$-breather solution. To the best of our knowledge, the
%rigorous proof for the asymptotic analysis of $N$-breather solution is the
%first time. Last but not least, we give the general higher order rogue wave
%solutions.
Recently the modulational instability has been also
considered as a wave breaking mechanism \cite{breaking}. Indeed, if the
initial steepness of the monochromatic wave is large, during the process of
modulational instability, one wave will start growing and will soon reach
the limiting steepness, and break before becoming a rogue wave. The NLS
theory does not predict the breaking or overturning of the waves \cite%
{Onorato}. Different from previous research regarding the rogue wave solution to the NLS
equation, we find that there exists the wave breaking phenomenon in the
rogue wave theory of the CSP equation \eqref{CSP}. These results could deepen
our understanding about the MI mechanism \cite{Zakharov}.

The outline of the present paper is organized as follows. In section \ref%
{section2}, the generalized Darboux transformation \cite{Matveev,Guo1,Guo2}
of the CCD equation was derived through loop group method
\cite{loop-group}. Based on the generalized Darboux transformation, we can
obtain the general soliton formulas for the CCD equation. Further, by integrating the reciprocal transformation
exactly, we can construct the general soliton formulas for the CSP equation. In section \ref{section3}, the
$N$-bright soliton solution and the $N$-breather solution are constructed, and their asymptotic
analyses are performed. In section \ref{section4}, we construct
the rogue wave solution including the first-order and general higher order rogue wave solution.  Section \ref{section4} is devoted to conclusions and some discussions. In Appendices, we give the details involving the proofs of asymptotic analysis and the modulational instability analysis.
\section{Generalized Darboux transformation for the CSP equation}
\label{section2}
Prior to giving the Darboux transformation (DT) for the CSP equation \eqref{CSP}%
, we briefly review the link between the CSP equation and the CCD equation.
It is known that the CCD equation \eqref{cd} admits the following Lax pair
\begin{equation}
\begin{split}
\Psi _{y}=& U(\rho ,q;\lambda )\Psi , \\
\Psi _{s}=& V(q;\lambda )\Psi ,
\end{split}
\label{cd-lax}
\end{equation}%
where
\begin{equation}
U(\rho ,q;\lambda )=%
\begin{bmatrix}
-\frac{\mathrm{i}\rho }{\lambda } & -\frac{q_{y}^{\ast }}{\lambda } \\[8pt]
\frac{q_{y}}{\lambda } & \frac{\mathrm{i}\rho }{\lambda }
\end{bmatrix}%
,\,\,V(q;\lambda )=%
\left( \frac{\mathrm{i}}{4}\lambda\sigma_{3}+\frac{%
\mathrm{i}}{2}Q\right),\,\,\sigma_3=\mathrm{diag}(1,-1),\,\,Q=
\begin{bmatrix}
0 \,& \,q^{\ast } \\[8pt]
q \,& \,0
\end{bmatrix}\,,
\end{equation}%
and $^{\ast }$ represents the complex conjugate. Through the reciprocal
transformation \eqref{reciproc1},
%\begin{equation}
%\mathrm{d}x=\rho \mathrm{d}y-\frac{1}{2}|q|^{2}\mathrm{d}s,\,\,\mathrm{d}t=-%
%\mathrm{d}s,  \label{reciproc1}
%\end{equation}%
one can obtain the CSP equation \eqref{CSP} and its Lax pair:
\begin{equation}
\begin{split}
\Psi _{x}=&
\begin{bmatrix}
-\frac{\mathrm{i}}{\lambda } & -\frac{q_{x}^{\ast }}{\lambda } \\[8pt]
\frac{q_{x}}{\lambda } & \frac{\mathrm{i}}{\lambda }
\end{bmatrix}%
\Psi , \\
\Psi _{t}=&
\begin{bmatrix}
-\frac{\mathrm{i}}{4}\lambda +\frac{\mathrm{i}|q|^{2}}{2\lambda } & -\frac{%
\mathrm{i}q^{\ast }}{2}+\frac{|q|^{2}q_{x}^{\ast }}{2\lambda } \\[8pt]
-\frac{\mathrm{i}q}{2}-\frac{|q|^{2}q_{x}}{2\lambda } & \frac{\mathrm{i}}{4}%
\lambda -\frac{\mathrm{i}|q|^{2}}{2\lambda }
\end{bmatrix}%
\Psi .
\end{split}
\label{csp-lax}
\end{equation}%
On the contrary, the CSP equation \eqref{CSP} can be transformed into the
CCD equation \eqref{cd}. Note that the CSP equation \eqref{CSP} can
be rewritten as the following conservative form
\begin{equation}
\left( \sqrt{1+|q_{x}|^{2}}\right) _{t}+\frac{1}{2}\left( |q|^{2}\sqrt{%
1+|q_{x}|^{2}}\right) _{x}=0\,,
\end{equation}
thus, by letting $\rho ^{-1}=\sqrt{1+|q_{x}|^{2}}$ and defining
an inverse reciprocal transformation
\begin{equation}
\mathrm{d}y=\rho ^{-1}\mathrm{d}x-\frac{1}{2}\rho ^{-1}|q|^{2}\mathrm{d}%
t,\,\,\mathrm{d}s=-\mathrm{d}t,
\end{equation}%
we can convert system \eqref{csp-lax} into system \eqref{cd-lax}.
The equivalence between the CSP and the CCD equations is kind of formal
under the reciprocal and inverse reciprocal transformations. The rigorous
equivalence is valid only if $\rho \neq 0$ for $(y,s)\in \mathbb{R}^{2},$ or
$|u_{x}|\neq \infty $ for $(x,t)\in \mathbb{R}^{2}.$

To construct the soliton and rogue wave solutions for the CSP equation %
\eqref{CSP}, we give the following proposition
\begin{prop}
The Darboux matrix
\begin{equation}
T=I+\frac{\lambda _{1}^{\ast }-\lambda _{1}}{\lambda -\lambda _{1}^{\ast }}%
P_{1},\,\,P_{1}=\frac{|y_{1}\rangle \langle y_{1}|}{\langle
y_{1}|y_{1}\rangle },\langle y_{1}|=|y_{1} \rangle ^{\dag
},\,\,|y_{1}\rangle =%
\begin{bmatrix}
\psi _{1}(y,s;\lambda _{1}) \\
\phi _{1}(y,s;\lambda _{1}) \\
\end{bmatrix}
\label{DT}
\end{equation}%
where $|y_1\rangle$ is a special solution for linear system \eqref{cd-lax} with $\lambda=\lambda_1$, can convert system \eqref{cd-lax} into a new system
\begin{equation}
\begin{split}
\Psi \lbrack 1]_{y}=& U(\rho \lbrack 1],q[1];\lambda )\Psi \lbrack 1], \\
\Psi \lbrack 1]_{s}=& V(\rho \lbrack 1],q[1];\lambda )\Psi \lbrack 1].
\end{split}%
\end{equation}%
The B\"{a}cklund transformations between $(\rho \lbrack 1],q[1])$ and $(\rho
,q)$ are given through
\begin{equation}
\begin{split}
\rho \lbrack 1]=& \rho -2\ln _{ys}\left( \frac{\langle y_{1}|y_{1}\rangle }{%
\lambda _{1}^{\ast }-\lambda _{1}}\right) , \\
q[1]=& q+\frac{(\lambda _{1}^{\ast }-\lambda _{1})\psi _{1}^{\ast }\phi _{1}%
}{\langle y_{1}|y_{1}\rangle }, \\
|q[1]|^{2}=& |q|^{2}+4\ln _{ss}\left( \frac{\langle y_{1}|y_{1}\rangle }{%
\lambda _{1}^{\ast }-\lambda _{1}}\right) .
\end{split}
\label{backlund}
\end{equation}
\end{prop}

\textbf{Proof:} The Darboux transformation for the system \eqref{cd-lax} is
a standard one for the AKNS system with $SU(2)$ symmetry. The rest of the
proposition is to prove the formulas \eqref{backlund}, in which carry on
some ideas from the classical monograph \cite{algebraic}.

Suppose there is a holomorphic solution for Lax pair equation %
\eqref{cd-lax} in some punctured neighborhood of infinity on the Riemann
surface, smoothing depending on $y$ and $s$. Thus, we may assume the following asymptotical
expansion as $\lambda \rightarrow \infty .$
\begin{equation}
\begin{bmatrix}
\psi _{1} \\
\phi _{1}%
\end{bmatrix}%
=\left[
\begin{bmatrix}
1 \\
0%
\end{bmatrix}%
+\sum_{i=1}^{\infty }\Psi _{i}\lambda ^{-i}\right] \exp {\left( \frac{%
\mathrm{i}}{4}\lambda s\right)}\,,  \label{infinity-asy}
\end{equation}
for the wave function $\Psi$ and
\begin{equation} \label{infinity-asy-T}
T=I+\sum_{i=1}^{\infty }T^{[i]}\lambda ^{-i}.
\end{equation}
for the Darboux matrix $T$. Since $T$ is the Darboux matrix, it satisfies the following relations
\begin{equation}
T_{y}+TU=U[1]T.
\end{equation}
By comparing the entries of the matrices, we get
\begin{equation}
\label{hodo}
\begin{split}
q_{y}[1]& =q_{y}+\left( T_{2,1}^{[1]}\right) _{y}, \\
\rho \lbrack 1]& =\rho +\mathrm{i}\left( T_{1,1}^{[1]}\right) _{y}\,.
\end{split}%
\end{equation}
Integrating the first equation with respect to $y$, we have the second
equation in \eqref{backlund}. Let
\begin{equation*}
H\equiv q^{\ast }\frac{\phi _{1}}{\psi _{1}}=\sum_{i=1}^{\infty
}H_{i}\lambda ^{-i},
\end{equation*}
%from
%\begin{equation*}
%\begin{split}
%\psi _{1,s}=& \frac{\mathrm{i}}{4}\lambda \psi _{1}+\frac{\mathrm{i}}{2}%
%q^{\ast }\phi _{1}, \\
%\phi _{1,s}=& \frac{\mathrm{i}}{2}q\psi _{1}-\frac{\mathrm{i}}{4}\lambda
%\phi _{1}.
%\end{split}%
%\end{equation*}
we then have
\begin{equation*}
  (\ln H)_{s} =\frac{\phi _{1,s}}{\phi _{1}}-\frac{\psi _{1,s}}{\psi _{1}}%
+(\ln q^{\ast })_{s} =-\frac{\mathrm{i}}{2}\lambda -\frac{\mathrm{i}}{2}H+\frac{\mathrm{i}}{2}%
|q|^{2}H^{-1}+(\ln q^{\ast })_{s}
\end{equation*}
from the first equation of (\ref{cd-lax}). Thus
\begin{equation*}
H_{s}=\frac{\mathrm{i}}{2}|q|^{2}-\frac{\mathrm{i}}{2}\lambda H-\frac{%
\mathrm{i}}{2}H^{2}+(\ln q^{\ast })_{s}H.
\end{equation*}%
Then the coefficient $H_{i}$ can be determined as following:
\begin{equation*}
\begin{split}
H_{1}=& |q|^{2},\,\,H_{2}=2\mathrm{i}q_{s}q^{\ast }, \\
H_{i+1}=& 2\mathrm{i}q^{\ast }\left( \frac{H_{i}}{q^{\ast }}\right)
_{s}-\sum_{j=1}^{i-1}H_{j}H_{i-j},\,\,i\geq 2.
\end{split}%
\end{equation*}
%Suppose there is a holomorphic solution for Lax pair equation \eqref{cd-lax}
%in some punctured neighborhood of infinity on the Riemann surface, smoothly
%depending on $y$ and $s$,the following asymptotical expansion at
%infinity can be assumed
%\begin{equation}
%\begin{bmatrix}
%\psi _{1} \\
%\phi _{1} \\
%\end{bmatrix}%
%=\left[
%\begin{bmatrix}
%1 \\
%0 \\
%\end{bmatrix}%
%+\sum_{i=1}^{\infty }\Psi _{i}\lambda ^{-i}\right] \exp {\left( \frac{%
%\mathrm{i}}{4}\lambda s\right) },\,\,\text{as }\lambda \rightarrow \infty.
%\label{infinity-asy}
%\end{equation}%
On the one hand, the first equation of \eqref{cd-lax} can be rewritten as
\begin{equation*}
\psi _{1,s}=\left( \frac{\mathrm{i}}{4}\lambda +\frac{\mathrm{i}}{2}%
\sum_{i=1}^{\infty }H_{i}\lambda ^{-i}\right) \psi _{1}.
\end{equation*}%
Substituting the asymptotical expansion \eqref{infinity-asy}
\begin{equation*}
\psi _{1}=\left( 1+\sum_{i=1}^{\infty }\Psi _{i}^{[1]}\lambda ^{-i}\right)
\exp {\left( \frac{\mathrm{i}}{4}\lambda s\right) },
\end{equation*}%
into above equation, where superscript $^{[1]}$ represents the first component of the vector,
we then have
\begin{equation}
\Psi _{1,s}^{[1]}=\frac{\mathrm{i}}{2}H_{1}=\frac{\mathrm{i}}{2}|q|^{2}.
\label{add1}
\end{equation}
Similarly, by assuming an asymptotical expansion
\begin{equation}
\begin{bmatrix}
\psi _{1}[1] \\
\phi _{1}[1] \\
\end{bmatrix}%
=\left[
\begin{bmatrix}
1 \\
0 \\
\end{bmatrix}%
+\sum_{i=1}^{\infty }\Psi \lbrack 1]_{i}\lambda ^{-i}\right] \exp {\left(
\frac{\mathrm{i}}{4}\lambda s\right) },\,\,\lambda \rightarrow \infty\,,
\end{equation}
we have
\begin{equation}
\Psi \lbrack 1]_{1,s}^{[1]}=\frac{\mathrm{i}}{2}|q[1]|^{2}.
\end{equation}
Moreover, by Darboux transformation
\begin{equation*}
\begin{bmatrix}
\psi _{1}[1] \\
\phi _{1}[1] \\
\end{bmatrix}%
=\left( I+\sum_{i=1}^{\infty }T^{[i]}\lambda ^{-i}\right) \left[
\begin{bmatrix}
1 \\
0 \\
\end{bmatrix}%
+\sum_{i=1}^{\infty }\Psi _{i}\lambda ^{-i}\right] \exp {\left( \frac{%
\mathrm{i}}{4}\lambda s\right) },
\end{equation*}%
one can obtain
\begin{equation}
\left( T_{1,1}^{[1]}\right) _{s}+\Psi _{1,s}^{[1]}=\frac{\mathrm{i}}{2}%
|q[1]|^{2},  \label{add2}
\end{equation}%
where the element $T_{i,j}^{[1]}$ denotes the $(i,j)$-th entry of matrix $%
T^{[1]}.$ Together with \eqref{add1}, we can obtain that
\begin{equation}
|q[1]|^{2}=|q|^{2}-2\mathrm{i}\left( T_{1,1}^{[1]}\right) _{s}.
\label{result1}
\end{equation}
Next, we proceed to the calculation of $\left( T_{1,1}^{[1]}\right) _{s}$
and $\left( T_{1,1}^{[1]}\right) _{y}$. Since
\begin{equation*}
|y_{1}\rangle _{s}=\left( \frac{\mathrm{i}}{4}\lambda _{1}\sigma _{3}+\frac{%
\mathrm{i}}{2}Q\right) |y_{1}\rangle ,\,\,-\langle y_{1}|_{s}\sigma
_{3}=\langle y_{1}|\sigma _{3}\left( \frac{\mathrm{i}}{4}\lambda^*%
_{1}\sigma _{3}+\frac{\mathrm{i}}{2}Q\right) ,
\end{equation*}
which originates from the Lax pair of the CSP equation (\ref{CSP}), we then have
%{\bf The notions of $\sigma _{3}$ and $Q$ are firstly introduced. It need to be explained here or to be added when we introduce the Lax pair}
\begin{equation}
\left( \frac{\langle y_{1}|y_{1}\rangle }{\lambda _{1}^{\ast }-\lambda _{1}}%
\right) _{s}=\frac{\mathrm{i}}{4}(-|\psi _{1}|^{2}+|\phi _{1}|^{2}).
\end{equation}%
On the other hand,
\begin{equation*}
\langle y_{1}|y_{1}\rangle =|\psi _{1}|^{2}+|\phi _{1}|^{2},
\end{equation*}
which implies
\begin{equation*}
\left( \frac{|\psi _{1}|^{2}}{\langle y_{1}|\sigma _{3}|y_{1}\rangle }%
\right) _{y}=-\left( \frac{|\phi _{1}|^{2}}{\langle y_{1}|\sigma
_{3}|y_{1}\rangle }\right) _{y}.
\end{equation*}
Thus, we have
\begin{eqnarray*}
(T_{1,1}^{[1]})_{y} &=&\left( \frac{|\psi _{1}|^{2}}{{\displaystyle\frac{%
\langle y_{1}|y_{1}\rangle }{\lambda _{1}^{\ast }-\lambda _{1}}}}\right)
_{y} =\left( \frac{|\psi _{1}|^{2}-|\phi _{1}|^{2}}{{\displaystyle2\frac{%
\langle y_{1}|y_{1}\rangle }{\lambda _{1}^{\ast }-\lambda _{1}}}}\right)
_{y}=2\mathrm{i}\ln _{ys}\left( \frac{\langle
y_{1}|y_{1}\rangle }{\lambda _{1}^{\ast }-\lambda _{1}}\right).
\end{eqnarray*}
%{\bf why above holds}
%Therefore, we have
%\begin{equation}
%(T_{1,1}^{[1]})_{y}=
%\end{equation}
Similarly, we could derive
\begin{equation}
(T_{1,1}^{[1]})_{s}=2\mathrm{i}\ln _{ss}\left( \frac{\langle
y_{1}|y_{1}\rangle }{\lambda _{1}^{\ast }-\lambda _{1}}\right) .
\end{equation}%
Finally, combining Eqs. \eqref{result1} and \eqref{hodo}, we
obtain the last two formulas in \eqref{backlund}. This completes the proof. $\square $ \\

To construct a general Darboux matrix, the following identities will be
used. Suppose $M$ is a $N\times N$ matrix, $\phi $, $\psi $ are $1\times N$
column vectors, then we have the following identities
\begin{equation}
\begin{split}
& \phi M^{-1}\psi ^{\dag }=%
\frac{\begin{vmatrix}
M & \psi ^{\dag } \\
-\phi & 0
\end{vmatrix}}{|M|}, \\
& 1+\phi M^{-1}\psi ^{\dag }=%
\frac{\begin{vmatrix}
M & \psi ^{\dag } \\
-\phi & 1
\end{vmatrix}}{|M|}=\frac{\det (M+\psi ^{\dag }\phi )}{\det (M)},
\end{split}
\label{linalglem}
\end{equation}
where $^{\dag}$ represents the Hermite conjugate.
Then we have the following proposition gives the N-fold Darboux
transformation and the generalized N-fold Darboux transformation for the CSP
equation
\begin{prop}
The N-fold Darboux transformation for the CCD equation can be represented as
\begin{equation}
T_{N}=I+YM^{-1}D^{-1}Y^{\dag },  \label{n-fod-dt}
\end{equation}%
where $Y=\left[ |y_{1}\rangle ,|y_{2}\rangle ,\cdots ,|y_{N}\rangle \right]
, $ and
\begin{equation*}
M=\left( \frac{\langle y_{i}|y_{j}\rangle }{\lambda _{i}^{\ast }-\lambda _{j}%
}\right) _{1\leq i,j\leq N},\,\,D=\mathrm{diag}\left( \lambda -\lambda
_{1}^{\ast },\lambda -\lambda _{2}^{\ast },\cdots ,\lambda -\lambda
_{N}^{\ast }\right) .
\end{equation*}%
Moreover, the general Darboux matrix is
\begin{equation}
T_{N}=I+YM^{-1}D^{-1}Y^{\dag },  \label{gDT}
\end{equation}%
where
\begin{equation*}
\begin{split}
Y=& \left[ |y_{1}^{[0]}\rangle ,|y_{1}^{[1]}\rangle ,\cdots
,|y_{1}^{[n_{1}-1]}\rangle ,\cdots ,|y_{r}^{[0]}\rangle ,|y_{r}^{[1]}\rangle
,\cdots ,|y_{r}^{[n_{r}-1]}\rangle \right] , \\
M=&
\begin{bmatrix}
M_{11} & M_{12} & \cdots & M_{1r} \\
M_{21} & M_{22} & \cdots & M_{2r} \\
\vdots & \vdots & \ddots & \vdots \\
M_{21} & M_{22} & \cdots & M_{2r}
\end{bmatrix}%
,\,\,M_{ij}=%
\begin{bmatrix}
M_{ij}^{[1,1]} & M_{ij}^{[1,2]} & \cdots & M_{ij}^{[1,n_{j}]} \\
M_{ij}^{[2,1]} & M_{ij}^{[2,2]} & \cdots & M_{ij}^{[2,n_{j}]} \\
\vdots & \vdots & \ddots & \vdots \\
M_{ij}^{[n_{i},1]} & M_{ij}^{[n_{i},2]} & \cdots & M_{ij}^{[n_{i},n_{j}]}
\end{bmatrix}%
, \\
D=& \mathrm{diag}\left( D_{1},D_{2}\cdots ,D_{r}\right) ,\,\,D_{i}=%
\begin{bmatrix}
D_{i}^{[0]} & \cdots & D_{i}^{[n_{i}-1]} \\
0 & \ddots & \vdots \\
0 & 0 & D_{i}^{[0]}
\end{bmatrix}%
\end{split}%
\end{equation*}%
and
\begin{equation*}
\begin{split}
|y_{i}(\lambda _{i}+\alpha _{i}\epsilon _{i})\rangle =&
\sum_{k=0}^{n_{i}-1}|y_{i}^{[k]}\rangle \epsilon _{i}^{k}+O(\epsilon
_{i}^{n_{i}}),\,\,\frac{1}{\lambda -\lambda _{i}^{\ast }-\alpha _{i}\epsilon
_{i}^{\ast }}=\sum_{k=0}^{n_{i}-1}D_{i}^{[k]}\epsilon _{i}^{\ast
k}+O(\epsilon _{i}^{\ast n_{i}}) \\
\frac{\langle y_{i}(\lambda _{i}+\alpha _{i}\epsilon _{i})|y_{j}(\lambda
_{j}+\alpha _{j}\epsilon _{j})\rangle }{\lambda _{i}^{\ast }-\lambda
_{j}+\alpha _{i}^{\ast }\epsilon _{i}^{\ast }-\alpha _{j}\epsilon _{j}}=&
\sum_{k=1}^{n_{i}}\sum_{l=1}^{n_{j}}M_{ij}^{[k,l]}\epsilon _{i}^{\ast
k}\epsilon _{j}^{l}+O(\epsilon _{i}^{\ast n_{i}},\epsilon _{j}^{n_{j}}).
\end{split}%
\end{equation*}%
The general B\"{a}cklund transformations are
\begin{equation}
\begin{split}
\rho \lbrack N]=& \rho -2\ln _{ys}(\det (M)), \\
q[N]=& q+\frac{\det (G)}{\det (M)}, \\
|q[N]|^{2}=& |q|^{2}+4\ln _{ss}(\det (M))
\end{split}
\label{gBT}
\end{equation}%
where $G=%
\begin{bmatrix}
M & Y_{1}^{\dag } \\
-Y_{2} & 0
\end{bmatrix}%
$, $Y_{k}$ represents the $k$-th row of matrix $Y$.
\end{prop}
\par \textbf{Proof:} Through the standard iterated step for DT \cite{Guo2}, we
can obtain the $N$-fold DT. Next, by using the following equalities
\begin{equation*}
\begin{split}
(T_{1,1}^{[1]})_{y}& =\left( Y_{1}M^{-1}Y_{1}^{\dag }\right) _{y}=\left(
-Y_{2}M^{-1}Y_{2}^{\dag }\right) _{y}=\left( \frac{Y_{1}M^{-1}Y_{1}^{\dag
}-Y_{2}M^{-1}Y_{2}^{\dag }}{2}\right) _{y}=2\mathrm{i}\ln _{ys}\left( \det
(M)\right) , \\
(T_{1,1}^{[1]})_{s}& =2\mathrm{i}\ln _{ss}\left( \det (M)\right) ,
\end{split}%
\end{equation*}%
we can obtain the formula \eqref{gBT} from the above $N$-fold DT %
\eqref{n-fod-dt}. To complete the generalized DT, we set
\begin{equation*}
\begin{split}
\lambda _{r+1}& =\lambda _{1}+\alpha _{1}\varepsilon
_{1,1},\,|y_{r+1}\rangle =|y_{1}(\lambda _{r+1})\rangle ;\,\,\cdots ;\lambda
_{r+n_{1}-1}=\lambda _{1}+\alpha _{1}\varepsilon
_{1,n_{1}-1},\,|y_{r+n_{1}-1}\rangle =|y_{1}(\lambda _{r+n_{1}-1})\rangle ;
\\
\lambda _{r+n_{1}}& =\lambda _{2}+\alpha _{2}\varepsilon
_{2,1},\;|y_{r+n_{1}}\rangle =|y_{2}(\lambda _{r+n_{1}})\rangle ;\,\,\cdots
,\lambda _{r+n_{1}+n_{2}-2}=\lambda _{2}+\alpha _{2}\varepsilon
_{2,n_{2}-1},\,|y_{r+n_{1}+n_{2}-2}\rangle =|y_{2}(\lambda
_{r+n_{1}+n_{2}-2})\rangle ; \\
& \vdots \\
\lambda _{N-n_{r}+1}& =\lambda _{r}+\alpha _{r}\varepsilon
_{r,1},\,|y_{N-n_{r}+1}\rangle =|y_{r}(\lambda _{N-n_{r}+1})\rangle
;\,\,\cdots ;\lambda _{N}=\lambda _{r}+\alpha _{r}\varepsilon
_{r,n_{r}-1},\,|y_{N}\rangle =|y_{r}(\lambda _{N})\rangle . \\
&
\end{split}%
\end{equation*}%
Taking limit $\varepsilon _{i,j}\rightarrow 0$, we can obtain the
generalized DT \eqref{gDT} and formulas \eqref{gBT}. $\square $

Recently the generalized DT for the AB system without the first and third relation in \eqref{gBT} was given in ref \cite{wangxin} in a different form. Actually, the first and third relation in \eqref{gBT} are the key procedures to construct the exact solution for the CSP equation.
In summary, with the aid of reciprocal transformation \eqref{reciproc1}, we
obtain the general expression for $N$-soliton solution of the CSP equation %
\eqref{CSP}:
\begin{equation}
\begin{split}
q[N]=& q+\frac{\det (G)}{\det (M)}, \\
x=&\int \rho (y,s)\mathrm{d}y- \frac{1}{2} \int|q(y,s)|^{2}\mathrm{d}s%
 -2\ln _{s}(\det (M)),\,\ t=-s.
\end{split}
\label{csp-gene}
\end{equation}
\section{Multi-soliton and Multi-breather solutions to the CSP equation}
\label{section3}
In this section, we provide multi-soliton and
multi-breather solutions to the CSP equation by using formula %
\eqref{csp-gene}.
\subsection{Single soliton solution and $N$-soliton solution}
We start with a seed solution
\begin{equation}
\rho \lbrack 0]=-\frac{\gamma }{2},\,\,q[0]=0,\,\,\gamma >0.
\end{equation}%
Solving the Lax pair equation \eqref{cd-lax} with $(\rho ,q;\lambda )=(\rho
\lbrack 0],q[0];\lambda _{i})$, we arrive at
\begin{equation}
\Psi _{i}=%
\begin{bmatrix}
\mathrm{e}^{\theta _{i}} \\
\mathrm{e}^{-\theta _{i}} \\
\end{bmatrix}%
,\,\,\theta _{i}=\frac{\mathrm{i}\gamma }{2\lambda _{i}}y+\frac{\mathrm{i}%
\lambda _{i}}{4}s+a_{i},  \label{thetai}
\end{equation}%
from which, we can obtain the single soliton solution through the
formula \eqref{csp-gene}:
\begin{equation}
\begin{split}
q[1]=& \lambda _{1,I}\mathrm{sech}(2\theta _{1,R})\mathrm{e}^{-2\mathrm{i}%
\theta _{1,I}-\frac{\pi \mathrm{i}}{2}}, \\
x=& -\frac{\gamma }{2}y+\lambda _{1,I}\tanh (2\theta _{1,R}),\,\,t=-s,
\end{split}%
\end{equation}%
where $\lambda_1=\lambda_{1,R}+{\rm i}\lambda_{1,I}$, $\theta _{1}=\theta _{1,R}+\mathrm{i}\theta _{1,I}$. We comment here that $\lambda _{1}$ is the reciprocal of the wave number $p_1$ in \cite{Feng1}.  As discussed in \cite{Feng1},
if $\lambda _{1,R}^{2}>\lambda _{1,I}^{2}$, one has the smooth soliton
solution; if $\lambda _{1,R}^{2}=\lambda _{1,I}^{2}$, ones has the
cusponed soliton solution; if $\lambda _{1,R}^{2}<\lambda _{1,I}^{2}$, one obtains
the loop soliton solution.

Furthermore, by using the $N$-fold DT, we could drive the $N$-soliton solution
through the formula \eqref{csp-gene}:
\begin{equation}
\begin{split}
q[N]& =\frac{\det (G)}{\det (M)}, \\
x& =-\frac{\gamma }{2}y-2\ln _{s}(\det (M)),\,\,\, t=-s,
\end{split}
\label{nsoliton}
\end{equation}%
where
\begin{equation}
\begin{split}
M& =\left( \frac{\mathrm{e}^{\theta _{i}^{\ast }+\theta _{j}}+\mathrm{e}%
^{-\theta _{i}^{\ast }-\theta _{j}}}{\lambda _{i}^{\ast }-\lambda _{j}}%
\right) _{1\leq i,j\leq N},\quad\,G=%
\begin{bmatrix}
M & Y_{1}^{\dag } \\
-Y_{2} & 0
\end{bmatrix}%
, \\
Y_{1}& =%
\begin{bmatrix}
\mathrm{e}^{\theta _{1}}, & \mathrm{e}^{\theta _{2}}, & \cdots , & \mathrm{e}%
^{\theta _{N}}
\end{bmatrix}%
,\quad \,Y_{2}=%
\begin{bmatrix}
\mathrm{e}^{-\theta _{1}}, & \mathrm{e}^{-\theta _{2}}, & \cdots , & \mathrm{%
e}^{-\theta _{N}}
\end{bmatrix}%
,
\end{split}%
\end{equation}%
the expressions $\theta _{i}$'s are given in \eqref{thetai}. The dynamics for two soliton is shown in ref. \cite{Feng2}. It should be pointed out that the interaction of
two smooth solitons could yield the singularity. The condition to avoid singularity for multi-soliton can not obtained through an analytical way. Finally, to understand
the dynamics of above $N$-soliton solution \eqref{nsoliton}, we give the
following asymptotic analysis and its proof
\begin{prop} Suppose $0<v_{1}<v_{2}<\cdots <v_{N}$.
When $s\rightarrow \pm \infty $, we have
\begin{equation}
q[N]=\sum_{k=1}^{N}\lambda _{k,I}\mathrm{sech}(2\theta _{k,R}^{\pm })\mathrm{%
e}^{-2\mathrm{i}\theta _{k,I}^{\pm }-\frac{\pi \mathrm{i}}{2}}+O(\mathrm{e}%
^{-c|s|}),  \label{asym-soliton}
\end{equation}%
where
\begin{equation}
\begin{split}
\theta _{k,R}^{\pm }=& \frac{\gamma \lambda _{k,I}}{2|\lambda _{k}|^{2}}%
(y-v_{k}s)+a_{k,R}\pm \Delta _{k,R}^{\pm },\,\,\Delta _{k,R}^{\pm }=\frac{1}{2%
}\left( \sum_{l=1}^{k-1}\left\vert \frac{\lambda _{l}^{\ast }-\lambda _{k}}{%
\lambda _{l}-\lambda _{k}}\right\vert -\sum_{l=k+1}^{N}\left\vert \frac{%
\lambda _{l}^{\ast }-\lambda _{k}}{\lambda _{l}-\lambda _{k}}\right\vert
\right) , \\
\theta _{k,I}^{\pm }=& \frac{\gamma \lambda _{k,R}}{2|\lambda _{k}|^{2}}y+%
\frac{\lambda _{k,R}}{4}s+a_{k,I}\mp \Delta _{k,I}^{\pm },\,\,\Delta
_{k,I}^{\pm }=\frac{1}{2}\left( \sum_{l=1}^{k-1}\arg \left( \frac{\lambda
_{l}-\lambda _{k}}{\lambda _{l}^{\ast }-\lambda _{k}}\right)
-\sum_{l=k+1}^{N}\arg \left( \frac{\lambda _{l}-\lambda _{k}}{\lambda
_{l}^{\ast }-\lambda _{k}}\right) \right) ,
\end{split}
\label{para1}
\end{equation}%
and $c=\mathrm{min}\left( \left\vert \frac{\gamma \lambda _{k,I}}{2|\lambda
_{k}|^{2}}\right\vert \right) \mathrm{min}_{i\neq j}(|v_{i}-v_{j}|),$ $v_{i}=%
\frac{|\lambda _{i}|^{2}}{2\gamma }.$
\end{prop}
The proof is given in Appendix A. Next we analyze the coordinates transformation: as $s\rightarrow\pm \infty$%
, along the line $\theta_{k,R}^{\pm}=0$, we have
\begin{equation*}
x=-\frac{\gamma}{2}y-2\ln_s(M)\rightarrow -\frac{\gamma}{2}y\pm\left[%
\sum_{i=1}^{k-1}\lambda_{i,I}-\sum_{j=k+1}^{N}\lambda_{j,I}\right],
\end{equation*}
it follows that
\begin{prop}
When $t\rightarrow \mp\infty$, along the trajectory $\theta_{k,R}^{\pm}=0$,
we have
\begin{equation*}
q[N]=\sum_{k=1}^{N}\lambda_{k,I}\mathrm{sech}(2\theta_{k,R}^{\pm})\mathrm{e}%
^{-2\mathrm{i}\theta_{k,I}^{\pm}-\frac{\pi\mathrm{i}}{2}}+O(\mathrm{e}%
^{-c|t|}),
\end{equation*}
where
\begin{equation*}
\begin{split}
\theta_{k,R}^{\pm}=&-\frac{\lambda_{k,I}}{|\lambda_k|^2}x+\frac{\lambda_{k,I}%
}{4}t+a_{k,R}\pm\frac{\lambda_{k,I}}{|\lambda_k|^2}\left(\sum_{i=1}^{k-1}%
\lambda_{i,I} -\sum_{j=k+1}^{N}\lambda_{j,I}\right)\pm\Delta_{k,R}^{\pm}, \\
\theta_{k,I}^{\pm}=&-\frac{\lambda_{k,R}}{|\lambda_k|^2}x-\frac{\lambda_{k,R}%
}{4}t+a_{k,I}\pm\frac{\lambda_{k,I}}{|\lambda_k|^2}\left(\sum_{i=1}^{k-1}%
\lambda_{i,I} -\sum_{j=k+1}^{N}\lambda_{j,I}\right)\mp\Delta_{k,I}^{\pm}.
\end{split}%
\end{equation*}
\end{prop}
\subsection{Single breather and multi-breather solutions}
To find a single breather solution, we depart from a seed solution
\begin{equation}
\rho \lbrack 0]=-\frac{\gamma }{2},\,\,q[0]=\frac{\beta }{2}\mathrm{e}^{%
\mathrm{i}\theta },\,\,\theta =y+\frac{\gamma }{2}s,\,\,\gamma >0,\,\,\beta
\geq 0.  \label{seed2}
\end{equation}%
Then we have the solution for the Lax pair equation \eqref{cd-lax} with $%
(q,\rho ;\lambda )=(q[0],\rho \lbrack 0];\lambda _{i})$,
\begin{equation}
|y_{i}\rangle =KL_{i}E_{i},\,\,K=\mathrm{diag}\left( \mathrm{e}^{-\frac{%
\mathrm{i}}{2}\theta },\mathrm{e}^{\frac{\mathrm{i}}{2}\theta }\right)
,\,\,\lambda _{i}\neq -\gamma +\mathrm{i}\beta ,  \label{sol-vec}
\end{equation}%
where
\begin{equation*}
L_{i}=%
\begin{bmatrix}
1 & 1 \\[10pt]
{\displaystyle\frac{\beta }{\gamma +\xi _{i}}} & {\displaystyle\frac{\beta }{%
\gamma +\chi _{i}}}
\end{bmatrix}%
,\,\,E_{i}=%
\begin{bmatrix}
\mathrm{e}^{\theta _{i}} \\
\mathrm{e}^{-\theta _{i}}
\end{bmatrix}%
\end{equation*}%
and
\begin{equation*}
\begin{split}
\theta _{i}=& \frac{\mathrm{i}}{4}\sqrt{\beta ^{2}+(\lambda _{i}+\gamma )^{2}%
}\left( s+\frac{2}{\lambda _{i}}y\right) +a_{i}, \\
\xi _{i}=& \lambda _{i}+\sqrt{\beta ^{2}+(\lambda _{i}+\gamma )^{2}}%
,\,\,\chi _{i}=\lambda _{i}-\sqrt{\beta ^{2}+(\lambda _{i}+\gamma )^{2}}.
\end{split}%
\end{equation*}%
To avoid the inconvenience of involving the square root of a complex number,
we introduce the following transformation:
\begin{equation*}
\lambda _{i}+\gamma =\beta \sinh (\varphi _{i,R}+\mathrm{i}\varphi
_{i,I}),\,\,(\varphi _{i,R},\varphi _{i,I})\in \Omega ,
\end{equation*}%
where $\Omega =\{(\varphi _{R},\varphi _{I})|0<\varphi _{I}<\pi ,\,\text{and
}0<\varphi _{R}<\infty ,\text{or }\varphi _{R}=0,\text{and }\frac{\pi }{2}%
\leq \varphi _{I}<\pi \}$, then
\begin{equation*}
\xi _{i}+\gamma =\beta \mathrm{e}^{\varphi _{i,R}+\mathrm{i}\varphi
_{i,I}},\,\,\chi _{i}+\gamma =-\beta \mathrm{e}^{-\varphi _{i,R}-\mathrm{i}%
\varphi _{i,I}}.
\end{equation*}%
By some tedious calculations, the single breather solution can be
constructed from the formula \eqref{csp-gene} by using the technique \cite{Ling}
\begin{equation}
\begin{split}
q[1]=& \frac{\beta }{2}\left[ \frac{\cosh (2\theta _{1,R}-2\mathrm{i}\varphi
_{1,I})\cosh (\varphi _{1,R})+\sin (2\theta _{1,I}+2\mathrm{i}\varphi
_{1,R})\sin (\varphi _{1,I})}{\cosh (2\theta _{1,R})\cosh (\varphi
_{1,R})-\sin (2\theta _{1,I})\sin (\varphi _{1,I})}\right] \mathrm{e}^{%
\mathrm{i}\theta}\,, \\
x=& -\frac{\gamma }{2}y-\frac{\beta ^{2}}{8}s-2\ln _{s}\left[ \cosh (2\theta
_{1,R})\cosh (\varphi _{1,R})-\sin (2\theta _{1,I})\sin (\varphi _{1,I})%
\right] ,\,\,t=-s,
\end{split}%
\end{equation}%
where
\begin{equation*}
\begin{split}
\theta _{1,R}& =\delta _{1}\left( y-\frac{2}{\gamma }v_{1}s\right) -\varphi
_{1,R}+a_{1,R}, \\
\theta _{1,I}& =\epsilon _{1}\left( y-\frac{2}{\gamma }w_{1}s\right)
-\varphi _{1,I}+a_{1,I},
\end{split}%
\end{equation*}%
and
\begin{equation*}
\begin{split}
v_{1}& =\frac{\alpha _{1}\gamma \sinh \left( \varphi _{1,R}\right) }{4\left(
\gamma \sinh \left( \varphi _{1,R}\right) +\beta \cos \left( \varphi
_{1,I}\right) \right) },\,\,\delta _{1}=\frac{2\beta }{\alpha _{1}}\sin
\left( \varphi _{1,I}\right) \left( \gamma \sinh \left( \varphi
_{1,R}\right) +\beta \cos \left( \varphi _{1,I}\right) \right) , \\
w_{1}& =\frac{-\alpha _{1}\gamma \cos (\varphi _{1,I})}{4\left( \beta \sinh
\left( \varphi _{1,R}\right) -\gamma \cos (\varphi _{1,I})\right) }%
,\,\,\epsilon _{1}=\frac{2\beta }{{\alpha _{1}}}\cosh (\varphi _{1,R})\left(
\beta \sinh \left( \varphi _{1,R}\right) -\gamma \cos (\varphi
_{1,I})\right) , \\
\alpha _{1}& =\left( \beta \sinh \left( \varphi _{1,R}\right) \cos \left(
\varphi _{1,I}\right) -\gamma \right) ^{2}+{\beta }^{2}\cosh ^{2}\left(
\varphi _{1,R}\right) \sin ^{2}\left( \varphi _{1,I}\right) .
\end{split}%
\end{equation*}%
If $(\varphi _{1,R},\varphi _{1,I})\in \Omega _{1}\equiv \{(\varphi
_{R},\varphi _{I})|0\leq \varphi _{R}<\mathrm{arcsinh}(\frac{\beta }{\gamma }%
),\,\,\arccos (-\frac{\gamma }{\beta }\sinh (\varphi _{R}))<\varphi _{I}<\pi
\}$, then the single breather $|q[1]|^{2}$ propagates with velocity $\frac{2%
}{\gamma }v_{1}\leq 0$. If $(\varphi _{1,R},\varphi _{1,I})\in \Omega
_{2}\equiv \{(\varphi _{R},\varphi _{I})|\frac{\pi }{2}\leq \varphi _{I}<\pi
,\,\,\mathrm{arcsinh}(-\frac{\gamma }{\beta }\cos (\varphi _{I}))<\varphi
_{R}\}$, then the single breather $|q[1]|^{2}$ propagates with velocity $%
\frac{2}{\gamma }v_{1}>0$. An example of this case is illustrated in Fig. 1
(a). If $\gamma \sinh \left( \varphi _{1,R}\right) +\beta \cos \left(
\varphi _{1,I}\right) =0$, then we can obtain the so-called Akhmediev
breather, which is periodic in time and localized in space. Fig. 1 (b) shows
an example of Akhmediev breather.

\begin{figure}[htb]
\centering
\subfigure[$|q|^2$]{\includegraphics[height=50mm,width=50mm]{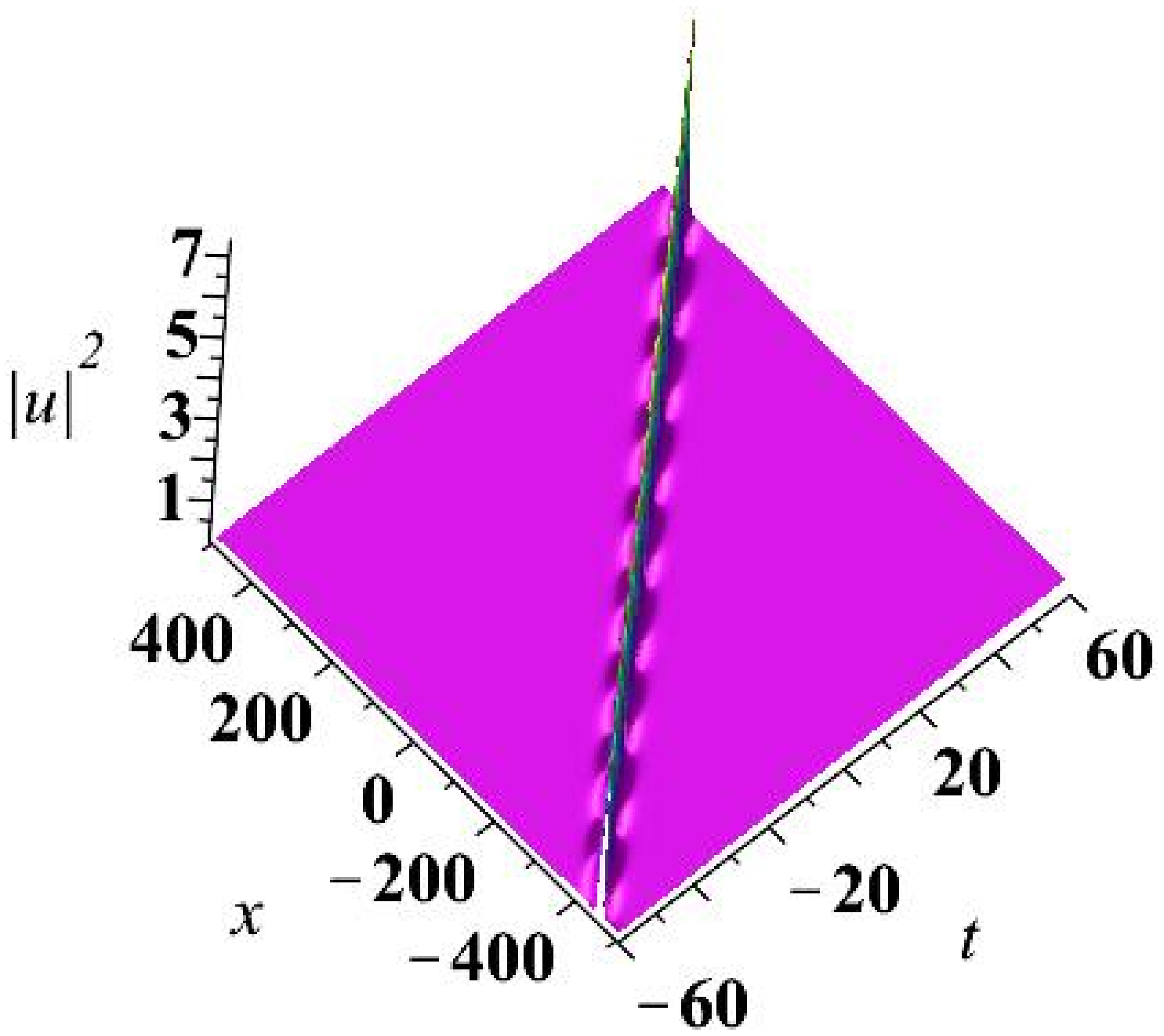}}
\hfil
\subfigure[$|q|^2$]{\includegraphics[height=50mm,width=50mm]{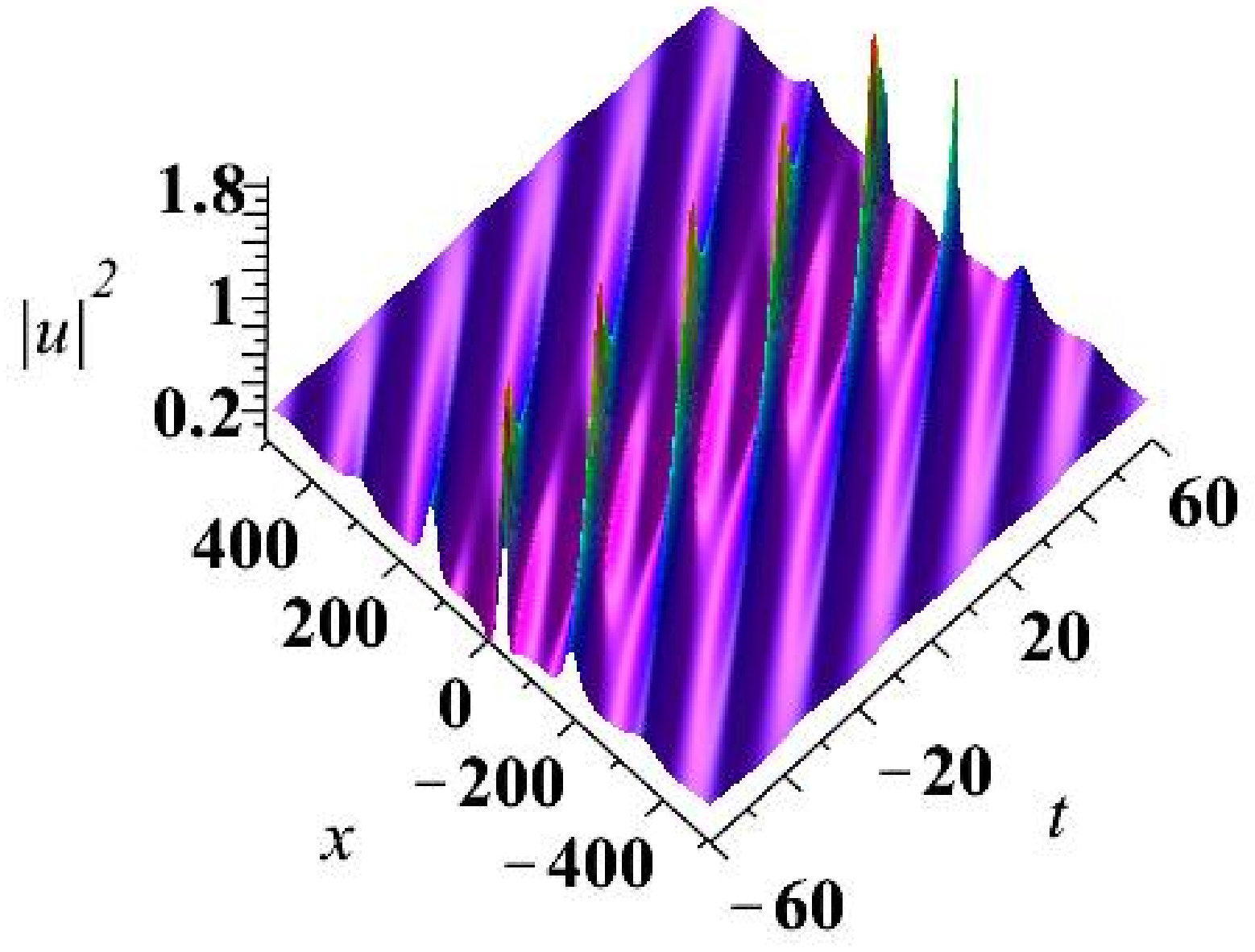}}
\hfil
\subfigure[$|q|^2$]{%
\includegraphics[height=50mm,width=50mm]{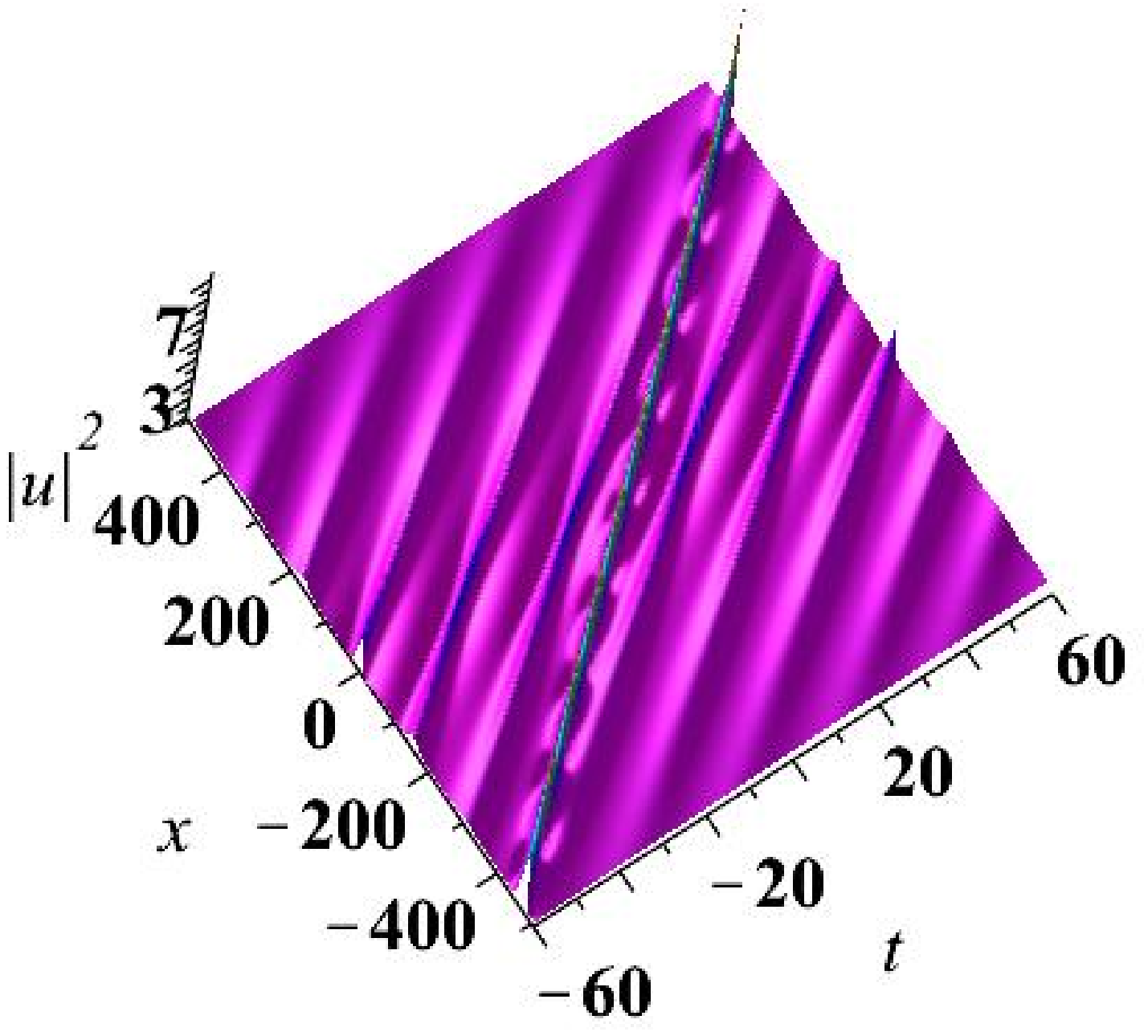}}
\caption{(color online): Parameters $\protect\beta=1$, $\protect\gamma=5$:
(a) Single breather solution. Parameters: $a_1=0$, $\protect\varphi_1=\ln(2+%
\protect\sqrt{5})+\frac{\protect\pi}{2}\mathrm{i}$, (b) Akhmediev breather
solution. Parameters: $a_1=0$, $\protect\varphi_1=\frac{\protect\pi}{3}%
\mathrm{i}$, (c)Two breather solution. Parameters: $\protect\varphi_1=\ln(2+%
\protect\sqrt{5})+\frac{\protect\pi}{2}\mathrm{i}$, $\protect\varphi_2=\frac{%
\protect\pi}{3}\mathrm{i}$, $a_1=a_2=0$. }
\label{fig1}
\end{figure}

To analyze the dynamics of the breather solution for the CSP equation \eqref{CSP}%
, we need to solve the relation between $(x,t)$ and $(y,s)$. Although, it
is not possible in general, we can obtain the relation at special location $\theta
_{1,R}=0$ and $\theta _{1,I}=k\pi +\frac{\pi }{4}$, that is, $s=-t$ and $y=-%
\frac{2}{\gamma }(x-\frac{\beta ^{2}}{8}t).$ It follows that
\begin{equation*}
\begin{split}
\theta _{1,R}& =-\frac{\gamma }{2}\delta _{1}\left[ x-\left( v_{1}+\frac{%
\beta ^{2}}{8}\right) t\right] -\varphi _{1,R}+a_{1,R}, \\
\theta _{1,I}& =-\frac{\gamma }{2}\epsilon _{1}\left[ x-\left( w_{1}+\frac{%
\beta ^{2}}{8}\right) t\right] -\varphi _{1,I}+a_{1,I}\,.
\end{split}%
\end{equation*}%
The breather solution $|q[1]|^{2}$ propagates with the velocity $v_{1}+\frac{%
\beta ^{2}}{8}$ (Fig.\ref{fig1} a). If $\delta _{1}=0$, we can obtain the
Akhmediev breather (Fig.\ref{fig1}b). The periodic in $x$ direction is $%
\frac{2\pi }{\gamma |\epsilon _{1}|},$ the periodic in $t$ direction is $%
\frac{2\pi }{\gamma |\epsilon _{1}|\left( w_{1}+\frac{\beta ^{2}}{8}\right) }%
.$ The peak value of $|q[1]|^{2}$ is located at
\begin{equation*}
\begin{split}
x=& \frac{1}{v_{1}-w_{1}}\left[ -\frac{2\left( v_{1}+\frac{\beta ^{2}}{8}%
\right) }{\gamma \epsilon _{1}}\left( \frac{\pi }{4}+k\pi +(\varphi
_{1,I}-a_{1,I})\right) -\frac{2}{\gamma }\frac{a_{1,R}-\varphi _{1,R}}{%
\delta _{1}}\left( w_{1}+\frac{\beta ^{2}}{8}\right) \right] , \\
t=& \frac{1}{v_{1}-w_{1}}\left[ -\frac{2}{\gamma \epsilon _{1}}\left( \frac{%
\pi }{4}+k\pi +(\varphi _{1,I}-a_{1,I})\right) -\frac{2}{\gamma }\frac{%
a_{1,R}-\varphi _{1,R}}{\delta _{1}}\right] .
\end{split}%
\end{equation*}
Similar to three cases of the single soliton solution, we can classify the
single breather solution by defining
\begin{equation}
\begin{split}
M_{1}=& \frac{1}{2}{\beta }^{3}\sinh (2\varphi _{1,R})\sin (2\varphi
_{1,I})-\gamma \left[ (\cosh (\varphi _{1,R})-\cos (\varphi
_{1,I}))^{2}+2\cosh (\varphi _{1,R})\sin (\varphi _{1,I})\right] {\beta }^{2}
\\
& -2\beta {\gamma }^{2}\sinh (\varphi _{1,R})\cos (\varphi _{1,I})+{\gamma }%
^{3}.
\end{split}%
\end{equation}%
It can be shown that if $M_{1}>0$, the breather solution is a smooth one; if
$M_{1}=0$, the breather becomes a cusponed one, in which $%
|q_{x}|\rightarrow \infty $ at the peak point; if $M_{1}<0$, then we have a
looped breather, which is a multi-valued solution.

Generally, through the formula \eqref{csp-gene} we have the following $N$%
-breather solution:
\begin{equation}
\begin{split}
q[N]& =\frac{\beta }{2}\left[ \frac{\det (G)}{\det (M)}\right] \mathrm{e}^{%
\mathrm{i}\theta }, \\
x& =-\frac{\gamma }{2}y-\frac{\beta ^{2}}{8}s-2\ln _{s}(\det (M)),\,\,t=-s,
\end{split}
\label{nbreather}
\end{equation}%
where
\begin{equation*}
\begin{split}
M& =\left( \left[ \frac{\mathrm{e}^{2(\theta _{i}^{\ast }+\theta _{j})}}{\xi
_{i}^{\ast }-\xi _{j}}+\frac{\mathrm{e}^{2\theta _{i}^{\ast }}}{\xi
_{i}^{\ast }-\chi _{j}}+\frac{\mathrm{e}^{2\theta _{j}}}{\chi _{i}^{\ast
}-\xi _{j}}+\frac{1}{\chi _{i}^{\ast }-\chi _{j}}\right] \mathrm{e}%
^{-(\theta _{i}^{\ast }+\theta _{j})}\right) _{1\leq i,j\leq N},\,\, \\
G& =\left( \left[ \frac{\xi _{i}^{\ast }+\gamma }{\xi _{j}+\gamma }\frac{%
\mathrm{e}^{2(\theta _{i}^{\ast }+\theta _{j})}}{\xi _{i}^{\ast }-\xi _{j}}+%
\frac{\xi _{i}^{\ast }+\gamma }{\chi _{j}+\gamma }\frac{\mathrm{e}^{2\theta
_{i}^{\ast }}}{\xi _{i}^{\ast }-\chi _{j}}+\frac{\chi _{i}^{\ast }+\gamma }{%
\xi _{j}+\gamma }\frac{\mathrm{e}^{2\theta _{j}}}{\chi _{i}^{\ast }-\xi _{j}}%
+\frac{\chi _{i}^{\ast }+\gamma }{\chi _{j}+\gamma }\frac{1}{\chi _{i}^{\ast
}-\chi _{j}}\right] \mathrm{e}^{-(\theta _{i}^{\ast }+\theta _{j})}\right)
_{1\leq i,j\leq N}.
\end{split}%
\end{equation*}%
The dynamics for $N$-breather solution is a very interesting topic.
%Although
%we can exhibit some figures of two-breather (Fig.\ref{fig1}c) or
%three-breather, we can not obtain the figure for the general $N$-breather.
It is naturally to conjecture that the $N$-breather solution possesses the
same law as the $N$-bright soliton solution.
%Up to now, to the best of our
%knowledge, there is no result for the asymptotic analysis for the $N$%
%-breather solution. In what follows, we will tackle this problem.
To understand the $N$-breather solution for the CSP equation \eqref{nbreather}, we
first give the following asymptotical analysis for the CCD equation \eqref{cd}:
\begin{prop}
\label{prop1}Suppose $v_1<v_2<\cdots<v_l\leq 0<v_N<v_{N-1}<\cdots<v_{l+1}$%
.  When $s\rightarrow -\infty $, we have
\begin{equation}
\begin{split}
q[N]=& \frac{\beta }{2}\left[ q_{1}^{-}+\left( q_{2}^{-}-\mathrm{e}^{-2%
\mathrm{i}\varphi _{1,I}}\Theta _{1}^{-}\right) +\cdots +\left( q_{k}^{-}-%
\mathrm{e}^{-2\mathrm{i}\varphi _{k-1,I}}\Theta _{k-1}^{-}\right) +\left(
q_{N}^{-}-\mathrm{e}^{-2\mathrm{i}\varphi _{k,I}}\Theta _{k}^{-}\right)
\right.  \\
& \left. +\left( q_{N-1}^{-}-\mathrm{e}^{2\mathrm{i}\varphi _{N,I}}\Theta
_{N}^{-}\right) +\cdots +\left( q_{k+1}^{-}-\mathrm{e}^{2\mathrm{i}\varphi
_{k+1,I}}\Theta _{k+2}^{-}\right) \right] \mathrm{e}^{\mathrm{i}\theta }+O(%
\mathrm{e}^{-c|s|}),
\end{split}
\label{asym1}
\end{equation}%
where $c=\frac{2}{\gamma }\mathrm{min}(\delta _{i})\mathrm{min}_{i\neq
j}(|v_{i}-v_{j}|)$. When $s\rightarrow +\infty $, we have
\begin{equation}
\begin{split}
q[N]=& \frac{\beta }{2}\left[ q_{1}^{+}+\left( q_{2}^{+}-\mathrm{e}^{2%
\mathrm{i}\varphi _{1,I}}\Theta _{1}^{+}\right) +\cdots +\left( q_{k}^{+}-%
\mathrm{e}^{2\mathrm{i}\varphi _{k-1,I}}\Theta _{k-1}^{+}\right) +\left(
q_{N}^{+}-\mathrm{e}^{2\mathrm{i}\varphi _{k,I}}\Theta _{k}^{+}\right)
\right.  \\
& \left. +\left( q_{N-1}^{+}-\mathrm{e}^{-2\mathrm{i}\varphi _{N,I}}\Theta
_{N}^{+}\right) +\cdots +\left( q_{k+1}^{+}-\mathrm{e}^{-2\mathrm{i}\varphi
_{k+1,I}}\Theta _{k+2}^{+}\right) \right] \mathrm{e}^{\mathrm{i}\theta }+O(%
\mathrm{e}^{-c|s|}),
\end{split}
\label{asym2}
\end{equation}%
where
\begin{equation*}
q_{k}^{\pm }=\Theta _{k}^{\pm }\left[ \frac{\cosh (2\theta _{k,R}^{\pm }-2%
\mathrm{i}\varphi _{k,I})\cosh (\varphi _{k,R})+\sin (2\theta _{k,I}^{\pm }+2%
\mathrm{i}\varphi _{k,I})\sin (\varphi _{k,I})}{\cosh (2\theta _{k,R}^{\pm
})\cosh (\varphi _{k,R})-\sin (2\theta _{k,I}^{\pm })\sin (\varphi _{k,I})}%
\right] ,
\end{equation*}%
and
\begin{equation}
\begin{split}
\theta _{k,R}^{+}=& \theta _{k,R}+\Delta _{k,R}^{+},\,\,\Delta _{k,R}^{+}=%
\frac{1}{2}\left[ \bigstar \ln \left\vert \frac{\chi _{n}-\xi _{k}}{\chi
_{n}^{\ast }-\xi _{k}}\right\vert \left\vert \frac{\chi _{n}^{\ast }-\chi
_{k}}{\chi _{n}-\chi _{k}}\right\vert +\blacklozenge \ln \left\vert \frac{%
\xi _{n}-\xi _{k}}{\xi _{n}^{\ast }-\xi _{k}}\right\vert \left\vert \frac{%
\xi _{n}^{\ast }-\chi _{k}}{\xi _{n}-\chi _{k}}\right\vert \right] , \\
\theta _{k,I}^{+}=& \theta _{k,I}+\Delta _{k,I}^{+},\,\,\Delta _{k,I}^{+}=%
\frac{1}{2}\left[ \bigstar \arg \left( \frac{\chi _{n}-\xi _{k}}{\chi
_{n}^{\ast }-\xi _{k}}\frac{\chi _{k}^{\ast }-\chi _{n}^{\ast }}{\chi
_{k}^{\ast }-\chi _{n}}\right) +\blacklozenge \arg \left( \frac{\xi _{n}-\xi
_{k}}{\xi _{n}^{\ast }-\xi _{k}}\frac{\chi _{k}^{\ast }-\xi _{n}^{\ast }}{%
\chi _{k}^{\ast }-\xi _{n}}\right) \right] , \\
\theta _{k,R}^{-}=& \theta _{k,R}+\Delta _{k,R}^{-},\,\,\Delta _{k,R}^{-}=%
\frac{1}{2}\left[ \bigstar \ln \left\vert \frac{\xi _{n}-\xi _{k}}{\xi
_{n}^{\ast }-\xi _{k}}\right\vert \left\vert \frac{\xi _{n}^{\ast }-\chi _{k}%
}{\xi _{n}-\chi _{k}}\right\vert +\blacklozenge \ln \left\vert \frac{\chi
_{n}-\xi _{k}}{\chi _{n}^{\ast }-\xi _{k}}\right\vert \left\vert \frac{\chi
_{n}^{\ast }-\chi _{k}}{\chi _{n}-\chi _{k}}\right\vert \right] , \\
\theta _{k,I}^{-}=& \theta _{k,I}+\Delta _{k,I}^{-},\,\,\Delta _{k,I}^{-}=%
\frac{1}{2}\left[ \bigstar \arg \left( \frac{\xi _{n}-\xi _{k}}{\xi
_{n}^{\ast }-\xi _{k}}\frac{\chi _{k}^{\ast }-\xi _{n}^{\ast }}{\chi
_{k}^{\ast }-\xi _{n}}\right) +\blacklozenge \arg \left( \frac{\chi _{n}-\xi
_{k}}{\chi _{n}^{\ast }-\xi _{k}}\frac{\chi _{k}^{\ast }-\chi _{n}^{\ast }}{%
\chi _{k}^{\ast }-\chi _{n}}\right) \right] , \\
\Theta _{k}^{+}=& \exp \left( \bigstar 2\mathrm{i}\varphi
_{n,I}-\blacklozenge 2\mathrm{i}\varphi _{n,I}\right) ,\,\,\Theta
_{k}^{-}=\exp \left( -\bigstar 2\mathrm{i}\varphi _{n,I}+\blacklozenge 2%
\mathrm{i}\varphi _{n,I}\right) ,
\end{split}
\label{para2}
\end{equation}%
if $1\leq k\leq l$, then $\bigstar =\left(
\sum_{n=1}^{k-1}+\sum_{n=l+1}^{N}\right) ,\,\,\blacklozenge
=\sum_{n=k+1}^{l};$ if $l<k\leq N$, then $\bigstar
=\sum_{n=1}^{k-1},\,\,\blacklozenge =\sum_{n=k+1}^{N};$ and
\begin{equation*}
\theta _{i,R}=\delta _{i}\left( y-\frac{2}{\gamma }v_{i}s\right) -\varphi
_{i,R}+a_{i,R},\,\,\theta _{i,I}=\epsilon _{i}\left( y-\frac{2}{\gamma }%
w_{i}s\right) -\varphi _{i,I}+a_{i,I},
\end{equation*}%
and
\begin{equation*}
\begin{split}
v_{i}& =\frac{\alpha _{i}\gamma \sinh \left( \varphi _{i,R}\right) }{4\left(
\gamma \sinh \left( \varphi _{i,R}\right) +\beta \cos \left( \varphi
_{i,I}\right) \right) },\,\,\delta _{i}=\frac{2\beta }{\alpha _{i}}\sin
\left( \varphi _{i,I}\right) \left( \gamma \sinh \left( \varphi
_{i,R}\right) +\beta \cos \left( \varphi _{i,I}\right) \right) , \\
w_{i}& =\frac{-\alpha _{i}\gamma \cos (\varphi _{i,I})}{4\left( \beta \sinh
\left( \varphi _{i,R}\right) -\gamma \cos (\varphi _{i,I})\right) }%
,\,\,\epsilon _{i}=\frac{2\beta }{{\alpha _{i}}}\cosh (\varphi _{i,R})\left(
\beta \sinh \left( \varphi _{i,R}\right) -\gamma \cos (\varphi
_{i,I})\right) , \\
\alpha _{i}& =\left( \beta \sinh \left( \varphi _{i,R}\right) \cos \left(
\varphi _{i,I}\right) -\gamma \right) ^{2}+{\beta }^{2}\cosh ^{2}\left(
\varphi _{i,R}\right) \sin ^{2}\left( \varphi _{i,I}\right) .
\end{split}%
\end{equation*}
\end{prop}
Based on the above proposition, we can obtain the dynamics of $%
N$-breather solution for the CSP equation \eqref{CSP}. In general, the
dynamics of $N$-breather solution for the CSP equation \eqref{CSP} cannot be
solved analytically. However, in some special location, we can analyze them
by the coordinate transformation. When $s\rightarrow \pm \infty $, $\theta
_{k,R}^{\pm }=0$ and $\theta _{k,I}^{\pm }=\frac{\pi }{4}+k\pi $, we have
\begin{equation*}
\begin{split}
x& =-\frac{\gamma }{2}y-\frac{\beta ^{2}}{8}s-2\ln _{s}(\det (M)) \\
& \rightarrow -\frac{\gamma }{2}y-\frac{\beta ^{2}}{8}s-2\ln _{s}\left( \exp %
\left[ \pm 2\left( -\bigstar \theta _{n,R}+\blacklozenge \theta
_{n,R}\right) \right] \exp \left[ -2\theta _{k,R}\right] \det (M_{k})\right)
, \\
& =-\frac{\gamma }{2}y-\frac{\beta ^{2}}{8}s\pm \tau _{k},
\end{split}%
\end{equation*}%
where
\begin{equation*}
\tau _{k}=\beta \left( \bigstar \sin (\varphi _{n,I})\sinh (\varphi
_{n,R})-\blacklozenge \sin (\varphi _{n,I})\sinh (\varphi _{n,R})\right) .
\end{equation*}
\begin{prop}
When $t\rightarrow \mp \infty $, along the trajectory $\theta _{k,R}^{\pm
}=0 $ and $\theta _{k,I}^{\pm }=\frac{\pi }{4}+k\pi $, we have
\begin{equation*}
q[N]=\frac{\beta }{2}q_{k}^{\pm }\mathrm{e}^{\mathrm{i}\theta }+O(\mathrm{e}%
^{-c|t|}),
\end{equation*}%
where
\begin{equation*}
\begin{split}
\theta _{k,R}^{\pm }=& -\frac{\gamma }{2}\delta _{k}\left[ x-\left( v_{k}+%
\frac{\beta ^{2}}{8}\right) t\mp \tau _{k}\right] -\varphi
_{k,R}+a_{k,R}+\Delta _{k,R}^{\pm }, \\
\theta _{k,I}^{\pm }=& -\frac{\gamma }{2}\epsilon _{k}\left[ x-\left( w_{k}+%
\frac{\beta ^{2}}{8}\right) t\mp \tau _{k}\right] -\varphi
_{k,I}+a_{k,I}+\Delta _{k,I}^{\pm }.
\end{split}%
\end{equation*}
\end{prop}
\section{General rogue wave solution to the CSP equation}
\label{section4}
In previous section, we solved the linear system \eqref{cd-lax} with plane
wave seed solution under the restriction $\lambda _{i}\neq -\gamma +\mathrm{i%
}\beta $. It is natural to ask what happens if $\lambda _{i}=-\gamma +%
\mathrm{i}\beta$. Actually, we can obtain the rogue wave solution and high
order rogue wave solutions under this special condition. The general
procedure to yield these solutions was proposed in \cite{Guo1,Guo2}.

Starting from the linear system \eqref{cd-lax} with $(q,\rho ,\lambda
)=(q[0],\rho \lbrack 0],-\gamma +\mathrm{i}\beta )$, where $q[0]$ and $\rho
\lbrack 0]$ are given in equation \eqref{seed2}, then one can firstly
obtain the quasi-rational solution, from which the first order rogue wave
solution can be obtained through formula \eqref{csp-gene}. However, the higher
order RW solution cannot be constructed in the same way. To find the general
higher order rogue wave solution, we need to solve the linear system %
\eqref{cd-lax} with $(q,\rho ,\lambda )=(q[0],\rho \lbrack 0],-\gamma +%
\mathrm{i}\beta -\frac{\mathrm{i}\epsilon ^{2}}{2\beta })$, where $\epsilon $
is a small parameter.

To this end, we give the following Lemma.
\begin{lem}
\label{lem2} Denote
\begin{equation}
\lambda _{1}=-\gamma +\mathrm{i}\beta -\frac{\mathrm{i}\epsilon ^{2}}{2\beta
},\,\,\mu _{1}=\epsilon \sqrt{1-\left( \frac{\epsilon }{2\beta }\right) ^{2}}%
,\,\,\xi _{1}=\lambda _{1}+\mu _{1},  \label{subs1}
\end{equation}%
then the following parameters can be expanded in terms of a small parameter $\epsilon $
\begin{equation*}
\begin{split}
\mu _{1}& =\sum_{n=0}^{\infty }\mu _{1}^{[n]}\epsilon ^{2n+1}, \\
\frac{1}{\xi _{1}^{\ast }-\xi _{1}}& =\sum_{i=0,j=0}^{\infty ,\infty
}F^{[i,j]}\epsilon ^{\ast i}\epsilon ^{j}, \\
\frac{1}{\xi _{1}+\gamma }& \equiv \frac{1}{\mathrm{i}\beta (\sqrt{%
1-\epsilon ^{2}}-\mathrm{i}\epsilon )^{2}}=\sum_{i=0}^{\infty
}J^{[i]}\epsilon ^{i},
\end{split}%
\end{equation*}%
where
\begin{equation*}
\begin{split}
\mu _{1}^{[n]}& =%
\begin{pmatrix}
\frac{1}{2} \\[5pt]
n \\
\end{pmatrix}%
\left( \frac{-1}{4\beta ^{2}}\right) ^{n},\,\,%
\begin{pmatrix}
\frac{1}{2} \\[5pt]
n \\
\end{pmatrix}%
=\frac{\frac{1}{2}(\frac{1}{2}-1)\cdots (\frac{1}{2}-n+1)}{n!}, \\
F^{[i,j]}& =\frac{\mathrm{i}}{i!j!\beta }\frac{\partial ^{i+j}}{\partial
\epsilon ^{\ast i}\partial \epsilon ^{j}}\left( \left[ \exp \left( 2\mathrm{i%
}\arcsin \left( \frac{\epsilon ^{\ast }}{2\beta }\right) \right) +\exp
\left( -2\mathrm{i}\arcsin \left( \frac{\epsilon }{2\beta }\right) \right) %
\right] ^{-1}\right) _{|_{\epsilon ^{\ast }=0,\epsilon =0}}, \\
J^{[0]}=& \frac{1}{\mathrm{i}\beta },\,\,J^{[1]}=\frac{1}{\beta ^{2}}%
,\,\,J^{[2]}=\frac{\mathrm{i}}{2\beta ^{3}},\,\,J^{[2i+1]}=\frac{(-1)^{i}}{%
\beta ^{2}}%
\begin{pmatrix}
\frac{1}{2} \\[8pt]
i \\
\end{pmatrix}%
\left( \frac{1}{2\beta }\right) ^{2i},\,\,J^{[2i+2]}=0,\,\,i\geq 1.
\end{split}%
\end{equation*}
\end{lem}
With the aid of above lemma, we have the following expansion
\begin{equation*}
\begin{split}
X_{1}& \equiv \frac{\mathrm{i}}{4}\mu _{1}\left( s+{\frac{2y}{\lambda }}%
+\sum_{i=1}^{\infty }(a_{i}+\mathrm{i}b_{i})\epsilon ^{2i}\right) +\frac{1}{2%
}\ln \left( {\frac{\mu _{1}+\lambda _{1}+\gamma }{\mathrm{i}\beta }}\right),
\\
& =\mathrm{i}\epsilon \left( \sum_{i=1}^{\infty }\frac{\mu _{1}^{[i]}}{4}%
\epsilon ^{2i}\right) \left( \sum_{i=0}^{\infty }K^{[i]}\epsilon
^{2i}\right) -\mathrm{i}\arcsin \left( \frac{\epsilon }{2\beta }\right)  \\
& =\mathrm{i}\epsilon \sum_{k=0}^{\infty }X_{1}^{[2k+1]}\epsilon ^{2k},
\end{split}%
\end{equation*}%
where
\begin{equation*}
\begin{split}
X_{1}^{[2k+1]}& =\left[ \sum_{j=0}^{k}\frac{1}{4}K^{[j]}\mu _{1}^{[k-j]}-%
\frac{(-1)^{k}}{2k+1}%
\begin{pmatrix}
-\frac{1}{2} \\[5pt]
k \\
\end{pmatrix}%
\right] , \\
K^{[k]}& =\left\{
\begin{array}{ll}
{\displaystyle s-\frac{2(\gamma +\mathrm{i}\beta )y}{\beta ^{2}+\gamma ^{2}}}%
, & k=0, \\[8pt]
{\displaystyle\left( \frac{-2(\gamma +\mathrm{i}\beta )y}{\gamma ^{2}+\beta
^{2}}\right) \left( \frac{(\gamma +\mathrm{i}\beta )}{2\beta (\gamma
^{2}+\beta ^{2})\mathrm{i}}\right) ^{k}}+a_{k}+\mathrm{i}b_{k}, & k\geq 1.%
\end{array}%
\right.
\end{split}%
\end{equation*}%
Furthermore we have
\begin{equation*}
\mathrm{e}^{X_{1}}=\sum_{i=0}^{\infty }S_{i}(\mathbf{X}_{1})\epsilon
^{i},\,\,\mathbf{X}_{1}=\left( X_{1}^{[1]},X_{1}^{[2]},\cdots \right)
,\,\,X_{1}^{[2k]}=0,\,\,k\geq 1, \\
\end{equation*}%
where $S_{i}(\mathbf{X}_{1})$ are elementary Schur polynomials
\begin{equation*}
\begin{split}
S_{0}(\mathbf{X}_{1})=& 1,\,\,S_{1}(\mathbf{X}_{1})=X_{1}^{[1]},\,\,S_{2}(%
\mathbf{X}_{1})=X_{1}^{[2]}+\frac{(X_{1}^{[1]})^{2}}{2},\,\,S_{3}(\mathbf{X}%
_{1})=X_{1}^{[3]}+X_{1}^{[1]}X_{1}^{[2]}+\frac{(X_{1}^{[1]})^{3}}{6},\cdots
\\
S_{i}(\mathbf{X}_{1})=& \sum_{l_{1}+2l_{2}+\cdots +kl_{k}=i}\frac{%
(X_{1}^{[1]})^{l_{1}}(X_{1}^{[2]})^{l_{2}}\cdots (X_{1}^{[k]})^{l_{k}}}{%
l_{1}!l_{2}!\cdots l_{k}!}.
\end{split}%
\end{equation*}
Since $KE_1(\epsilon)$ satisfies the Lax equation \eqref{cd-lax}, then $%
KE_1(-\epsilon)$ also satisfies the Lax equation \eqref{cd-lax}. To obtain
the general higher order rogue wave solution, we choose the general special
solution
\begin{equation*}
|y_1\rangle=\frac{K}{2\epsilon}\left[E_1(\epsilon)-E_1(-\epsilon)\right]%
\equiv K%
\begin{bmatrix}
\varphi_1 \\
\beta\psi_{1} \\
\end{bmatrix}%
, \quad E_1=%
\begin{bmatrix}
\mathrm{e}^{X_1} \\
{\displaystyle \frac{\beta\mathrm{e}^{X_1}}{\xi_1+\gamma}} \\
\end{bmatrix}\,.
\end{equation*}
Finally, we have
\begin{equation}  \label{expansion1}
\frac{\langle y_1|y_1\rangle}{2(\lambda_1^*-\lambda_1)}=\frac{1}{4}\left[\frac{%
\mathrm{e}^{X_{1}^*+X_{1}}}{\xi_{1}^*-\xi_{1}}-\frac{\mathrm{e}%
^{X_{1}^*-X_{1}}}{\xi_{1}^*-\chi_{1}}-\frac{\mathrm{e}^{-X_{1}^*+X_{1}}}{%
\chi_{1}^*-\xi_{1}}+\frac{\mathrm{e}^{-X_{1}^*-X_{1}}}{\chi_{1}^*-\chi_{1}}%
\right]=\sum_{m=1,n=1}^{\infty,\infty}M^{[m,n]} {\epsilon}%
^{*2(m-1)}\epsilon^{2(n-1)},
\end{equation}
where $\chi_1=\xi_1(-\epsilon),$
\begin{equation*}
M^{[m,n]}=\sum_{i=0}^{2m-1}%
\sum_{j=0}^{2n-1}F^{[i,j]}S_{2n-i-1}(X_1)S_{2m-j-1}(X_1^*).
\end{equation*}
On the other hand, by using lemma \ref{lem2}, we have the following
expansion
\begin{equation}
\varphi _{1}=\frac{1}{2}\left( \mathrm{e}^{X_{1}}-\mathrm{e}^{-X_{1}}\right)
=\sum_{n=1}^{\infty }\varphi _{1}^{[n]}\epsilon ^{2(n-1)},\,\,\psi _{1}=%
\frac{1}{2}\left( \frac{\mathrm{e}^{X_{1}}}{\xi _{1}+\gamma }-\frac{\mathrm{e%
}^{-X_{1}}}{\chi _{1}+\gamma }\right) =\sum_{n=1}^{\infty }\psi
_{1}^{[n]}\epsilon ^{2(n-1)}\,,  \label{expansion2}
\end{equation}%
where
\begin{equation*}
\varphi _{1}^{[n]}=S_{2n-1}(\mathbf{X}_{1}),\,\,\psi
_{1}^{[n]}=\sum_{k=0}^{2n-1}S_{k}(\mathbf{X}_{1})J_{1}^{[2n-1-k]}.
\end{equation*}%
Based on the expansion equations \eqref{expansion1}-\eqref{expansion2}, and
formulas \eqref{gBT}-\eqref{csp-gene}-\eqref{linalglem}, we can obtain the
general rogue wave solutions:
\begin{prop}
\label{prop3} The general higher order rogue wave solution for the CSP
equation \eqref{CSP} can be represented as
\begin{equation}
\begin{split}
q[N]& =\frac{\beta }{2}\left[ \frac{\det (G)}{\det (M)}\right] \mathrm{e}^{%
\mathrm{i}\theta }, \\
x& =-\frac{\gamma }{2}y-\frac{{\beta }^{2}}{8}s-2\ln _{s}(\det (M)),\,\,t=-s,
\end{split}
\label{gene-rogue}
\end{equation}%
where
\begin{equation*}
M=\left( M^{[m,n]}\right) _{1\leq m,n\leq N},\,\,G=\left( M^{[m,n]}+\varphi
_{1}^{[m]\ast }\psi _{1}^{[n]}\right) _{1\leq m,n\leq N}.
\end{equation*}
\end{prop}

\begin{figure}[tbh]
\centering
\subfigure[$|q|^2$]{\includegraphics[height=50mm,width=50mm]{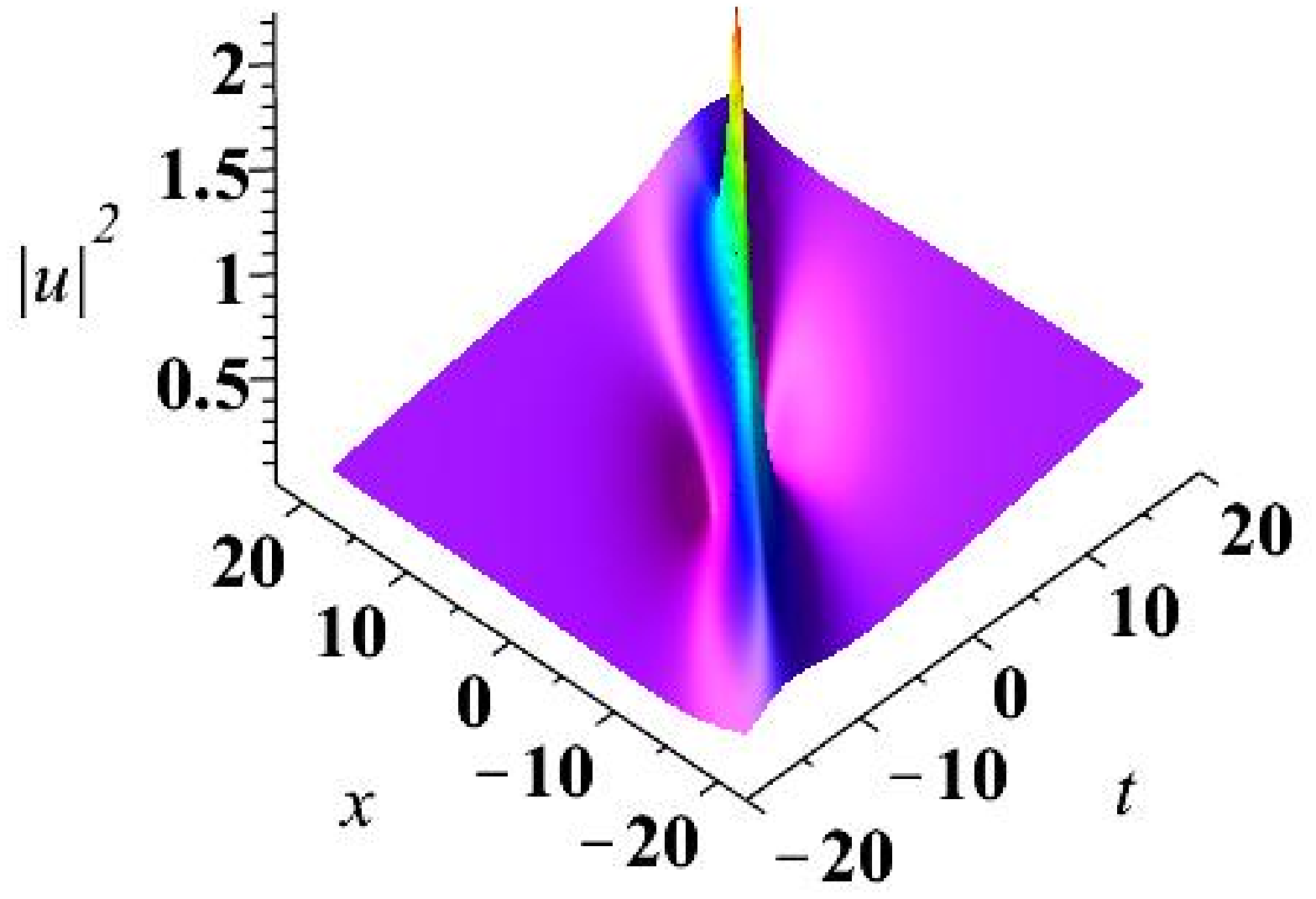}}
\hfil
\subfigure[$|q|^2$]{\includegraphics[height=50mm,width=50mm]{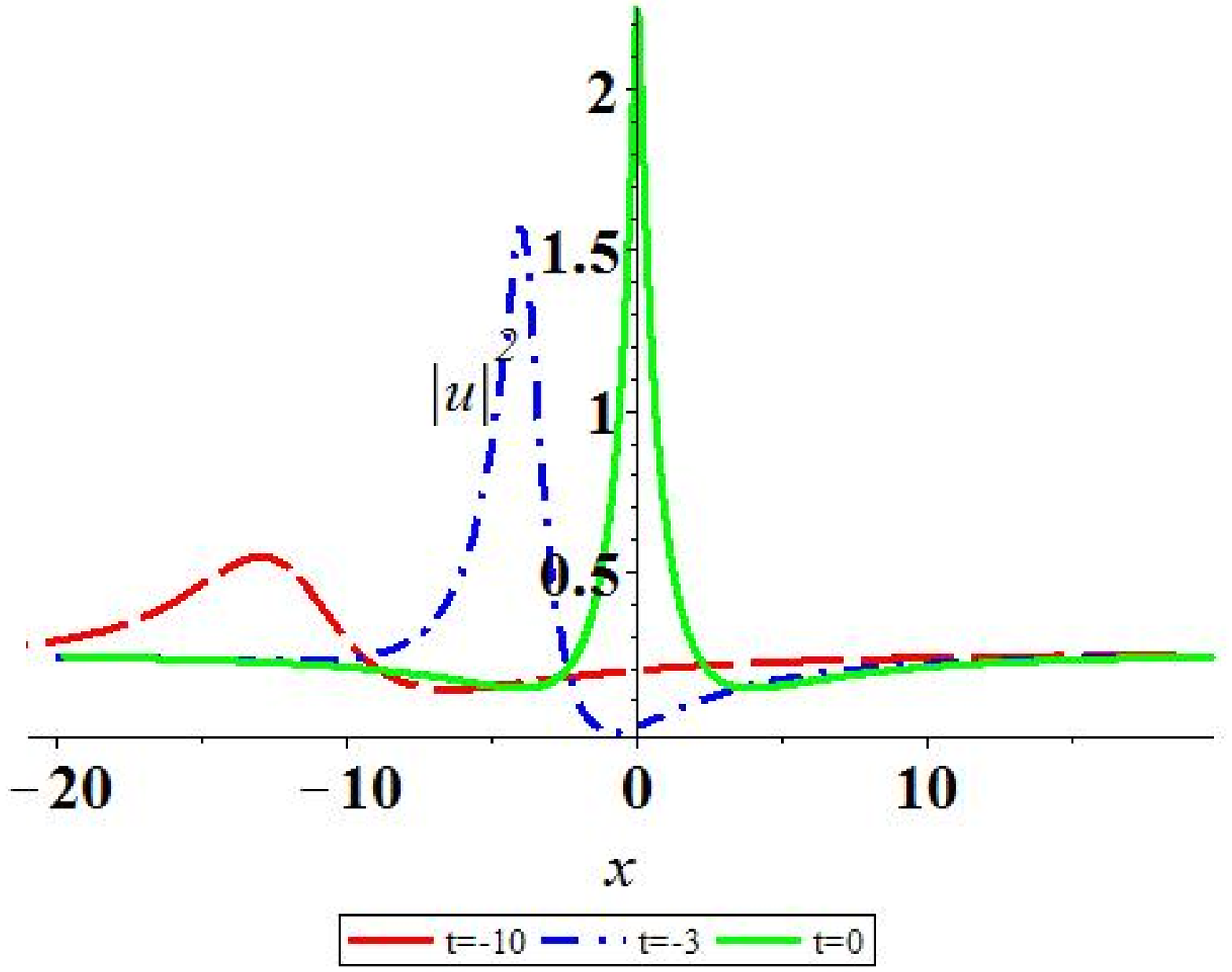}}
\caption{(color online): Parameters: $\protect\beta =1$, $\protect\gamma =2$%
, (a) The spatio-temporal pattern for the regular first order RW, (b) The
figure of $|q[1]|^{2}$ for different time, it is seen that the amplitude is
variation. }
\label{fig2}
\end{figure}

\begin{figure}[tbh]
\centering
\subfigure[$|q|^2$]{\includegraphics[height=50mm,width=50mm]{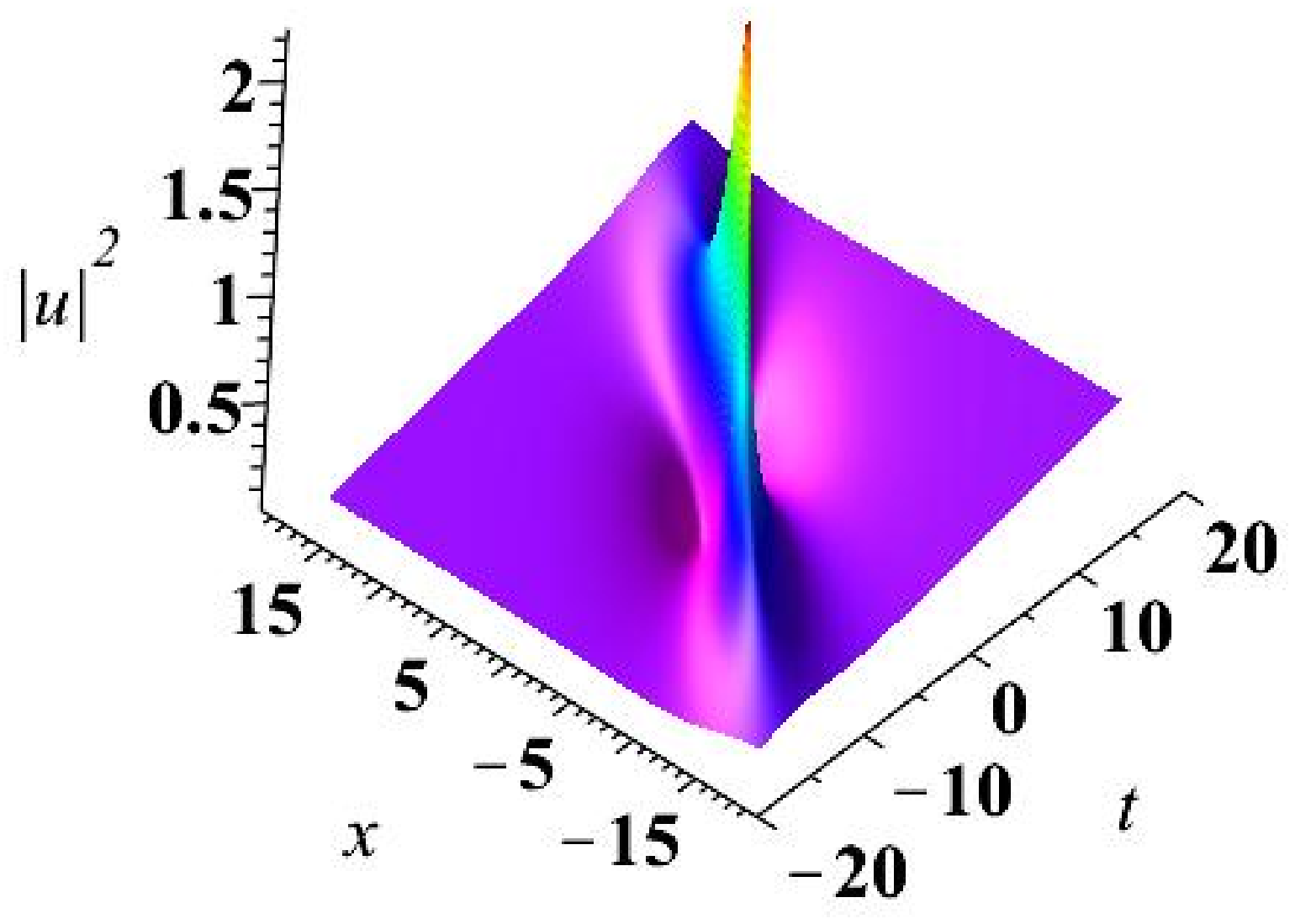}}
\hfil
\subfigure[$|q|^2$]{\includegraphics[height=50mm,width=50mm]{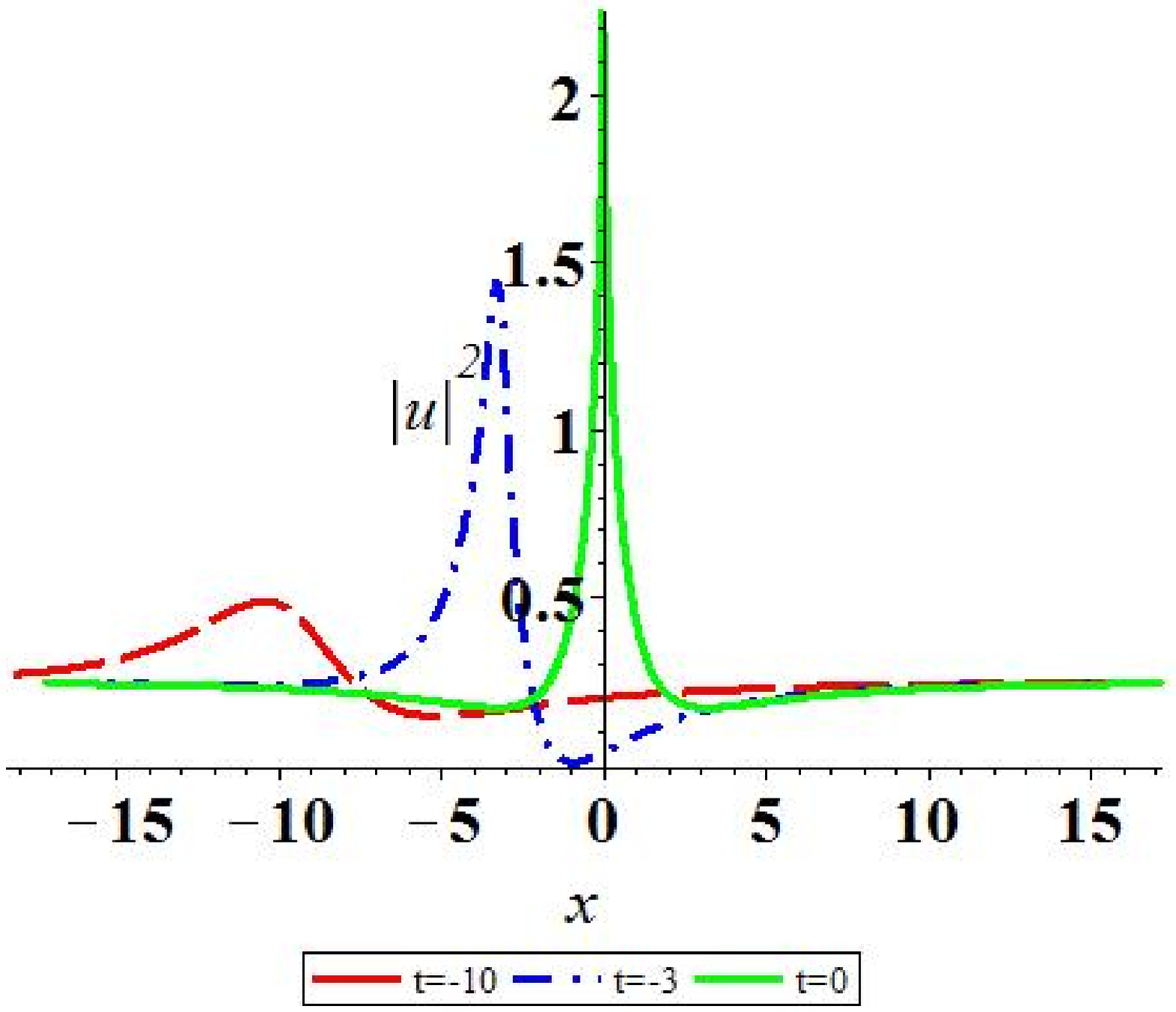}}
\caption{(color online): Parameters: $\protect\beta =1$, $\protect\gamma =%
\protect\sqrt{3}$, (a) The spatio-temporal pattern for the first order
wave-breaking RW, (b) The figure of $|q[1]|^{2}$ for different time, it is
seen that the derivative for the amplitude $|q|^{2}$ at $(x,t)=(0,0)$ is
very large.}
\label{fig3}
\end{figure}

Specifically, the first order rogue wave solution can be written explicitly
through formula \eqref{gene-rogue}
\begin{equation}
\begin{split}
q[1]=& \frac{\beta }{2}\left[ 1+\frac{16(\mathrm{i}{\beta }^{2}y-{\beta }%
^{2}-{\gamma }^{2})}{\beta ^{2}\left( 2y-\gamma s\right) ^{2}+{\beta }^{4}{s}%
^{2}+4\gamma ^{2}+4{\beta }^{2}}\right] \mathrm{e}^{\mathrm{i}\theta }, \\
x=& -\frac{\gamma }{2}y-\frac{{\beta }^{2}}{8}s-\frac{4{\beta }^{2}\left( {%
\gamma }^{2}s+{\beta }^{2}s-2\gamma y\right) }{\beta ^{2}\left( 2y-\gamma
s\right) ^{2}+{\beta }^{4}{s}^{2}+4\gamma ^{2}+4{\beta }^{2}},\,\,t=-s.
\end{split}%
\end{equation}%
It can be shown that if $\beta ^{2}<\frac{\gamma ^{2}}{3}$, then one has the regular rogue wave solution (Fig. \ref{fig2}); if $\beta ^{2}=%
\frac{\gamma ^{2}}{3}$, then one obtains the cusponed rogue wave
solution, in which $|q_{x}|\rightarrow \infty$ at the peak point (Fig. \ref{fig3}); if $\beta ^{2}>\frac{\gamma ^{2}}{3}$, then we has the looped rogue wave solution (Fig. \ref{fig4}).  Although
both the NLS and the CSP equations possess the modulational instability (see
the Appendix), the rogue wave solution of the CSP equation \eqref{CSP} could
yield the singularity which is   different from the NLS equation. This
solution may be related to the wave breaking in the CSP equation.
\begin{figure}[tbh]
\centering
\subfigure[$|q|^2$]{\includegraphics[height=50mm,width=50mm]{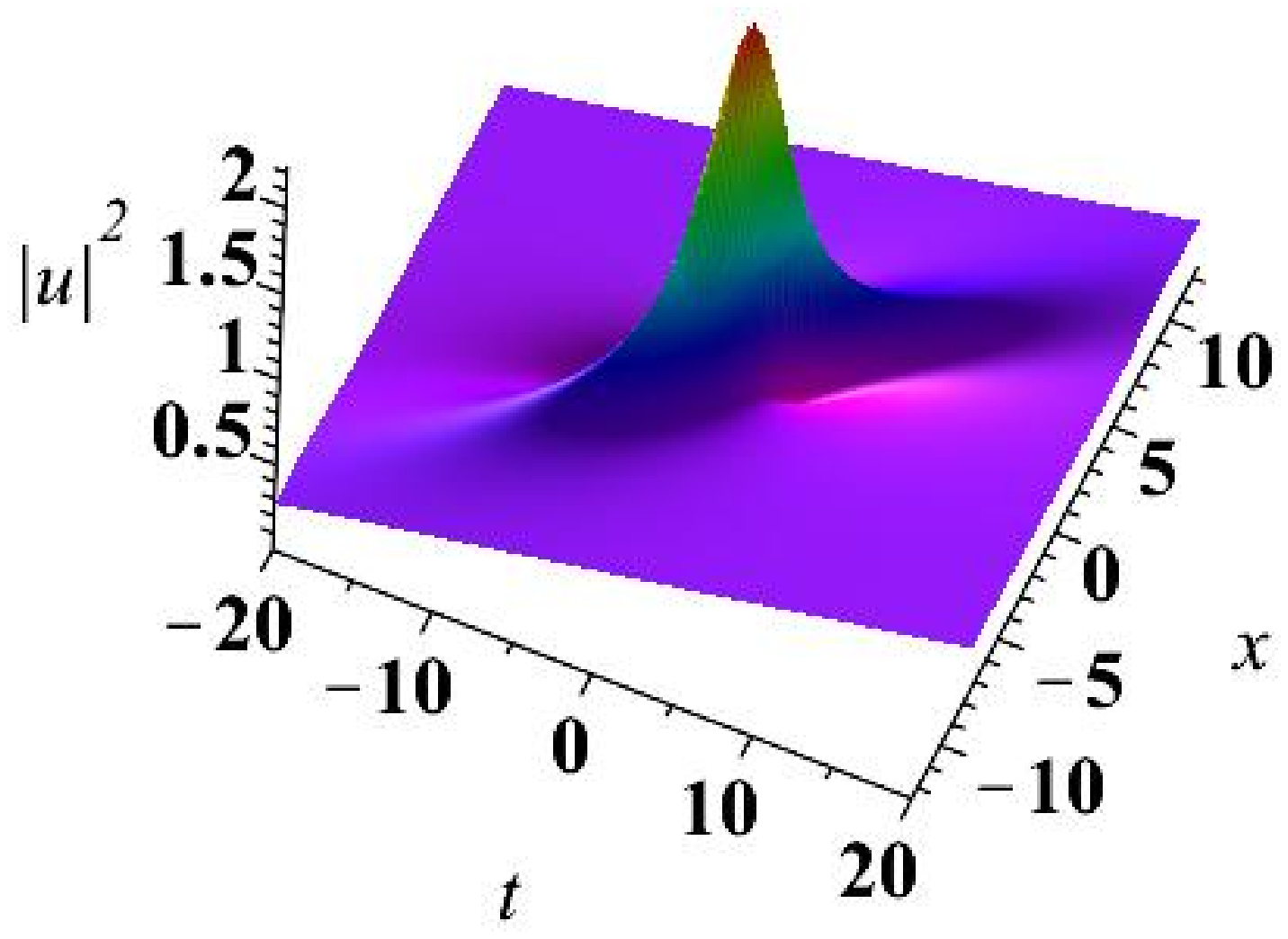}}
\hfil
\subfigure[$|q|^2$]{\includegraphics[height=50mm,width=50mm]{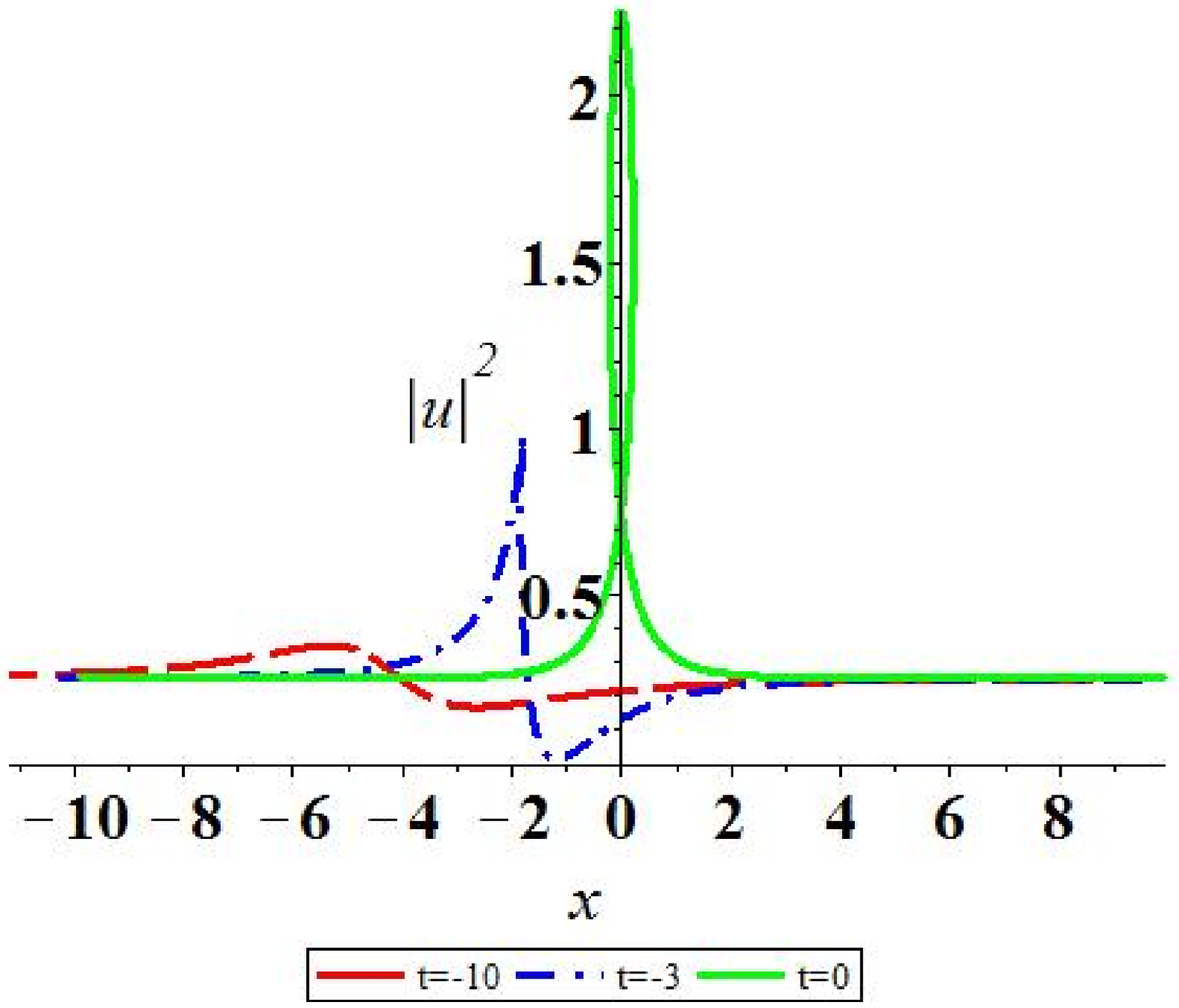}}
\caption{(color online): Parameters: $\protect\beta =1$, $\protect\gamma =1$%
, (a) The spatio-temporal pattern for the first order loop RW, (b) The
figure of $|q[1]|^{2}$ for different time, it is seen that the amplitude $%
|q|^{2}$ at $t=-10$ is regular, but when $t=0$, it appears a loop.}
\label{fig4}
\end{figure}
By the formula \eqref{csp-gene},  the second order rogue wave solution can
be calculated as
\begin{equation}
\begin{split}
q[2]=& \frac{\beta }{2}\left[ 1+\frac{G_{2}}{M_{2}}\right] \mathrm{e}^{%
\mathrm{i}\theta }, \\
x=& -\frac{\gamma }{2}y-\frac{{\beta }^{2}}{8}s-2\ln _{s}(M_{2}),\,\,t=-s,
\end{split}%
\end{equation}%
where
\begin{equation}
\begin{split}
M_{2}=& \beta ^{6}A\hat{y}^{6}+\left[ 3{\beta }^{6}A\hat{s}^{2}+12{\beta }%
^{6}+108{\beta }^{4}{\gamma }^{2}\right] \hat{y}^{4}+\left[ -288\gamma {%
\beta }^{5}\hat{s}-96b_{{1}}A{\beta }^{6}\right] \hat{y}^{3} \\
& +\left[ 3{\beta }^{6}A\hat{s}^{4}+\left( -72{\beta }^{4}{\gamma }^{2}+216{%
\beta }^{6}\right) \hat{s}^{2}-288a_{{1}}{\beta }^{6}A\hat{s}+432{\beta }%
^{4}+1584{\gamma }^{2}{\beta }^{2}\right] \hat{y}^{2} \\
& +\left[ 96\gamma {\beta }^{5}\hat{s}^{3}+288Ab_{{1}}{\beta }^{6}\hat{s}%
^{2}-1152\gamma {\beta }^{3}\hat{s}+4608a_{{1}}{\beta }^{5}\gamma +1152b_{{1}%
}{\beta }^{6}-3456b_{{1}}{\beta }^{4}{\gamma }^{2}\right] \hat{y} \\
& +A\left[ {\beta }^{6}\hat{s}^{6}+12{\beta }^{4}\hat{s}^{4}+96a_{{1}}{\beta
}^{6}\hat{s}^{3}+432{\beta }^{2}\hat{s}^{2}-1152a_{{1}}{\beta }^{4}\hat{s}%
+576+2304\left( {a_{1}}^{2}+{b_{1}}^{2}\right) {\beta }^{6}\right] , \\
G_{2}=& a_{{1}}{\beta }^{4}A\left( 2304\mathrm{i}\beta \hat{y}+4068\right)
\hat{s}+A\left[ 1152\mathrm{i}{\beta }^{5}\left( {\hat{y}}^{2}-\hat{s}%
^{2}\right) +4068{\beta }^{4}\hat{y}-4608\mathrm{i}{\beta }^{3}\right] b_{{1}%
} \\
& -24\mathrm{i}A{\beta }^{5}{\hat{y}}^{5}-240{\beta }^{4}A{\hat{y}}^{4}+%
\left[ -48\mathrm{i}A{\beta }^{5}\hat{s}^{2}-192\mathrm{i}{\gamma }^{2}{%
\beta }^{3}+960\mathrm{i}{\beta }^{5}\right] \hat{y}^{3} \\
& +\left[ -288{\beta }^{4}A\hat{s}^{2}+2304\mathrm{i}\gamma {\beta }^{4}\hat{%
s}-3456{\gamma }^{2}{\beta }^{2}+1152{\beta }^{4}\right] {\hat{y}}^{2} \\
& +\left[ -24\mathrm{i}A{\beta }^{5}\hat{s}^{4}+576\mathrm{i}\left( {\gamma }%
^{2}-{\beta }^{2}\right) {\beta }^{3}\hat{s}^{2}+4608{\beta }^{3}\gamma \hat{%
s}+5760\mathrm{i}{\gamma }^{2}\beta +1152\mathrm{i}{\beta }^{3}\right] \hat{y%
} \\
& -A(48{\beta }^{4}\hat{s}^{4}+1152{\beta }^{2}\hat{s}^{2}-2304)
\end{split}%
\end{equation}%
and
\begin{equation*}
\hat{s}=s-\frac{2\gamma y}{A},\,\,\hat{y}=\frac{-2\beta y}{A},\,\,A=\beta
^{2}+\gamma ^{2}.
\end{equation*}
The spatio-temporal pattern of the second order RW solution is
similar to the ones of the NLS equation \cite{Guo1} or derivative NLS equation \cite{Guo2}.
An example is shown in (Fig. \ref{fig5}b). For the general case, it is
impossible to describe their dynamics analytically.
%Also we can not obtain the condition to determine the types of rogue wave.
However for the standard case
$a_{1}=b_{1}=0$, it is shown that if $\beta ^{2}<\left( 1-\frac{2\sqrt{5}}{5}%
\right) \gamma ^{2}$, one obtains the regular rogue wave (Fig. \ref{fig5}a);
if $\beta ^{2}=\left( 1-\frac{2\sqrt{5}}{5}\right) \gamma ^{2}$, one can
obtain the cuspon-type rogue wave; if $\beta ^{2}>\left( 1-\frac{2\sqrt{5}}{5%
}\right) \gamma ^{2}$, one arrives at the loop-type rogue wave (Fig. \ref%
{fig6}).

\begin{figure}[htb]
\centering
\subfigure[$|q|^2$]{%
\includegraphics[height=50mm,width=50mm]{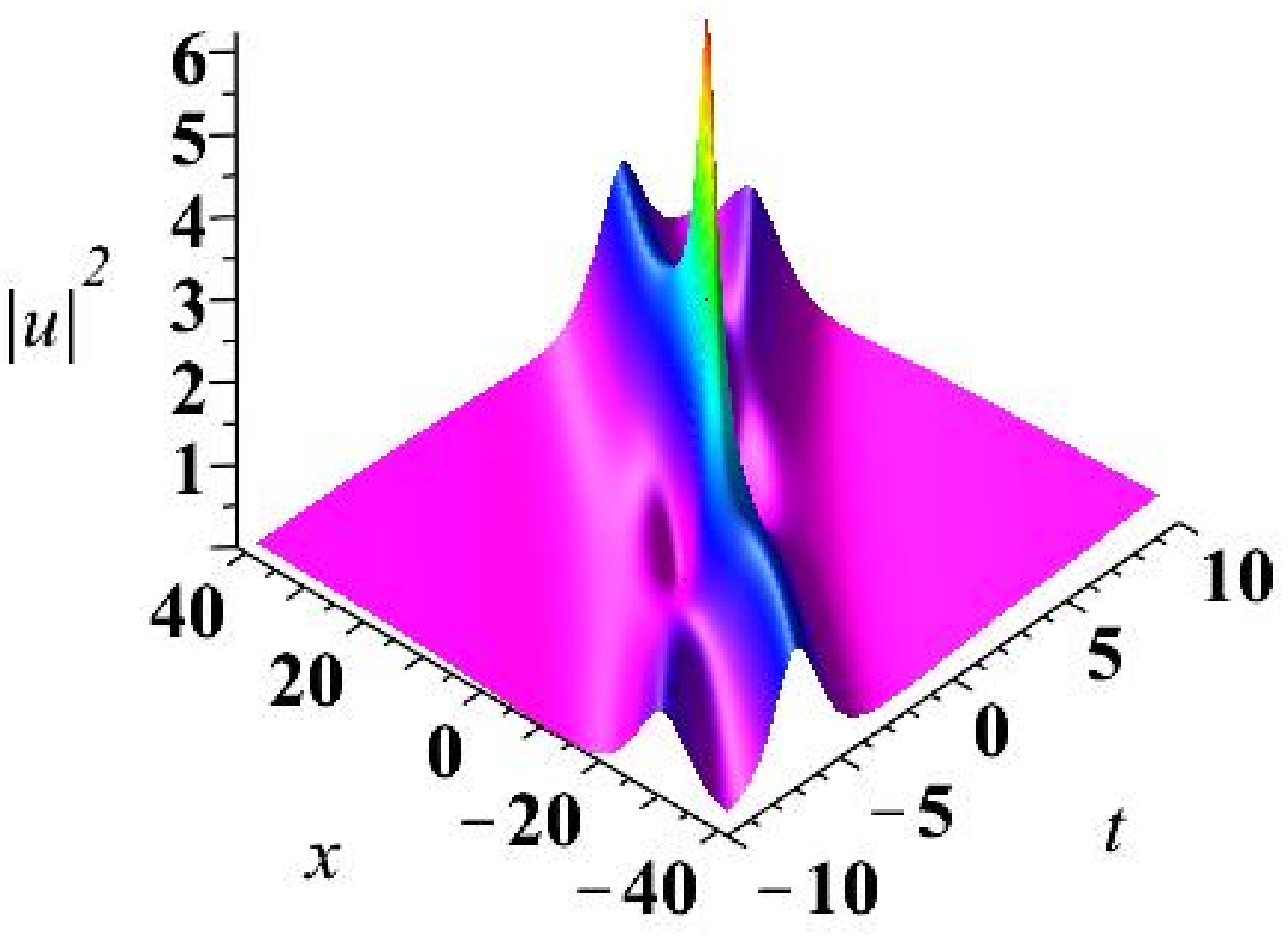}} \hfil
\subfigure[$|q|^2$]{%
\includegraphics[height=50mm,width=50mm]{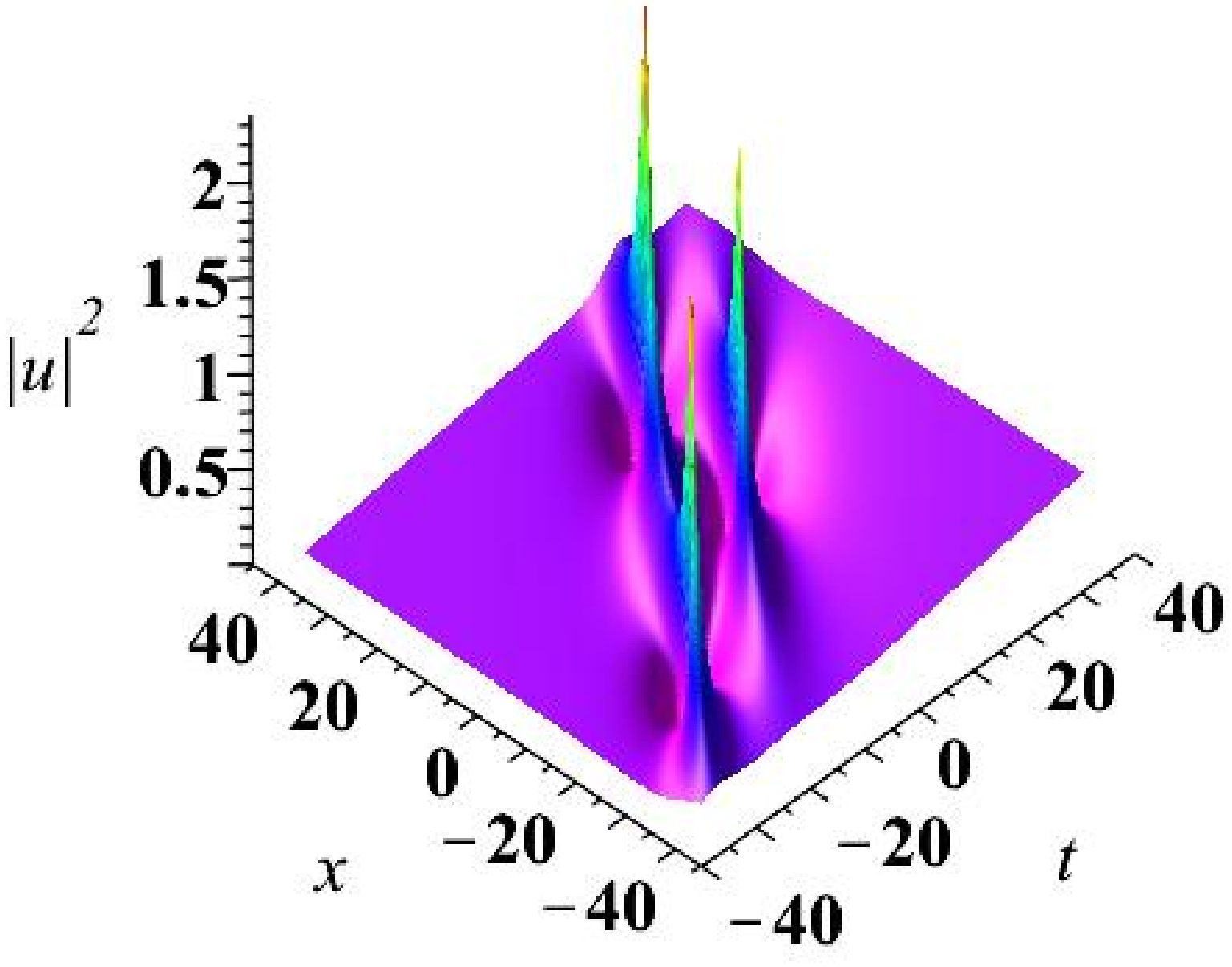}}
\caption{(color online): (a)The second order regular RW; Parameters: $%
\protect\beta=1$, $\protect\gamma=4$, $a_1=b_1=0$, (b) The second order
regular RW; Parameters: $\protect\beta=1$, $\protect\gamma=2$, $a_1=20,b_1=0$%
,}
\label{fig5}
\end{figure}
\begin{figure}[htb]
\centering
\subfigure[$|q|^2$]{%
\includegraphics[height=50mm,width=50mm]{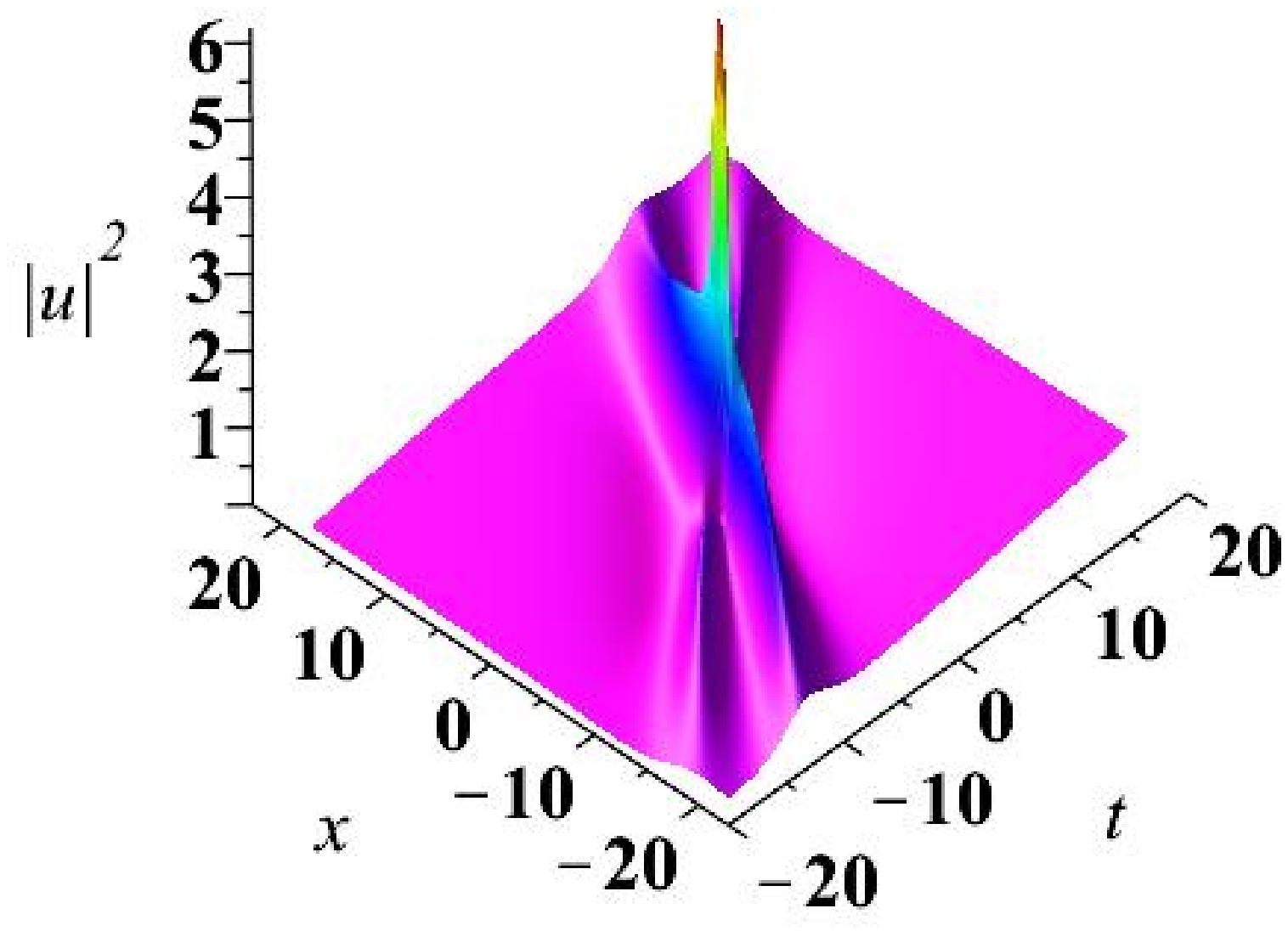}} \hfil
\subfigure[$|q|^2$]{%
\includegraphics[height=50mm,width=50mm]{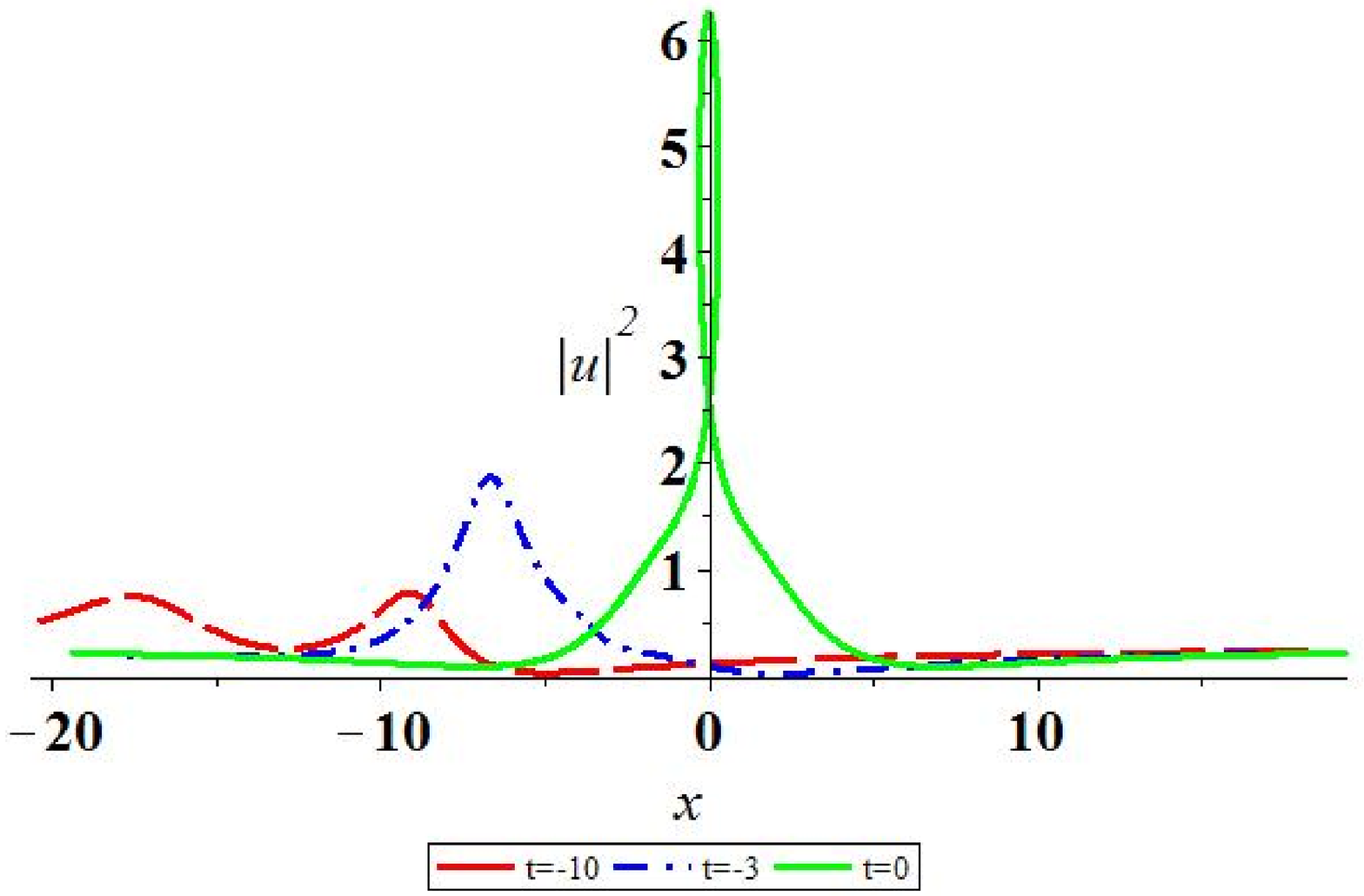}}
\caption{(color online): Parameters: $\protect\beta=1$, $\protect\gamma=2$, $%
a_1=b_1=0$ (a): The second order loop RW; (b) The different time of $%
|q[2]|^2 $.}
\label{fig6}
\end{figure}
\begin{figure}[htb]
\centering
\subfigure[$|q|^2$]{%
\includegraphics[height=50mm,width=50mm]{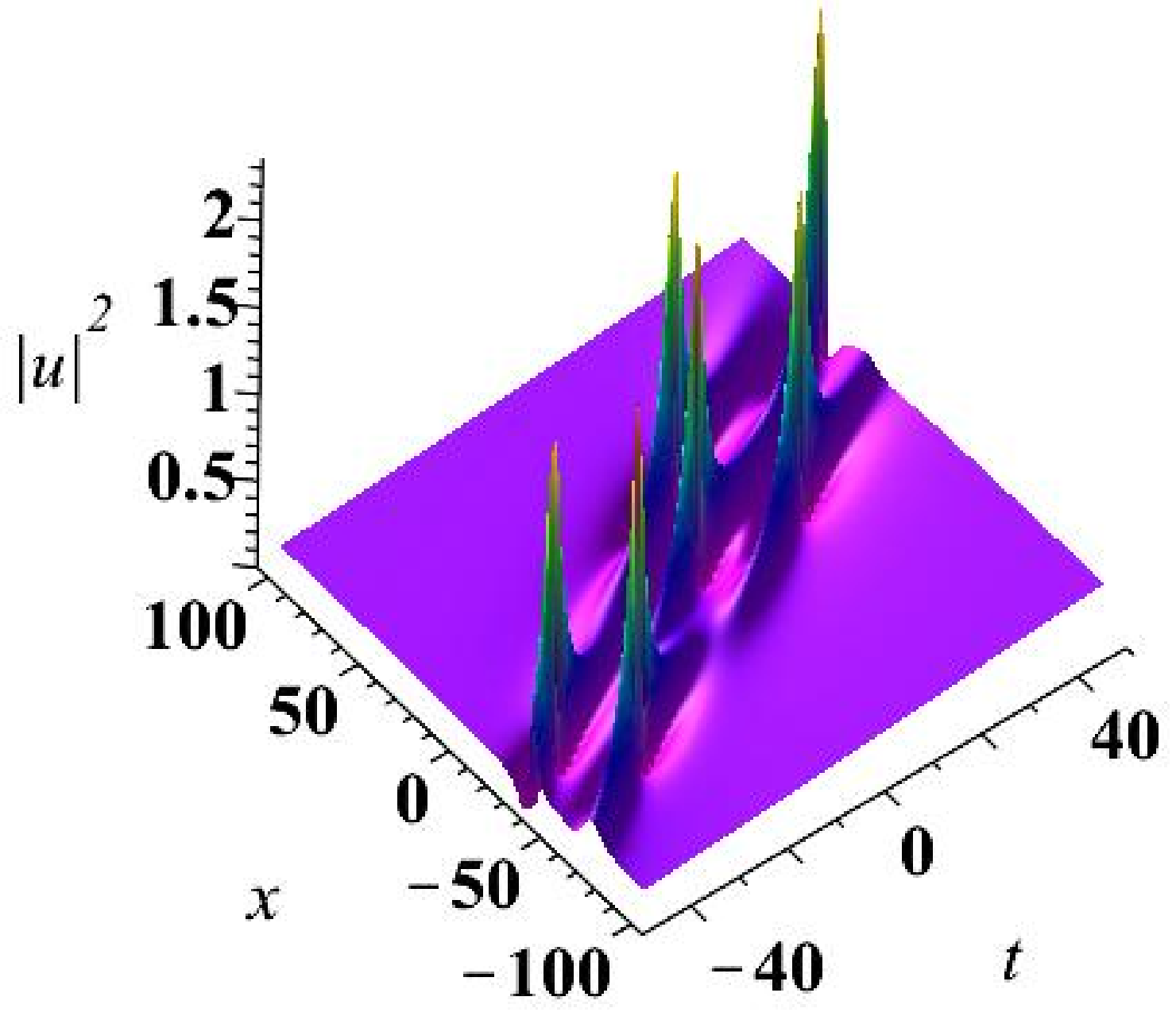}} \hfil
\subfigure[$|q|^2$]{%
\includegraphics[height=50mm,width=50mm]{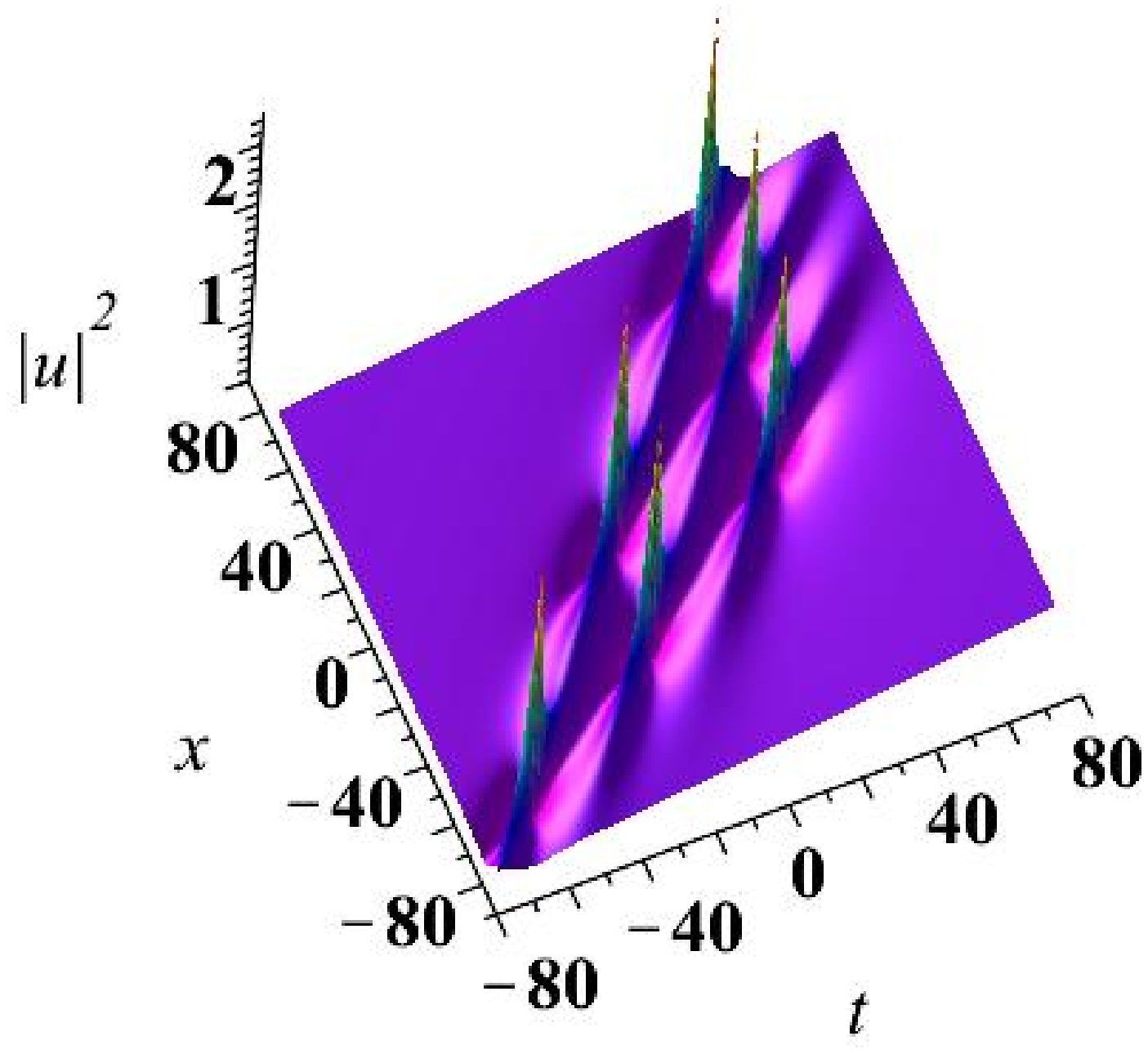}}
\caption{(color online): (a)The third order RW with pentagon arrangement.
Parameters: $\protect\beta=1$, $\protect\gamma=2$, $a_2=500$, $a_1=0$, $%
b_1=b_2=0$, (b) The third order RW with triangle arrangement. $\protect\beta%
=1$, $\protect\gamma=2$, $a_2=0$, $a_1=100$, $b_1=b_2=0$, }
\label{fig7}
\end{figure}
The expression for the higher order rogue wave solution $N\geq 3$ becomes very complicated.
Here, we only illustrate a third-order rogue wave solution (Fig. \ref{fig7})
without providing an analytical expression.
\section{Conclusions and discussions}
\label{section5}
In the present paper, we study the complex short pulse (CSP) equation by Darboux transformation method.
We firstly develop a generalized Darboux transformation (DT) and associated B\"acklund transformation for the complex coupled dispersionless (CCD) equation, which leads to a general soliton formulas for the CCD equation. Then by integrating the reciprocal
transformation exactly, the $N$-bright soliton solution in a compact determinant form is constructed. Furthermore, 
the $N$-breather solution and higher order rogue wave solution to the CSP equation are constructed by a delicate limiting process.
 
The $N$-bright soliton solution should be equivalent to the ones found by one of the authors \cite{Feng2,shen}, 
the $N$-breather solution and higher order rogue wave solution are found to the CSP equation for the first time and deserve further study.
Especially, this is the first example for the existence of rogue wave solution in a nonlinear wave equation possessing reciprocal (hodograph) transformation. Due to this reciprocal transformation, all the localized solutions including the bright, breather and rogue wave ones can be either smooth, cupson or loop ones.
Based on the compact determinant form of the solutions, we perform an asymptotic analysis
for the $N$-bright soliton and $N$-breather solutions. It should be pointed
out that the method for the asymptotical analysis can be extended to other
integrable equations as well.
In compared to the NLS equation, the rogue wave solution for the
CSP equation \eqref{CSP} could develop into wave-breaking. This illustrates that
the modulational instability for the CSP equation \eqref{CSP} is stronger than
the NLS equation.

Finally, the CSP equation could be of defocusing type, which admits the dark soliton solution, same as the NLS equation. It turns out that this is indeed the case. The complex short pulse equation of both focusing and defocusing types can be derived from the context of nonlinear optics. The results are summarized in a separate work \cite{LingFengdCSP}.
\section*{Acknowledgments}
%The work of BF is partially supported by the National Natural Science Foundation of China under grant 11428102, that of ZNZ by the NSFC under grant 11271254, and in part by the Ministry of
%Economy and Competitiveness of Spain under contract MTM2012-37070.
This work is partially supported by National Natural Science Foundation of China
(Nos. 11401221,11271254,11428102), Fundamental Research Funds for the Central
Universities (No. 2014ZB0034) and by the Ministry of Economy and Competitiveness of Spain under contract MTM2012-37070.

\section*{Appendix A: Proof of Proposition 3}
\textbf{Proof:} Fixed $y-v_{k}s=%
\mathrm{const}$, and $s\rightarrow -\infty $, it follows that $\theta
_{1},\theta _{2},\cdots ,\theta _{k-1}\rightarrow -\infty $; $\theta
_{k+1},\theta _{k+2},\cdots ,\theta _{N}\rightarrow +\infty .$ On the other
hand, $q[N]$ can be rewritten as
\begin{equation}
q[N]=-\frac{\det (\widehat{G})}{\det (\widehat{M})},  \label{n-soliton}
\end{equation}%
where
\begin{equation*}
\begin{split}
\widehat{M}& =\left( \frac{\mathrm{e}^{2(\theta _{i}^{\ast }+\theta _{j})}+1%
}{\lambda _{i}^{\ast }-\lambda _{j}}\right) _{1\leq i,j\leq N},\,\,\widehat{G%
}=%
\begin{bmatrix}
\widehat{M} & \widehat{Y_{1}}^{\dag } \\
\widehat{Y_{2}} & 0 \\
\end{bmatrix}%
, \\
\widehat{Y_{1}}& =%
\begin{bmatrix}
\mathrm{e}^{2\theta _{1}} & \mathrm{e}^{2\theta _{2}} & \cdots & \mathrm{e}%
^{2\theta _{N}} \\
\end{bmatrix}%
,\,\,\widehat{Y_{2}}=%
\begin{bmatrix}
1 & 1 & \cdots & 1 \\
\end{bmatrix}%
.
\end{split}%
\end{equation*}%
It follows that
\begin{equation*}
\begin{split}
\det (\widehat{M})=& \mathrm{e}^{2(\theta _{k+1}+\theta _{k+2}+\cdots
+\theta _{N})}\left[ \det (M_{k})+O(\mathrm{e}^{-c|s|})\right] , \\
\det (\widehat{G})=& \mathrm{e}^{2(\theta _{k+1}+\theta _{k+2}+\cdots
+\theta _{N})}\left[ \det (G_{k})+O(\mathrm{e}^{-c|s|})\right] ,
\end{split}%
\end{equation*}%
where
\begin{equation*}
\begin{split}
M_{k}& =%
\begin{bmatrix}
\frac{1}{\lambda _{1}^{\ast }-\lambda _{1}} & \cdots & \frac{1}{\lambda
_{1}^{\ast }-\lambda _{k-1}} & \frac{1}{\lambda _{1}^{\ast }-\lambda _{k}} &
0 & \cdots & 0 \\
\vdots & \ddots & \vdots & \vdots & \vdots & \ddots & \vdots \\
\frac{1}{\lambda _{k-1}^{\ast }-\lambda _{1}} & \cdots & \frac{1}{\lambda
_{k-1}^{\ast }-\lambda _{k-1}} & \frac{1}{\lambda _{k-1}^{\ast }-\lambda _{k}%
} & 0 & \cdots & 0 \\
\frac{1}{\lambda _{k}^{\ast }-\lambda _{1}} & \cdots & \frac{1}{\lambda
_{k}^{\ast }-\lambda _{k-1}} & \frac{\mathrm{e}^{2(\theta _{k}^{\ast
}+\theta _{k})}+1}{\lambda _{k}^{\ast }-\lambda _{k}} & \frac{\mathrm{e}%
^{2\theta _{k}^{\ast }}}{\lambda _{k}^{\ast }-\lambda _{k+1}} & \cdots &
\frac{\mathrm{e}^{2\theta _{k}^{\ast }}}{\lambda _{k}^{\ast }-\lambda _{N}}
\\
0 & \cdots & 0 & \frac{\mathrm{e}^{2\theta _{k}}}{\lambda _{k+1}^{\ast
}-\lambda _{k}} & \frac{1}{\lambda _{k+1}^{\ast }-\lambda _{k+1}} & \cdots &
\frac{1}{\lambda _{k+1}^{\ast }-\lambda _{N}} \\
\vdots & \ddots & \vdots & \vdots & \vdots & \ddots & \vdots \\
0 & \cdots & 0 & \frac{\mathrm{e}^{2\theta _{k}}}{\lambda _{N}^{\ast
}-\lambda _{k}} & \frac{1}{\lambda _{N}^{\ast }-\lambda _{k+1}} & \cdots &
\frac{1}{\lambda _{N}^{\ast }-\lambda _{N}} \\
&  &  &  &  &  &
\end{bmatrix}%
, \\
G_{k}& =%
\begin{bmatrix}
\frac{1}{\lambda _{1}^{\ast }-\lambda _{1}} & \cdots & \frac{1}{\lambda
_{1}^{\ast }-\lambda _{k-1}} & \frac{1}{\lambda _{1}^{\ast }-\lambda _{k}} &
0 & \cdots & 0 & 0 \\
\vdots & \ddots & \vdots & \vdots & \vdots & \ddots & \vdots & \vdots \\
\frac{1}{\lambda _{k-1}^{\ast }-\lambda _{1}} & \cdots & \frac{1}{\lambda
_{k-1}^{\ast }-\lambda _{k-1}} & \frac{1}{\lambda _{k-1}^{\ast }-\lambda _{k}%
} & 0 & \cdots & 0 & 0 \\
\frac{1}{\lambda _{k}^{\ast }-\lambda _{1}} & \cdots & \frac{1}{\lambda
_{k}^{\ast }-\lambda _{k-1}} & \frac{\mathrm{e}^{2(\theta _{k}^{\ast
}+\theta _{k})}+1}{\lambda _{k}^{\ast }-\lambda _{k}} & \frac{\mathrm{e}%
^{2\theta _{k}^{\ast }}}{\lambda _{k}^{\ast }-\lambda _{k+1}} & \cdots &
\frac{\mathrm{e}^{2\theta _{k}^{\ast }}}{\lambda _{k}^{\ast }-\lambda _{N}}
& \mathrm{e}^{2\theta _{k}^{\ast }} \\
0 & \cdots & 0 & \frac{\mathrm{e}^{2\theta _{k}}}{\lambda _{k+1}^{\ast
}-\lambda _{k}} & \frac{1}{\lambda _{k+1}^{\ast }-\lambda _{k+1}} & \cdots &
\frac{1}{\lambda _{k+1}^{\ast }-\lambda _{N}} & 1 \\
\vdots & \ddots & \vdots & \vdots & \vdots & \ddots & \vdots & \vdots \\
0 & \cdots & 0 & \frac{\mathrm{e}^{2\theta _{k}}}{\lambda _{N}^{\ast
}-\lambda _{k}} & \frac{1}{\lambda _{N}^{\ast }-\lambda _{k+1}} & \cdots &
\frac{1}{\lambda _{N}^{\ast }-\lambda _{N}} & 1 \\
1 & \cdots & 1 & 1 & 0 & \cdots & 0 & 0 \\
&  &  &  &  &  &  &
\end{bmatrix}%
.
\end{split}%
\end{equation*}%
By direct calculation, we have
\begin{equation*}
\begin{split}
\det (G_{k})=& (-1)^{k+N+1}%
\begin{vmatrix}
\frac{1}{\lambda _{1}^{\ast }-\lambda _{1}} & \cdots & \frac{1}{\lambda
_{1}^{\ast }-\lambda _{k-1}} & \frac{1}{\lambda _{1}^{\ast }-\lambda _{k}}
\\
\vdots & \ddots & \vdots & \vdots \\
\frac{1}{\lambda _{k-1}^{\ast }-\lambda _{1}} & \cdots & \frac{1}{\lambda
_{k-1}^{\ast }-\lambda _{k-1}} & \frac{1}{\lambda _{k-1}^{\ast }-\lambda _{k}%
} \\
1 & \cdots & 1 & 1 \\
&  &  &
\end{vmatrix}%
\begin{vmatrix}
\frac{\mathrm{e}^{2\theta _{k}^{\ast }}}{\lambda _{k}^{\ast }-\lambda _{k+1}}
& \cdots & \frac{\mathrm{e}^{2\theta _{k}^{\ast }}}{\lambda _{k}^{\ast
}-\lambda _{N}} & \mathrm{e}^{2\theta _{k}^{\ast }} \\
\frac{1}{\lambda _{k+1}^{\ast }-\lambda _{k+1}} & \cdots & \frac{1}{\lambda
_{k+1}^{\ast }-\lambda _{N}} & 1 \\
\vdots & \ddots & \vdots & \vdots \\
\ \frac{1}{\lambda _{N}^{\ast }-\lambda _{k+1}} & \cdots & \frac{1}{\lambda
_{N}^{\ast }-\lambda _{N}} & 1 \\
&  &  &
\end{vmatrix}
\\
=& -\prod_{l=1}^{k-1}\left( \frac{\lambda _{l}-\lambda _{k}}{\lambda
_{l}^{\ast }-\lambda _{k}}\right) \prod_{l=k+1}^{N}\left( \frac{\lambda
_{l}^{\ast }-\lambda _{k}^{\ast }}{\lambda _{l}-\lambda _{k}^{\ast }}\right)
C(\lambda _{1}^{\ast },\lambda _{2}^{\ast },\cdots ,\lambda _{k-1}^{\ast
})C(\lambda _{k+1}^{\ast },\lambda _{k+2}^{\ast },\cdots ,\lambda _{N}^{\ast
})\mathrm{e}^{2\theta _{k}^{\ast }};
\end{split}%
\end{equation*}%
and
\begin{equation*}
\begin{split}
& \det (M_{k}) \\
=&
\begin{vmatrix}
\frac{1}{\lambda _{1}^{\ast }-\lambda _{1}} & \cdots & \frac{1}{\lambda
_{1}^{\ast }-\lambda _{k}} & 0 & \cdots & 0 \\
\vdots & \ddots & \vdots & \vdots & \ddots & \vdots \\
\frac{1}{\lambda _{k-1}^{\ast }-\lambda _{1}} & \cdots & \frac{1}{\lambda
_{k-1}^{\ast }-\lambda _{k}} & 0 & \cdots & 0 \\
\frac{1}{\lambda _{k}^{\ast }-\lambda _{1}} & \cdots & \frac{1}{\lambda
_{k}^{\ast }-\lambda _{k}} & \frac{\mathrm{e}^{2\theta _{k}^{\ast }}}{%
\lambda _{k}^{\ast }-\lambda _{k+1}} & \cdots & \frac{\mathrm{e}^{2\theta
_{k}^{\ast }}}{\lambda _{k}^{\ast }-\lambda _{N}} \\
0 & \cdots & 0 & \frac{1}{\lambda _{k+1}^{\ast }-\lambda _{k+1}} & \cdots &
\frac{1}{\lambda _{k+1}^{\ast }-\lambda _{N}} \\
\vdots & \ddots & \vdots & \vdots & \ddots & \vdots \\
0 & \cdots & 0 & \frac{1}{\lambda _{N}^{\ast }-\lambda _{k+1}} & \cdots &
\frac{1}{\lambda _{N}^{\ast }-\lambda _{N}} \\
&  &  &  &  &
\end{vmatrix}%
+%
\begin{vmatrix}
\frac{1}{\lambda _{1}^{\ast }-\lambda _{1}} & \cdots & \frac{1}{\lambda
_{1}^{\ast }-\lambda _{k-1}} & 0 & 0 & \cdots & 0 \\
\vdots & \ddots & \vdots & \vdots & \vdots & \ddots & \vdots \\
\frac{1}{\lambda _{k-1}^{\ast }-\lambda _{1}} & \cdots & \frac{1}{\lambda
_{k-1}^{\ast }-\lambda _{k-1}} & 0 & 0 & \cdots & 0 \\
\frac{1}{\lambda _{k}^{\ast }-\lambda _{1}} & \cdots & \frac{1}{\lambda
_{k}^{\ast }-\lambda _{k-1}} & \frac{\mathrm{e}^{2(\theta _{k}^{\ast
}+\theta _{k})}}{\lambda _{k}^{\ast }-\lambda _{k}} & \frac{\mathrm{e}%
^{2\theta _{k}^{\ast }}}{\lambda _{k}^{\ast }-\lambda _{k+1}} & \cdots &
\frac{\mathrm{e}^{2\theta _{k}^{\ast }}}{\lambda _{k}^{\ast }-\lambda _{N}}
\\
0 & \cdots & 0 & \frac{\mathrm{e}^{2\theta _{k}}}{\lambda _{k+1}^{\ast
}-\lambda _{k}} & \frac{1}{\lambda _{k+1}^{\ast }-\lambda _{k+1}} & \cdots &
\frac{1}{\lambda _{k+1}^{\ast }-\lambda _{N}} \\
\vdots & \ddots & \vdots & \vdots & \vdots & \ddots & \vdots \\
0 & \cdots & 0 & \frac{\mathrm{e}^{2\theta _{k}}}{\lambda _{N}^{\ast
}-\lambda _{k}} & \frac{1}{\lambda _{N}^{\ast }-\lambda _{k+1}} & \cdots &
\frac{1}{\lambda _{N}^{\ast }-\lambda _{N}} \\
&  &  &  &  &  &
\end{vmatrix}
\\
=& C(\lambda _{1}^{\ast },\lambda _{2}^{\ast },\cdots ,\lambda _{k}^{\ast
})C(\lambda _{k+1}^{\ast },\lambda _{k+2}^{\ast },\cdots ,\lambda _{N}^{\ast
})+C(\lambda _{1}^{\ast },\lambda _{2}^{\ast },\cdots ,\lambda _{k-1}^{\ast
})C(\lambda _{k}^{\ast },\lambda _{k+1}^{\ast },\cdots ,\lambda _{N}^{\ast })%
\mathrm{e}^{2(\theta _{k}^{\ast }+\theta _{k})} \\
=& \frac{1}{\lambda _{k}^{\ast }-\lambda _{k}}C(\lambda _{1}^{\ast },\lambda
_{2}^{\ast },\cdots ,\lambda _{k-1}^{\ast })C(\lambda _{k+1}^{\ast },\lambda
_{k+2}^{\ast },\cdots ,\lambda _{N}^{\ast })\left(
\prod_{l=1}^{k-1}\left\vert \frac{\lambda _{l}-\lambda _{k}}{\lambda
_{l}^{\ast }-\lambda _{k}}\right\vert ^{2}+\prod_{l=k+1}^{N}\left\vert \frac{%
\lambda _{l}-\lambda _{k}}{\lambda _{l}^{\ast }-\lambda _{k}}\right\vert ^{2}%
\mathrm{e}^{2(\theta _{k}^{\ast }+\theta _{k})}\right) ,
\end{split}%
\end{equation*}%
where $C(a_1^*,a_2^*,\cdots,a_m^*)=\det(\frac{1}{a_i^*-a_j})_{1\leq i,j\leq m}$ represents the determinant of a Cauchy
matrix.
Thus, along the trajectory $y-v_{k}s=\mathrm{const}$, we have $q[N]=\lambda
_{k,I}\mathrm{sech}(2\theta _{k,R}^{-})\mathrm{e}^{-2\mathrm{i}\theta
_{k,I}^{-}-\frac{\pi \mathrm{i}}{2}}+O(\mathrm{e}^{-c|s|}).$

For the general
case $y-vs=\mathrm{const}$, $v\neq v_{k}(k=1,2,\cdots ,N)$, we have $q[N]=O(%
\mathrm{e}^{-c|s|}).$ Thus we obtain the asymptotic behavior %
\eqref{asym-soliton} when $s\rightarrow -\infty .$

By the same procedure as above, we can obtain the asymptotical behavior %
\eqref{asym-soliton} when $s\rightarrow +\infty .$ Finally, we have obtain
the $N$-soliton's asymptotic behavior \eqref{asym-soliton}. $\square$
\section*{Appendix B: Proof of Proposition 5}
\textbf{Proof:} Fixed $y-\frac{2}{\gamma}v_ks=\mathrm{const}$, $l<k\leq N$, and $%
s\rightarrow +\infty$, it follows that $\theta_1,\theta_2,\cdots,%
\theta_{k-1}\rightarrow -\infty$; $\theta_{k+1},\theta_{k+2},\cdots,%
\theta_{N}\rightarrow +\infty.$ It follows that
\begin{equation*}
\begin{split}
\det(M)=&\exp\!{\left[\sum_{i=k+1}^{N}(\theta_i+\theta_i^*)-\sum_{i=1}^{k}(%
\theta_i+\theta_i^*)\right]}\left(\det(M_k)+O(\mathrm{e}^{-c|s|})\right) \\
\det(G)=&\exp\!{\left[\sum_{i=k+1}^{N}(\theta_i+\theta_i^*)-\sum_{i=1}^{k}(%
\theta_i+\theta_i^*)\right]}\left(\det(G_k)+O(\mathrm{e}^{-c|s|})\right)
\end{split}%
\end{equation*}
where
\begin{equation*}
\begin{split}
M_k&=%
\begin{bmatrix}
\frac{1}{\chi_{1}^*-\chi_{1}} & \cdots & \frac{1}{\chi_{1}^*-\chi_{k-1}} &
\frac{\mathrm{e}^{2\theta_k}}{\chi_{1}^*-\xi_{k}}+\frac{1}{%
\chi_{1}^*-\chi_{k}} & \frac{1}{\chi_{1}^*-\xi_{k+1}} & \cdots & \frac{1}{%
\chi_{1}^*-\xi_{N}} \\
\vdots & \ddots & \vdots & \vdots & \vdots & \ddots & \vdots \\
\frac{1}{\chi_{k-1}^*-\chi_{1}} & \cdots & \frac{1}{\chi_{k-1}^*-\chi_{k-1}}
& \frac{\mathrm{e}^{2\theta_k}}{\chi_{k-1}^*-\xi_{k}}+\frac{1}{%
\chi_{k-1}^*-\chi_{k}} & \frac{1}{\chi_{k-1}^*-\xi_{k+1}} & \cdots & \frac{1%
}{\chi_{k-1}^*-\xi_{N}} \\
\frac{\mathrm{e}^{2\theta_k^*}}{\xi_{k}^*-\chi_{1}}+\frac{1}{%
\chi_{k}^*-\chi_{1}} & \cdots & \frac{\mathrm{e}^{2\theta_k^*}}{%
\xi_{k}^*-\chi_{k-1}}+\frac{1}{\chi_{k}^*-\chi_{k-1}} & m_k & \frac{\mathrm{e%
}^{2\theta_k^*}}{\xi_{k}^*-\xi_{k+1}}+\frac{1}{\chi_{k}^*-\xi_{k+1}} & \cdots
& \frac{\mathrm{e}^{2\theta_k^*}}{\xi_{k}^*-\xi_{N}}+\frac{1}{%
\chi_{k}^*-\xi_{N}} \\
\frac{1}{\xi^*_{k+1}-\chi_1} & \cdots & \frac{1}{\xi^*_{k+1}-\chi_{k-1}} &
\frac{\mathrm{e}^{2\theta_k}}{\xi_{k+1}^*-\xi_{k}}+\frac{1}{%
\xi_{k+1}^*-\chi_{k}} & \frac{1}{\xi_{k+1}^*-\xi_{k+1}} & \cdots & \frac{1}{%
\xi_{k+1}^*-\xi_{N}} \\
\vdots & \ddots & \vdots & \vdots & \vdots & \ddots & \vdots \\
\frac{1}{\xi^*_{N}-\chi_1} & \cdots & \frac{1}{\xi^*_{N}-\chi_{k-1}} & \frac{%
\mathrm{e}^{2\theta_k}}{\xi_{N}^*-\xi_{k}}+\frac{1}{\xi_{N}^*-\chi_{k}} &
\frac{1}{\xi_N^*-\xi_{k+1}} & \cdots & \frac{1}{\xi_N^*-\xi_{N}} \\
&  &  &  &  &  &
\end{bmatrix}%
,
\end{split}%
\end{equation*}
\begin{multline*}
G_k=\left[%
\begin{array}{cccccccc}
\frac{1}{\chi_{1}^*-\chi_{1}}\frac{\chi_{1}^*+\gamma}{\chi_{1}+\gamma} &
\cdots & \frac{1}{\chi_{1}^*-\chi_{k-1}} \frac{\chi_{1}^*+\gamma}{%
\chi_{k-1}+\gamma} & \frac{\mathrm{e}^{2\theta_k}}{\chi_{1}^*-\xi_{k}}\frac{%
\chi_{1}^*+\gamma}{\xi_{k}+\gamma}+\frac{1}{\chi_{1}^*-\chi_{k}} \frac{%
\chi_{1}^*+\gamma}{\chi_{k}+\gamma} &  &  &  &  \\
\vdots & \ddots & \vdots & \vdots &  &  &  &  \\
\frac{1}{\chi_{k-1}^*-\chi_{1}}\frac{\chi_{k-1}^*+\gamma}{\chi_{1}+\gamma} &
\cdots & \frac{1}{\chi_{k-1}^*-\chi_{k-1}}\frac{\chi_{k-1}^*+\gamma}{%
\chi_{k-1}+\gamma} & \frac{\mathrm{e}^{2\theta_k}}{\chi_{k-1}^*-\xi_{k}}%
\frac{\chi_{k-1}^*+\gamma}{\xi_{k}+\gamma}+\frac{1}{\chi_{k-1}^*-\chi_{k}}
\frac{\chi_{k-1}^*+\gamma}{\chi_{k}+\gamma} &  &  &  &  \\
\frac{\mathrm{e}^{2\theta_k^*}}{\xi_{k}^*-\chi_{1}}\frac{\xi_{k}^*+\gamma}{%
\chi_{1}+\gamma}+\frac{1}{\chi_{k}^*-\chi_{1}}\frac{\chi_{k}^*+\gamma}{%
\chi_{1}+\gamma} & \cdots & \frac{\mathrm{e}^{2\theta_k^*}}{%
\xi_{k}^*-\chi_{k-1}}\frac{\xi_{k}^*+\gamma}{\chi_{k-1}+\gamma}+\frac{1}{%
\chi_{k}^*-\chi_{k-1}} \frac{\chi_{k}^*+\gamma}{\chi_{k-1}+\gamma} & g_k &
&  &  &  \\
\frac{1}{\xi^*_{k+1}-\chi_1}\frac{\xi_{k+1}^*+\gamma}{\chi_{1}+\gamma} &
\cdots & \frac{1}{\xi^*_{k+1}-\chi_{k-1}} \frac{\xi_{k+1}^*+\gamma}{%
\chi_{k-1}+\gamma} & \frac{\mathrm{e}^{2\theta_k}}{\xi_{k+1}^*-\xi_{k}}\frac{%
\xi_{k+1}^*+\gamma}{\xi_{k}+\gamma}+\frac{1}{\xi_{k+1}^*-\chi_{k}}\frac{%
\xi_{k+1}^*+\gamma}{\chi_{k}+\gamma} &  &  &  &  \\
\vdots & \ddots & \vdots & \vdots &  &  &  &  \\
\frac{1}{\xi^*_{N}-\chi_1}\frac{\xi_{N}^*+\gamma}{\chi_{1}+\gamma} & \cdots
& \frac{1}{\xi^*_{N}-\chi_{k-1}}\frac{\xi_{N}^*+\gamma}{\chi_{k-1}+\gamma} &
\frac{\mathrm{e}^{2\theta_k}}{\xi_{N}^*-\xi_{k}}\frac{\xi_{N}^*+\gamma}{%
\xi_{k}+\gamma}+\frac{1}{\xi_{N}^*-\chi_{k}} \frac{\xi_{N}^*+\gamma}{%
\chi_{k}+\gamma} &  &  &  &  \\
&  &  &  &  &  &  &
\end{array}%
\right. \\
\left.%
\begin{array}{ccc}
\frac{1}{\chi_{1}^*-\xi_{k+1}}\frac{\chi_{1}^*+\gamma}{\xi_{k+1}+\gamma} &
\cdots & \frac{1}{\chi_{1}^*-\xi_{N}}\frac{\chi_{1}^*+\gamma}{\xi_{N}+\gamma}
\\
\vdots & \ddots & \vdots \\
\frac{1}{\chi_{k-1}^*-\xi_{k+1}}\frac{\chi_{k-1}^*+\gamma}{\xi_{k+1}+\gamma}
& \cdots & \frac{1}{\chi_{k-1}^*-\xi_{N}}\frac{\chi_{k-1}^*+\gamma}{%
\xi_{N}+\gamma} \\
\frac{\mathrm{e}^{2\theta_k^*}}{\xi_{k}^*-\xi_{k+1}} \frac{\xi_{k}^*+\gamma}{%
\xi_{k+1}+\gamma}+\frac{1}{\chi_{k}^*-\xi_{k+1}}\frac{\chi_{k}^*+\gamma}{%
\xi_{k+1}+\gamma} & \cdots & \frac{\mathrm{e}^{2\theta_k^*}}{%
\xi_{k}^*-\xi_{N}}\frac{\xi_{k}^*+\gamma}{\xi_{N}+\gamma}+\frac{1}{%
\chi_{k}^*-\xi_{N}}\frac{\chi_{k}^*+\gamma}{\xi_{N}+\gamma} \\
\frac{1}{\xi_{k+1}^*-\xi_{k+1}}\frac{\xi_{k+1}^*+\gamma}{\xi_{k+1}+\gamma} &
\cdots & \frac{1}{\xi_{k+1}^*-\xi_{N}}\frac{\chi_{k+1}^*+\gamma}{%
\xi_{N}+\gamma} \\
\vdots & \ddots & \vdots \\
\frac{1}{\xi_N^*-\xi_{k+1}}\frac{\xi_{N}^*+\gamma}{\xi_{k+1}+\gamma} & \cdots
& \frac{1}{\xi_N^*-\xi_{N}}\frac{\xi_{N}^*+\gamma}{\xi_{N}+\gamma} \\
&  &
\end{array}%
\right]
\end{multline*}
and $m_k=\frac{\mathrm{e}^{2(\theta_k^*+\theta_k)}}{\xi_k^*-\xi_k}+\frac{%
\mathrm{e}^{2\theta_k^*}}{\xi_k^*-\chi_k}+\frac{\mathrm{e}^{2\theta_k}}{%
\chi_k^*-\xi_k} +\frac{1}{\chi_k^*-\chi_k},$ $g_k=\frac{\mathrm{e}%
^{2(\theta_k^*+\theta_k)}}{\xi_k^*-\xi_k}\frac{\xi_k^*+\gamma}{\xi_k+\gamma}+%
\frac{\mathrm{e}^{2\theta_k^*}}{\xi_k^*-\chi_k}\frac{\xi_k^*+\gamma}{%
\chi_k+\gamma}+\frac{\mathrm{e}^{2\theta_k}}{\chi_k^*-\xi_k}\frac{%
\chi_k^*+\gamma}{\xi_k+\gamma} +\frac{1}{\chi_k^*-\chi_k}\frac{%
\chi_k^*+\gamma}{\chi_k+\gamma}.$ Directly calculation, we have
\begin{equation*}
\begin{split}
\det(M_k)=&\det(M_k^{[1]})\mathrm{e}^{2(\theta_k^*+\theta_k)}+\det(M_k^{[2]})%
\mathrm{e}^{2\theta_k^*}+\det(M_k^{[3]})\mathrm{e}^{2\theta_k}+%
\det(M_k^{[4]})
\end{split}%
\end{equation*}
where
\begin{equation*}
M_k^{[1]}=%
\begin{bmatrix}
\frac{1}{\chi_{1}^*-\chi_{1}} & \cdots & \frac{1}{\chi_{1}^*-\chi_{k-1}} &
\frac{1}{\chi_{1}^*-\xi_{k}} & \frac{1}{\chi_{1}^*-\xi_{k+1}} & \cdots &
\frac{1}{\chi_{1}^*-\xi_{N}} \\
\vdots & \ddots & \vdots & \vdots & \vdots & \ddots & \vdots \\
\frac{1}{\chi_{k-1}^*-\chi_{1}} & \cdots & \frac{1}{\chi_{k-1}^*-\chi_{k-1}}
& \frac{1}{\chi_{k-1}^*-\xi_{k}} & \frac{1}{\chi_{k-1}^*-\xi_{k+1}} & \cdots
& \frac{1}{\chi_{k-1}^*-\xi_{N}} \\[8pt]
\frac{1}{\xi_{k}^*-\chi_{1}} & \cdots & \frac{1}{\xi_{k}^*-\chi_{k-1}} &
\frac{1}{\xi_k^*-\xi_k} & \frac{1}{\xi_{k}^*-\xi_{k+1}} & \cdots & \frac{1}{%
\xi_{k}^*-\xi_{N}} \\[8pt]
\frac{1}{\xi^*_{k+1}-\chi_1} & \cdots & \frac{1}{\xi^*_{k+1}-\chi_{k-1}} &
\frac{1}{\xi_{k+1}^*-\xi_{k}} & \frac{1}{\xi_{k+1}^*-\xi_{k+1}} & \cdots &
\frac{1}{\xi_{k+1}^*-\xi_{N}} \\
\vdots & \ddots & \vdots & \vdots & \vdots & \ddots & \vdots \\
\frac{1}{\xi^*_{N}-\chi_1} & \cdots & \frac{1}{\xi^*_{N}-\chi_{k-1}} & \frac{%
1}{\xi_{N}^*-\xi_{k}} & \frac{1}{\xi_N^*-\xi_{k+1}} & \cdots & \frac{1}{%
\xi_N^*-\xi_{N}} \\
&  &  &  &  &  &
\end{bmatrix}%
,
\end{equation*}
\begin{equation*}
M_k^{[2]}=%
\begin{bmatrix}
\frac{1}{\chi_{1}^*-\chi_{1}} & \cdots & \frac{1}{\chi_{1}^*-\chi_{k-1}} &
\frac{1}{\chi_{1}^*-\chi_{k}} & \frac{1}{\chi_{1}^*-\xi_{k+1}} & \cdots &
\frac{1}{\chi_{1}^*-\xi_{N}} \\
\vdots & \ddots & \vdots & \vdots & \vdots & \ddots & \vdots \\
\frac{1}{\chi_{k-1}^*-\chi_{1}} & \cdots & \frac{1}{\chi_{k-1}^*-\chi_{k-1}}
& \frac{1}{\chi_{k-1}^*-\chi_{k}} & \frac{1}{\chi_{k-1}^*-\xi_{k+1}} & \cdots
& \frac{1}{\chi_{k-1}^*-\xi_{N}} \\[8pt]
\frac{1}{\xi_{k}^*-\chi_{1}} & \cdots & \frac{1}{\xi_{k}^*-\chi_{k-1}} &
\frac{1}{\xi_k^*-\chi_k} & \frac{1}{\xi_{k}^*-\xi_{k+1}} & \cdots & \frac{1}{%
\xi_{k}^*-\xi_{N}} \\[8pt]
\frac{1}{\xi^*_{k+1}-\chi_1} & \cdots & \frac{1}{\xi^*_{k+1}-\chi_{k-1}} &
\frac{1}{\xi_{k+1}^*-\chi_{k}} & \frac{1}{\xi_{k+1}^*-\xi_{k+1}} & \cdots &
\frac{1}{\xi_{k+1}^*-\xi_{N}} \\
\vdots & \ddots & \vdots & \vdots & \vdots & \ddots & \vdots \\
\frac{1}{\xi^*_{N}-\chi_1} & \cdots & \frac{1}{\xi^*_{N}-\chi_{k-1}} & \frac{%
1}{\xi_{N}^*-\chi_{k}} & \frac{1}{\xi_N^*-\xi_{k+1}} & \cdots & \frac{1}{%
\xi_N^*-\xi_{N}} \\
&  &  &  &  &  &
\end{bmatrix}%
,
\end{equation*}
\begin{equation*}
M_k^{[3]}=%
\begin{bmatrix}
\frac{1}{\chi_{1}^*-\chi_{1}} & \cdots & \frac{1}{\chi_{1}^*-\chi_{k-1}} &
\frac{1}{\chi_{1}^*-\xi_{k}} & \frac{1}{\chi_{1}^*-\xi_{k+1}} & \cdots &
\frac{1}{\chi_{1}^*-\xi_{N}} \\
\vdots & \ddots & \vdots & \vdots & \vdots & \ddots & \vdots \\
\frac{1}{\chi_{k-1}^*-\chi_{1}} & \cdots & \frac{1}{\chi_{k-1}^*-\chi_{k-1}}
& \frac{1}{\chi_{k-1}^*-\xi_{k}} & \frac{1}{\chi_{k-1}^*-\xi_{k+1}} & \cdots
& \frac{1}{\chi_{k-1}^*-\xi_{N}} \\[8pt]
\frac{1}{\chi_{k}^*-\chi_{1}} & \cdots & \frac{1}{\chi_{k}^*-\chi_{k-1}} &
\frac{1}{\chi_k^*-\xi_k} & \frac{1}{\chi_{k}^*-\xi_{k+1}} & \cdots & \frac{1%
}{\chi_{k}^*-\xi_{N}} \\[8pt]
\frac{1}{\xi^*_{k+1}-\chi_1} & \cdots & \frac{1}{\xi^*_{k+1}-\chi_{k-1}} &
\frac{1}{\xi_{k+1}^*-\xi_{k}} & \frac{1}{\xi_{k+1}^*-\xi_{k+1}} & \cdots &
\frac{1}{\xi_{k+1}^*-\xi_{N}} \\
\vdots & \ddots & \vdots & \vdots & \vdots & \ddots & \vdots \\
\frac{1}{\xi^*_{N}-\chi_1} & \cdots & \frac{1}{\xi^*_{N}-\chi_{k-1}} & \frac{%
1}{\xi_{N}^*-\xi_{k}} & \frac{1}{\xi_N^*-\xi_{k+1}} & \cdots & \frac{1}{%
\xi_N^*-\xi_{N}} \\
&  &  &  &  &  &
\end{bmatrix}%
,
\end{equation*}
\begin{equation*}
M_k^{[4]}=%
\begin{bmatrix}
\frac{1}{\chi_{1}^*-\chi_{1}} & \cdots & \frac{1}{\chi_{1}^*-\chi_{k-1}} &
\frac{1}{\chi_{1}^*-\chi_{k}} & \frac{1}{\chi_{1}^*-\xi_{k+1}} & \cdots &
\frac{1}{\chi_{1}^*-\xi_{N}} \\
\vdots & \ddots & \vdots & \vdots & \vdots & \ddots & \vdots \\
\frac{1}{\chi_{k-1}^*-\chi_{1}} & \cdots & \frac{1}{\chi_{k-1}^*-\chi_{k-1}}
& \frac{1}{\chi_{k-1}^*-\chi_{k}} & \frac{1}{\chi_{k-1}^*-\xi_{k+1}} & \cdots
& \frac{1}{\chi_{k-1}^*-\xi_{N}} \\[8pt]
\frac{1}{\chi_{k}^*-\chi_{1}} & \cdots & \frac{1}{\chi_{k}^*-\chi_{k-1}} &
\frac{1}{\chi_k^*-\chi_k} & \frac{1}{\chi_{k}^*-\xi_{k+1}} & \cdots & \frac{1%
}{\chi_{k}^*-\xi_{N}} \\[8pt]
\frac{1}{\xi^*_{k+1}-\chi_1} & \cdots & \frac{1}{\xi^*_{k+1}-\chi_{k-1}} &
\frac{1}{\xi_{k+1}^*-\chi_{k}} & \frac{1}{\xi_{k+1}^*-\xi_{k+1}} & \cdots &
\frac{1}{\xi_{k+1}^*-\xi_{N}} \\
\vdots & \ddots & \vdots & \vdots & \vdots & \ddots & \vdots \\
\frac{1}{\xi^*_{N}-\chi_1} & \cdots & \frac{1}{\xi^*_{N}-\chi_{k-1}} & \frac{%
1}{\xi_{N}^*-\chi_{k}} & \frac{1}{\xi_N^*-\xi_{k+1}} & \cdots & \frac{1}{%
\xi_N^*-\xi_{N}} \\
&  &  &  &  &  &
\end{bmatrix}%
.
\end{equation*}
On the other hand, we have
\begin{equation*}
\begin{split}
\det(M_k^{[1]})&=%
\begin{vmatrix}
\frac{1}{\chi_{1}^*-\chi_{1}}\frac{\chi_1-\xi_k}{\chi_{1}^*-\xi_{k}} & \cdots
& \frac{1}{\chi_{1}^*-\chi_{k-1}}\frac{\chi_{k-1}-\xi_k}{\chi_{1}^*-\xi_{k}}
& \frac{1}{\chi_{1}^*-\xi_{k}} & \frac{1}{\chi_{1}^*-\xi_{k+1}}\frac{%
\xi_{k+1}-\xi_k}{\chi_{1}^*-\xi_{k}} & \cdots & \frac{1}{\chi_{1}^*-\xi_{N}}%
\frac{\xi_{N}-\xi_k}{\chi_{1}^*-\xi_{k}} \\
\vdots & \ddots & \vdots & \vdots & \vdots & \ddots & \vdots \\
\frac{1}{\chi_{k-1}^*-\chi_{1}}\frac{\chi_{1}-\xi_{k}}{\chi_{k-1}^*-\xi_{k}}
& \cdots & \frac{1}{\chi_{k-1}^*-\chi_{k-1}}\frac{\chi_{k-1}-\xi_{k}}{%
\chi_{k-1}^*-\xi_{k}} & \frac{1}{\chi_{k-1}^*-\xi_{k}} & \frac{1}{%
\chi_{k-1}^*-\xi_{k+1}}\frac{\xi_{k+1}-\xi_{k}}{\chi_{k-1}^*-\xi_{k}} &
\cdots & \frac{1}{\chi_{k-1}^*-\xi_{N}}\frac{\xi_{N}-\xi_{k}}{%
\chi_{k-1}^*-\xi_{k}} \\[8pt]
\frac{1}{\xi_{k}^*-\chi_{1}}\frac{\chi_1-\xi_k}{\xi_k^*-\xi_k} & \cdots &
\frac{1}{\xi_{k}^*-\chi_{k-1}}\frac{\chi_{k-1}-\xi_k}{\xi_k^*-\xi_k} & \frac{%
1}{\xi_k^*-\xi_k} & \frac{1}{\xi_{k}^*-\xi_{k+1}}\frac{\xi_{k+1}-\xi_k}{%
\xi_k^*-\xi_k} & \cdots & \frac{1}{\xi_{k}^*-\xi_{N}}\frac{\xi_{N}-\xi_k}{%
\xi_k^*-\xi_k} \\[8pt]
\frac{1}{\xi^*_{k+1}-\chi_1}\frac{\chi_1-\xi_k}{\xi_{k+1}^*-\xi_{k}} & \cdots
& \frac{1}{\xi^*_{k+1}-\chi_{k-1}}\frac{\chi_{k-1}-\xi_k}{\xi_{k+1}^*-\xi_{k}%
} & \frac{1}{\xi_{k+1}^*-\xi_{k}} & \frac{1}{\xi_{k+1}^*-\xi_{k+1}} \frac{%
\xi_{k+1}-\xi_k}{\xi_{k+1}^*-\xi_{k}} & \cdots & \frac{1}{\xi_{k+1}^*-\xi_{N}%
}\frac{\xi_{N}-\xi_k}{\xi_{k+1}^*-\xi_{k}} \\
\vdots & \ddots & \vdots & \vdots & \vdots & \ddots & \vdots \\
\frac{1}{\xi^*_{N}-\chi_1}\frac{\chi_1-\xi_k}{\xi_{N}^*-\xi_{k}} & \cdots &
\frac{1}{\xi^*_{N}-\chi_{k-1}} \frac{\chi_{k-1}-\xi_k}{\xi_{N}^*-\xi_{k}} &
\frac{1}{\xi_{N}^*-\xi_{k}} & \frac{1}{\xi_N^*-\xi_{k+1}}\frac{%
\xi_{k+1}-\xi_k}{\xi_{N}^*-\xi_{k}} & \cdots & \frac{1}{\xi_N^*-\xi_{N}}%
\frac{\xi_{N}-\xi_k}{\xi_{N}^*-\xi_{k}} \\
&  &  &  &  &  &
\end{vmatrix}
\\
&= \frac{1}{\xi_k^*-\xi_k}\left(\prod_{l=1}^{k-1}\frac{\chi_l-\xi_k}{%
\chi_l^*-\xi_k}\right)\left(\prod_{l=k+1}^{N}\frac{\xi_l-\xi_k}{\xi_l^*-\xi_k%
}\right) \Delta_k^{[1]}
\end{split}%
\end{equation*}
where
\begin{equation*}
\Delta_k^{[1]}=%
\begin{vmatrix}
\frac{1}{\chi_{1}^*-\chi_{1}} & \cdots & \frac{1}{\chi_{1}^*-\chi_{k-1}} & 1
& \frac{1}{\chi_{1}^*-\xi_{k+1}} & \cdots & \frac{1}{\chi_{1}^*-\xi_{N}} \\
\vdots & \ddots & \vdots & \vdots & \vdots & \ddots & \vdots \\
\frac{1}{\chi_{k-1}^*-\chi_{1}} & \cdots & \frac{1}{\chi_{k-1}^*-\chi_{k-1}}
& 1 & \frac{1}{\chi_{k-1}^*-\xi_{k+1}} & \cdots & \frac{1}{%
\chi_{k-1}^*-\xi_{N}} \\[8pt]
\frac{1}{\xi_{k}^*-\chi_{1}} & \cdots & \frac{1}{\xi_{k}^*-\chi_{k-1}} & 1 &
\frac{1}{\xi_{k}^*-\xi_{k+1}} & \cdots & \frac{1}{\xi_{k}^*-\xi_{N}} \\[8pt]
\frac{1}{\xi^*_{k+1}-\chi_1} & \cdots & \frac{1}{\xi^*_{k+1}-\chi_{k-1}} & 1
& \frac{1}{\xi_{k+1}^*-\xi_{k+1}} & \cdots & \frac{1}{\xi_{k+1}^*-\xi_{N}}
\\
\vdots & \ddots & \vdots & \vdots & \vdots & \ddots & \vdots \\
\frac{1}{\xi^*_{N}-\chi_1} & \cdots & \frac{1}{\xi^*_{N}-\chi_{k-1}} & 1 &
\frac{1}{\xi_N^*-\xi_{k+1}} & \cdots & \frac{1}{\xi_N^*-\xi_{N}} \\
&  &  &  &  &  &
\end{vmatrix}
\\
\end{equation*}
moreover
\begin{equation*}
\begin{split}
\Delta_k^{[1]}&=%
\begin{vmatrix}
\frac{1}{\chi_{1}^*-\chi_{1}}\frac{\xi_{k}^*-\chi_{1}^*}{\xi_{k}^*-\chi_{1}}
& \cdots & \frac{1}{\chi_{1}^*-\chi_{k-1}} \frac{\xi_{k}^*-\chi_{1}^*}{%
\xi_{k}^*-\chi_{k-1}} & 0 & \frac{1}{\chi_{1}^*-\xi_{k+1}}\frac{%
\xi_{k}^*-\chi_{1}^*}{\xi_{k}^*-\xi_{k+1}} & \cdots & \frac{1}{%
\chi_{1}^*-\xi_{N}}\frac{\xi_{k}^*-\chi_{1}^*}{\xi_{k}^*-\xi_{N}} \\
\vdots & \ddots & \vdots & \vdots & \vdots & \ddots & \vdots \\
\frac{1}{\chi_{k-1}^*-\chi_{1}}\frac{\xi_{k}^*-\chi_{k-1}^*}{%
\xi_{k}^*-\chi_{1}} & \cdots & \frac{1}{\chi_{k-1}^*-\chi_{k-1}} \frac{%
\xi_{k}^*-\chi_{k-1}^*}{\xi_{k}^*-\chi_{k-1}} & 0 & \frac{1}{%
\chi_{k-1}^*-\xi_{k+1}}\frac{\xi_{k}^*-\chi_{k-1}^*}{\xi_{k}^*-\xi_{k+1}} &
\cdots & \frac{1}{\chi_{k-1}^*-\xi_{N}}\frac{\xi_{k}^*-\chi_{k-1}^*}{%
\xi_{k}^*-\xi_{N}} \\[8pt]
\frac{1}{\xi_{k}^*-\chi_{1}} & \cdots & \frac{1}{\xi_{k}^*-\chi_{k-1}} & 1 &
\frac{1}{\xi_{k}^*-\xi_{k+1}} & \cdots & \frac{1}{\xi_{k}^*-\xi_{N}} \\[8pt]
\frac{1}{\xi^*_{k+1}-\chi_1}\frac{\xi_{k}^*-\xi_{k+1}^*}{\xi_{k}^*-\chi_{1}}
& \cdots & \frac{1}{\xi^*_{k+1}-\chi_{k-1}} \frac{\xi_{k}^*-\xi_{k+1}^*}{%
\xi_{k}^*-\chi_{k-1}} & 0 & \frac{1}{\xi_{k+1}^*-\xi_{k+1}}\frac{%
\xi_{k}^*-\xi_{k+1}^*}{\xi_{k}^*-\xi_{k+1}} & \cdots & \frac{1}{%
\xi_{k+1}^*-\xi_{N}}\frac{\xi_{k}^*-\xi_{k+1}^*}{\xi_{k}^*-\xi_{N}} \\
\vdots & \ddots & \vdots & \vdots & \vdots & \ddots & \vdots \\
\frac{1}{\xi^*_{N}-\chi_1}\frac{\xi_{k}^*-\xi_{N}^*}{\xi_{k}^*-\chi_{1}} &
\cdots & \frac{1}{\xi^*_{N}-\chi_{k-1}} \frac{\xi_{k}^*-\xi_{N}^*}{%
\xi_{k}^*-\chi_{k-1}} & 0 & \frac{1}{\xi_N^*-\xi_{k+1}}\frac{%
\xi_{k}^*-\xi_{N}^*}{\xi_{k}^*-\xi_{k+1}} & \cdots & \frac{1}{\xi_N^*-\xi_{N}%
}\frac{\xi_{k}^*-\xi_{N}^*}{\xi_{k}^*-\xi_{N}} \\
&  &  &  &  &  &
\end{vmatrix}
\\
&=\left(\prod_{l=1}^{k-1}\frac{\xi_{k}^*-\chi_{l}^*}{\xi_{k}^*-\chi_{l}}%
\right)\left(\prod_{l=k+1}^{N}\frac{\xi_{k}^*-\xi_{l}^*}{\xi_{k}^*-\xi_{l}}%
\right)
C(\chi_1^*,\chi_2^*,\cdots,\chi_{k-1}^*)C(\xi_{k+1}^*,\xi_{k+2}^*,\cdots,%
\xi_{N}^*)
\end{split}%
\end{equation*}
where $C(\cdot,\cdot,\cdots,\cdot)$ represents the determinant of a Cauchy
matrix. Thus, we have
\begin{equation*}
\det(M_k^{[1]})=\frac{1}{\xi_k^*-\xi_k}\left(\prod_{l=1}^{k-1}\frac{%
\chi_l-\xi_k}{\chi_l^*-\xi_k}\frac{\xi_{k}^*-\chi_{l}^*}{\xi_{k}^*-\chi_{l}}%
\right) \left(\prod_{l=k+1}^{N}\frac{\xi_l-\xi_k}{\xi_l^*-\xi_k}\frac{%
\xi_{k}^*-\xi_{l}^*}{\xi_{k}^*-\xi_{l}}\right)
C(\chi_1^*,\chi_2^*,\cdots,\chi_{k-1}^*)C(\xi_{k+1}^*,\xi_{k+2}^*,\cdots,%
\xi_{N}^*).
\end{equation*}
Similarly, we have
\begin{equation*}
\begin{split}
\det(M_k^{[2]})&=\frac{1}{\xi_k^*-\chi_k}\left(\prod_{l=1}^{k-1}\frac{%
\chi_l-\chi_k}{\chi_l^*-\chi_k}\frac{\xi_{k}^*-\chi_{l}^*}{\xi_{k}^*-\chi_{l}%
}\right) \left(\prod_{l=k+1}^{N}\frac{\xi_l-\chi_k}{\xi_l^*-\chi_k}\frac{%
\xi_{k}^*-\xi_{l}^*}{\xi_{k}^*-\xi_{l}}\right)
C(\chi_1^*,\chi_2^*,\cdots,\chi_{k-1}^*)C(\xi_{k+1}^*,\xi_{k+2}^*,\cdots,%
\xi_{N}^*), \\
\det(M_k^{[3]})&=\frac{1}{\chi_k^*-\xi_k}\left(\prod_{l=1}^{k-1}\frac{%
\chi_l-\xi_k}{\chi_l^*-\xi_k}\frac{\chi_{k}^*-\chi_{l}^*}{\chi_{k}^*-\chi_{l}%
}\right) \left(\prod_{l=k+1}^{N}\frac{\xi_l-\xi_k}{\xi_l^*-\xi_k}\frac{%
\chi_{k}^*-\xi_{l}^*}{\chi_{k}^*-\xi_{l}}\right)
C(\chi_1^*,\chi_2^*,\cdots,\chi_{k-1}^*)C(\xi_{k+1}^*,\xi_{k+2}^*,\cdots,%
\xi_{N}^*), \\
\det(M_k^{[4]})&=\frac{1}{\chi_k^*-\chi_k}\left(\prod_{l=1}^{k-1}\frac{%
\chi_l-\chi_k}{\chi_l^*-\chi_k}\frac{\chi_{k}^*-\chi_{l}^*}{%
\chi_{k}^*-\chi_{l}}\right) \left(\prod_{l=k+1}^{N}\frac{\xi_l-\chi_k}{%
\xi_l^*-\chi_k}\frac{\chi_{k}^*-\xi_{l}^*}{\chi_{k}^*-\xi_{l}}\right)
C(\chi_1^*,\chi_2^*,\cdots,\chi_{k-1}^*)C(\xi_{k+1}^*,\xi_{k+2}^*,\cdots,%
\xi_{N}^*).
\end{split}%
\end{equation*}
Moreover, we have
\begin{equation*}
\begin{split}
\det(M_k)&=C(\chi_1^*,\chi_2^*,\cdots,\chi_{k-1}^*)C(\xi_{k+1}^*,%
\xi_{k+2}^*,\cdots,\xi_{N}^*)\left[\frac{1}{\xi_k^*-\xi_k}%
\left(\prod_{l=1}^{k-1}\left|\frac{\chi_l-\xi_k}{\chi_l^*-\xi_k}%
\right|^2\right) \left(\prod_{l=k+1}^{N}\left|\frac{\xi_l-\xi_k}{%
\xi_l^*-\xi_k}\right|^2\right)\mathrm{e}^{2(\theta_k^*+\theta_k)} \right. \\
&+\frac{1}{\xi_k^*-\chi_k}\left(\prod_{l=1}^{k-1}\frac{\chi_l-\chi_k}{%
\chi_l^*-\chi_k}\frac{\xi_{k}^*-\chi_{l}^*}{\xi_{k}^*-\chi_{l}}\right)
\left(\prod_{l=k+1}^{N}\frac{\xi_l-\chi_k}{\xi_l^*-\chi_k}\frac{%
\xi_{k}^*-\xi_{l}^*}{\xi_{k}^*-\xi_{l}}\right)\mathrm{e}^{2\theta_k^*} \\
&+\frac{1}{\chi_k^*-\xi_k}\left(\prod_{l=1}^{k-1}\frac{\chi_l-\xi_k}{%
\chi_l^*-\xi_k}\frac{\chi_{k}^*-\chi_{l}^*}{\chi_{k}^*-\chi_{l}}\right)
\left(\prod_{l=k+1}^{N}\frac{\xi_l-\xi_k}{\xi_l^*-\xi_k}\frac{%
\chi_{k}^*-\xi_{l}^*}{\chi_{k}^*-\xi_{l}}\right)\mathrm{e}^{2\theta_k} \\
&\left.+\frac{1}{\chi_k^*-\chi_k}\left(\prod_{l=1}^{k-1}\left|\frac{%
\chi_l-\chi_k}{\chi_l^*-\chi_k}\right|^2\right) \left(\prod_{l=k+1}^{N}\left|%
\frac{\xi_l-\chi_k}{\xi_l^*-\chi_k}\right|^2\right)\right].
\end{split}%
\end{equation*}
Similar procedure as above, we have
\begin{equation*}
\begin{split}
\det(G_k)=&\left(\prod_{l=1}^{k-1}\frac{\chi_l^*+\gamma}{\chi_l+\gamma}%
\right)\left(\prod_{l=k+1}^{N}\frac{\xi_l^*+\gamma}{\xi_l+\gamma}\right) %
\left[\frac{\xi_k^*+\gamma}{\xi_k+\gamma}\det(M_k^{[1]})\mathrm{e}%
^{2(\theta_k^*+\theta_k)}+ \frac{\xi_k^*+\gamma}{\chi_k+\gamma}%
\det(M_k^{[2]})\mathrm{e}^{2\theta_k^*}\right. \\
&\left.+\frac{\chi_k^*+\gamma}{\xi_k+\gamma}\det(M_k^{[3]})\mathrm{e}%
^{2\theta_k}+\frac{\chi_k^*+\gamma}{\chi_k+\gamma}\det(M_k^{[4]})\right].
\end{split}%
\end{equation*}
Finally, as $s\rightarrow +\infty $, along the trajectory $y-v_{k}s=\mathrm{%
const}$, we have
\begin{equation*}
q[N]=\frac{\beta }{2}q_{k}^{+}\mathrm{e}^{\mathrm{i}\theta }+O(\mathrm{e}%
^{-c|s|}),
\end{equation*}%
where $\theta _{k,R}^{+}$, $\theta _{k,I}^{+}$ are given in equations %
\eqref{para2}.

For the general case $y-vs=\mathrm{const},\,\,v\neq v_{k}$ $%
(k=1,2,\cdots ,N)$, if $v<v_{1}$ then $q[N]=\frac{\beta }{2}\Theta _{1}^{+}%
\mathrm{e}^{\mathrm{i}\theta -2\mathrm{i}\varphi _{1,I}}+O(\mathrm{e}%
^{-c|s|});$ if $v_{m-1}<v<v_{m}$$(m=2,3,\cdots ,l)$, then $q[N]=\frac{\beta
}{2}\Theta _{m}^{+}\mathrm{e}^{\mathrm{i}\theta -2\mathrm{i}\varphi
_{m,I}}+O(\mathrm{e}^{-c|s|});$ if $v_{l}<v<v_{N}$ then $q[N]=\frac{\beta }{2%
}\Theta _{l}^{+}\mathrm{e}^{\mathrm{i}\theta +2\mathrm{i}\varphi _{l,I}}+O(%
\mathrm{e}^{-c|s|});$ if $v_{m+1}<v<v_{m}$$(m=l+1,l+2,\cdots ,N-1)$, then $%
q[N]=\frac{\beta }{2}\Theta _{m}^{+}\mathrm{e}^{\mathrm{i}\theta +2\mathrm{i}%
\varphi _{m,I}}+O(\mathrm{e}^{-c|s|});$ if $v_{l+1}<v$ then $q[N]=\frac{%
\beta }{2}\Theta _{l+1}^{+}\mathrm{e}^{\mathrm{i}\theta -2\mathrm{i}\varphi
_{l+1,I}}+O(\mathrm{e}^{-c|s|}).$ Thus we have the asymptotic behavior %
\eqref{asym2}.
By the same procedure as above, we can obtain the asymptotic behavior %
\eqref{asym1} when $s\rightarrow -\infty .$ So we complete the proof. $%
\square$
\section*{Appendix C: Modulational instability analysis for plane wave solution}
The simplest exact solution to the CSP equation \eqref{CSP} is the plane
wave-a constant amplitude, exponential wavetrain,
\begin{equation}  \label{plane}
q_{0}=\frac{\beta}{2} {\mathrm{e}^{-\mathrm{i} \left({\frac {2}{\gamma}}%
x-\omega t \right) }},\,\, \omega=\frac{1}{4}\left({\frac {{\beta}^{2}}{{%
\gamma}^{2 }}}-2 \right) \gamma
\end{equation}
where $\beta$, $\gamma$ are real constants. The linearized stability of the
plane wave is easily obtained from Fourier analysis \cite{MI1}. It proves
most convenient to introduce the disturbance quantities $\tilde{q}$ as
multiplicative perturbations to the plane wave
\begin{equation}  \label{disturb}
q=\left(\frac{\beta}{2}+\tilde{q}\right) {\mathrm{e}^{-\mathrm{i} \left({%
\frac {2}{\gamma}}x-\omega t \right) }}
\end{equation}
since this results in a convenient simplification upon linearization.
Keeping only terms linear in $\tilde{q}$ after direct substitution of %
\eqref{disturb} into the CSP equation (2), the linearized disturbance
equations become
\begin{equation}  \label{liner}
\tilde{q}_{{t}}+\frac{{\beta}^{2}}{8}\tilde{q}^{*}_{{x}}+\frac{1}{4}\left( {%
\beta}^{2}+{\gamma}^{2} \right) \tilde{q}_{{x}}+\frac{\mathrm{i}\gamma}{2}%
\tilde{q}_{{\mathit{xt}}}+\frac{\mathrm{i}\gamma{\beta}^{ 2}}{16}\tilde{q}_{{%
\mathit{xx}}}-{\frac{\mathrm{i}{\beta}^{2}}{4\gamma} \left( \tilde{q}^*+%
\tilde{q} \right)=0. }
\end{equation}
Because of the
conjugates in \eqref{liner}, the eigenfunctions are most conveniently
expressed as linear combinations of pure Fourier modes,
\begin{equation}  \label{fourier}
\tilde{q}=f_+\mathrm{e}^{\mathrm{i}\kappa(x+\Omega t)}+f_-^*\mathrm{e}^{-%
\mathrm{i}\kappa(x+\Omega^* t)}.
\end{equation}
These eigenmodes are parameterized by the real wavenumber $\kappa$ of the
disturbance and the complex phase velocity $\Omega$, where a positive
imaginary part indicates a pure temporal growth mode of instability in
positive time. Substitution into the linearized PDEs \eqref{disturb} and
collection of resonant terms results in four linear homogeneous equations
for the Fourier amplitudes $f_{\pm}$,
\begin{equation}  \label{resonant}
\begin{bmatrix}
8{k}^{2}{\gamma}^{2}{\Omega}+ \left(4+{k}^{2}{\gamma}^{2} \right) {\beta}%
^{2} -4\gamma\kappa \left( 4\,\mathit{\Omega}+{\gamma}^{2}+{\beta}^{2}
\right) & 2{\beta}^{2}(2-\gamma\kappa) \\[10pt]
-2{\beta}^{2}(2+\gamma\kappa) & -8{k}^{2}{\gamma}^{2}{\Omega}- \left(4+{k}%
^{2}{\gamma}^{2} \right) {\beta}^{2} -4\gamma\kappa \left( 4\,\mathit{\Omega}%
+{\gamma}^{2}+{\beta}^{2} \right) \\
&
\end{bmatrix}%
\begin{bmatrix}
f_+ \\[10pt]
f_- \\
\end{bmatrix}%
=0.
\end{equation}
Solvability for this system requires that the determinant of the matrix of
coefficients vanish--this determines the dispersion relation for linearized
disturbances
\begin{equation*}
16\,{\gamma}^{2}{\kappa}^{2} \left( 4\,{\Omega}+{\gamma}^{2}+{\beta}^{2 }
\right) ^{2}+16\,{\beta}^{4}- \left[ 8\,{\kappa}^{2}{\gamma}^{2}{\Omega}+
\left( 4+{\kappa}^{2}{\gamma}^{2} \right) {\beta}^{2} \right] ^{2} -4\,{%
\gamma}^{2}{\beta}^{4}{\kappa}^{2}=0.
\end{equation*}
Ones can readily obtain that two roots for above square equation:
\begin{equation*}
\Omega={\frac {(8-{\beta}^{2}{\kappa}^{2}){\gamma}^{2}+4\,{\beta }%
^{2}\pm4\gamma\sqrt {{\gamma}^{2}{\kappa}^{2}({\beta}^{2}+{\gamma}^{2})-4{%
\beta}^ {2}}}{8({\kappa}^{2}{\gamma}^{2}-4)}}.
\end{equation*}
So when ${\kappa}^{2}<\frac{4{\beta}^{2}}{\gamma^{2}({\beta}^{2}+{\gamma}^{2})}$,
those roots with nonzero imaginary part correspond to linearly unstable
modes, with growth rate $\kappa|\mathrm{Im}(\Omega)|=\gamma\kappa\sqrt {4{\beta}^ {2}-{%
\gamma}^{2}{\kappa}^{2}({\beta}^{2}+{\gamma}^{2})}/[2|{\kappa}^{2}{\gamma}%
^{2}-4|]$. Then the baseband MI yields the rogue wave solution \cite{Baronio}.

%If $\beta^2>\gamma^2$, then $|\mathrm{Im}(\Omega)|_{max}=\frac{1}{%
%8}(\beta^2+\gamma^2)$ at $\kappa=\pm\frac{2}{\gamma}\sqrt{\,{\frac {{\beta}%
%^{2}-{\gamma}^{2}}{{\beta}^{2} +{\gamma}^{2}}}}$; if $\beta^2\leq\gamma^2$,
%then $|\mathrm{Im}(\Omega)|_{max}=\frac{|\beta\gamma|}{4}$ at $\kappa=0$.


\begin{thebibliography}{99}
\bibitem{Hasegawa} A.~Hasegawa, Y.~Kodama, 1995 \textit{Solitons in Optical
Communications}, Oxford University Press.

\bibitem{Agrawal} G.~P.~Agrawal, 2001  \textit{Nonlinear Fiber Optics}, Academic, San Diego.

\bibitem{MI} T.~B. Benjamin, J.~E. Feir, The disintegration of wave
trains on deep water Part 1. Theory, J. Fluid Mech. 27 (1967) 417.

\bibitem{Zakh} V.~E.~Zakharov, L.~A.~Ostrovsky, Modulation instability:
the beginning, Physica D, {238} (2009) 540-548.

\bibitem{AB} N. Akhmediev, V.~I. Korneev, Modulation instability and
periodic solutions of the nonlinear schr\"odinger equation, Theor. Math.
Phys. (USSR) 69(2) (1986) 1089.

\bibitem{Pregr} D.~H. Peregrine, Water waves, nonlinear Schr\"odinger
equations and their solutions, J. Aust. Math. Soc. Ser. B, Appl. Math 25
(1983) 16.

\bibitem{K-M} E.~A. Kuznetsov, Solitons in a parametrically unstable
plasma, Sov. Phys.---Dokl. 22 (1977) 507 (Engl. Transl.)

\bibitem{Dudley} J.~M. Dudley, F. Dias, M. Erkintalo, G. Genty, 
Instabilities, breathers and rogue waves in optics, Nat. Photonics 8 (2014)
755.

\bibitem{Kibler} B. Kibler, J. Fatome, C. Finot, G. Millot, F. Dias, G.
Genty, N. Akhmediev, J.~M. Dudley, The Peregrine soliton in nonlinear
fibre optics, Nat. Phys. 6 (2010) 790.

\bibitem{Kibler1} Kibler B, Fatome J, Finot C, et al.Observation of
Kuznetsov-Ma soliton dynamics in optical fibre. Scientific reports, 2012,
2.

\bibitem{Chabchoub} A. Chabchoub, N.~P. Hoffmann, N. Akhmediev, Rogue
Wave Observation in a Water Wave Tank, Phys. Rev. Lett. 106 (2011) 204502.

\bibitem{Bailung} H. Bailung, S.~K. Sharma, Y. Nakamura, Observation of
Peregrine Solitons in a Multicomponent Plasma with Negative Ions, Phys.
Rev. Lett. 107 (2011) 255005.

\bibitem{Roth} J.~E. Rothenberg, Space-time focusing: breakdown of the
slowly varying envelope approximation in the self-focusing of femtosecond
pulses, Optim. Lett. 17 (1992) 1340-1342.

\bibitem{Sch} T. Sch\"afer, C.~E. Wayne, Propagation of ultra-short
optical pulses in cubic nonlinear media, Physica D 196 (2004)
90-105.

\bibitem{Sko} S.~A. Skobelev, D.~V. Kartasholv, A.V. Kim, Solitary-wave
solutions for few-cycle optical pulses, Phys. Rev. Lett. 99 (2007) 203902.

\bibitem{Kim} A.~V. Kim, S.~A. Skobelev, D. Anderson, T. Hansson, M. Lisak,
Extreme nonlinear optics in a Kerr medium: Exact soliton solutions for a few
cycles, Phys. Rev. A 77 (2008) 043823.

\bibitem{Amir} S. Amiranashvili, A.~G. Vladimirov, U. Bandelow,
Few-optical-cycle solitons and pulse self-compression in a Kerr medium,
Phys. Rev. A 77 (2008) 063821.

\bibitem{Feng2} B.-F. Feng, Complex short pulse and coupled complex
short pulse equations, Physica D 297 (2015) 62-75.

\bibitem{Robelo} M.~L. Robelo, On equations which describe
pseudospherical surfaces, Stud. Appl. Math. 81 (1989) 221-248.

\bibitem{Beals} R. Beals, M. Rabelo, K. Tenenblat, \emph{B\"acklund
transformations and inverse scattering solutions for some pseudospherical
surface equations}, Stud. Appl. Math. 81 (1989) 125-151.

\bibitem{Sako} A. Sakovich, S. Sakovich, The short pulse equation is
integrable, J. Phys. Soc. Japan 74 (2005) 239-241.

\bibitem{Brun} J.~C. Brunelli, The short pulse hierarchy, J. Math.
Phys. 46 (2005) 123507.

\bibitem{Brun1} J.~C. Brunelli, The bi-Hamiltonian structure of the
short pulse equation, Phys. Lett. A 353 (2006) 475-478.

\bibitem{Sako1} A. Sakovich, S. Sakovich, Solitary wave solutions of
the short pulse equation, J. Phys. A 39 (2006) L361-367.

\bibitem{Kuet} V.~K. Kuetche, T.~B. Bouetou, T.~C. Kofane, On two-loop
soliton solution of the Sch\"afer-Wayne short-pulse equation using Hirotas
method and Hodnett-Moloney approach, J. Phys. Soc. Japan 76 (2007) 024004.

\bibitem{Parkes} E. Parkes, Some periodic and solitary
tralvelling-wave solutions of the short pulse equation, Chaos, Solitons
and Fractals 38 (2008) 154--159.

\bibitem{Matsuno} Y. Matsuno, Multisoliton and multibreather solutions
of the short pulse model equation, J. Phys. Soc. Japan 76 (2007) 084003.

\bibitem{Matsuno1} Y. Matsuno, Periodic solutions of the short pulse
model equation, J. Math. Phys. 49 (2008) 073508.

\bibitem{Hirota} R. Hirota, 2004 \textit{The Direct Method in Soliton Theory},
Cambridge University Press.

\bibitem{Feng} B.-F. Feng, K. Maruno, Y. Ohta, Integrable discretization of
the short pulse equation, J. Phys. A 43 (2010) 085203.

\bibitem{Feng1} B.-F. Feng, J. Inoguchi, K. Kajiwara, K. Maruno, Y. Ohta,
Discrete integrable systems and hodograph transformations arising from
motions of discrete plane curves, J. Phys. A 44 (2011) 395201.

\bibitem{Yariv} A. Yariv, P. Yeh, \textit{Optical Waves in Crystals:
Propagation and Control of Laser Radiation}, Wiley-Interscience, 1983.

\bibitem{shen} S. Shen, B.-F. Feng, and Y. Ohta, From the Real and
Complex Coupled Dispersionless Equations to the Real and Complex Short Pulse
Equations, Stud. Appl. Math (2015) DOI: 10.1111/sapm.12092.

\bibitem{cd} K. Konno and H. Oono, New coupled dispersionless
equations, J. Phys. Soc. Jpn. 63 (1994) 377--378.

\bibitem{WKI} M. Wadati, K. Konno and Y. Ichikawa, New integrable nonlinear evolution equations, J. Phys. Soc. Jpn. 47 (1979) 1698--1700.

\bibitem{Qiao} Z. Qiao, 2002 \textit{Finite-dimensional Integrable System and Nonlinear Evolution Equations}, Chinese National Higher Education Press, Beijing, PR China.

\bibitem{Zimerman} G.~S. Franca, J.~F. Gomes, and A.~H. Zimerman, The higher grading structure of the WKI hierarchy and the two-component short pulse equation, J. High Energy Phys. 8 (2012) 120.

\bibitem{breaking} A. Galchenko, A. Babanin, D. Chalikov, I. Young, T. Hsu,
Modulational instabilities and breaking strength for deep-water wave
groups, Journal of Physical Oceanography 40 (2010) 2313-2324.


\bibitem{Onorato} M.~Onorato,S.~Residori, U.~Bortolozzo, et al. Rogue
waves and their generating mechanisms in different physical contexts.
Physics Reports  528 (2013) 47-89.

\bibitem{Zakharov} A.~A. Gelash and V.~E. Zakharov, Superregular
solitonic solutions: a novel scenario for the nonlinear stage of modulation
instability, Nonlinearity 27 (2014) R1-R39

\bibitem{Matveev} V.~B. Matveev and M.~A. Salle, 1991 \textit{Darboux
transformations and solitons}, Springer, Berlin.

\bibitem{Guo1} B.~Guo, L.~Ling, Q.~P.~Liu,Nonlinear Schr\"odinger
equation: generalized Darboux transformation and rogue wave solutions,
Phys. Rev. E 85 (2012) 026607.

\bibitem{Guo2}  B.~Guo, L.~Ling, Q.~P.~Liu, High-Order Solutions and
Generalized Darboux Transformations of Derivative Nonlinear Schr\"odinger
Equations, Stud. Appl. Math. 130 (2013) 317-344.

\bibitem{loop-group} C.~L.~Terng, K.~Uhlenbeck, B\"acklund
transformations and loop group actions, Commun. Pure Appl. Math. 53
(2000) 1-75.

\bibitem{wangxin} X.~Wang, Y. Li, F. Huang and Y. Chen, Rogue wave solutions of AB system.,
Commun. Nonlinear Sci. Numer. Simulat., 20 (2015) 434-442.

\bibitem{MI1} M.~G. Forest, D.~W. McLaughlin, D. J. Muraki and O. C. Wright,
Nonfocusing Instabilities in Coupled, Integrable Nonlinear
Schr\"odinger pdes, J. Nonlinear Sci. 10 (2000) 291-331.

\bibitem{algebraic} E. Belokolos, A. Bobenko, V. Enol'skij, A. Its and V.~B.
Matveev, 1994 \textit{Algebro-geometric approach to nonlinear integrable equations.}
Springer.

\bibitem{Ling} L.~Ling, L.~C.~Zhao and B.~Guo, Darboux transformation and
multi-dark soliton for N-component nonlinear Schr\"odinger equations,
Nonlinearity 28 (2015) 3243--3261.

\bibitem{Baronio} F.~Baronio, M.~Conforti, A.~Degasperis et al., Vector rogue waves and baseband modulation instability in the defocusing regime. Phys. Rev. Lett.  113 (2014) 034101.
 \bibitem{LingFengdCSP}  B.-F. Feng, L. Ling, Z. Zhu,  A defocusing complex short pulse equation and its multi-dark soliton solution by Darboux transformation, arXiv:1511.00945 [nlin.SI].
\end{thebibliography}
\end{document}